\documentclass[useAMs,a4paper]{mn2e}
\usepackage{savesym}
\usepackage{graphicx}
\expandafter\let\csname equation*\endcsname\relax
  \expandafter\let\csname endequation*\endcsname\relax 
\usepackage{subfig}
\usepackage{amsmath}

\usepackage{hyperref}

\usepackage{amssymb}
\usepackage{verbatim}
\usepackage[yyyymmdd,hhmmss]{datetime}
\usepackage{array}
\usepackage{times}
\usepackage[total={17.8cm,24.0cm},centering]{geometry} 
\usepackage{color}

\newcommand{\beq}{\begin{equation}}
\newcommand{\eeq}{\end{equation}}

\newcommand\W {{W^r_\phi}}

\title[Extending the theory of propagating fluctuations]{Extending the theory of propagating fluctuations:  the first fully relativistic treatment and  analytical Fourier-Green's functions     }
\author [Andrew Mummery]{Andrew Mummery\thanks{E-mail:
andrew.mummery@physics.ox.ac.uk}
\\
Oxford Theoretical Physics, Beecroft Building,  Clarendon Laboratory, Parks Road, Oxford, OX1 3PU, United Kingdom}
\begin{document}

\date{}

\pagerange{\pageref{firstpage}--\pageref{lastpage}} \pubyear{2023}

\maketitle

\label{firstpage}

\begin{abstract} 
The aperiodic variability ubiquitously observed from accreting black hole X-ray binary systems is generally analysed within the framework of the so-called ``theory of propagating fluctuations''.  In this paper we derive the Fourier transforms of the Green's function solutions of the thin disc equations. These solutions suffice to describe all possible solutions through standard convolution techniques.   Solutions are found for both Newtonian discs and general relativistic solutions with a vanishing ISCO stress.  We use this new relativistic  theory to highlight the Kerr black hole spin dependence of a number of observable variability properties of black hole discs. The phase lags, coherence, and power density spectra of Kerr discs are shown to be strong functions of black hole spin.  Observations of the aperiodic variability of black hole accretion sources may now, at least in principle,  offer a new avenue to directly constrain black hole spins.  
\end{abstract}

\noindent

\section{Introduction}
Accreting black hole  systems generally exhibits pronounced temporal variability, a result of their fundamentally turbulent nature (Balbus \& Hawley 1991). The power spectral densities of the luminosity emergent from accretion discs reveal fluctuations whose root mean square variation on short timescales is found to be linearly proportional to its mean evolving over longer timescales, a property of the log-normal distribution (Uttley \& McHardy 2001, Uttley et al. 2005). 

Different accreting sources display a remarkable similarity in their variability properties,  despite the vast range of both length and time scales involved,  across a broad population of sources. In any individual source variability is observed over many temporal orders of magnitude: from timescales as rapid as the local dynamical timescale of the innermost disc edge up to many orders of magnitude longer than any physical process which is involved in the direct production of the disc luminosity. The length scales spanned by different sources which show similar variability structure is vast: ranging from the compact discs in Galactic X-ray binaries (e.g., Gleissner et al. 2004), to the very large discs in active galactic nuclei (e.g., Vaughan et al. 2011). 

This behaviour has a robust observational grounding, and has been confirmed with observations at various different frequencies in black hole accretion disc sources, including both AGN in X-ray (Gaskell 2004; Vaughan et al. 2011), AGN in optical (Lyutyi \& Oknyanskii 1987), and X-ray binaries at both X-ray (Gleissner et al. 2004) and optical (Gandhi 2009) frequencies. We also note that non black hole disc sources, e.g., cataclysmic variables (Scaringi et al. 2012), and young stellar objects (Scaringi et al. 2015) also show the same variability structure.

The typical variability structure of an accreting system is the following. The observed power density spectrum has a broad (aperiodic) component which is generally well described by a twice-broken power-law. In the case of some black hole binaries, these power density spectra display a narrow feature peaking in the range $\sim 0.1$--$10$ Hz's, which are known as (type C) quasi-periodic oscillations (QPOs). We will discuss only the aperiodic variability of black hole sources in this paper, and not QPOs, which will have different physical origins. Observed light curves of the same sources taken in different energy bands are found to correlate with each other, but  ``harder'' (higher energy)  X-ray variability usually lags ``softer'' (lower energy) X-ray variability (Priedhorsky et al. 1979; Nolan et al. 1981). The magnitude of the hard-soft time-lag depends on Fourier frequency, but  is typically found to be of order of 1 per cent of the variability time scale.  Some sources on the other hand show negative time lags, with the soft-band variability lagging the hard-band variability (McHardy et al. 2007; Emmanoulopoulos et al. 2011; De Marco et al. 2013). This type of behaviour has been detected in both supermassive and stellar mass black holes. Finally, the variability in emission from different energy bands is found to be coherent at low frequencies, but becomes increasingly incoherent at higher frequencies (Nowak et al. 1999). 

This observed aperiodic variability structure is rather naturally explained by the so-called ``theory of propagating fluctuations'' (first put forward by Lyubarskii 1997), in which fluctuations in the disc's alpha parameter (or equivalently surface density) are excited at all radii in the accretion flow. These fluctuations then evolve  throughout the disc and are observed as stochastic variability in the source's light curves.  As fluctuations are excited at every disc radii different time scales are injected into the accretion flow corresponding to the local evolution timescales of the individual fluctuations at each distance from the central object (Lyubarskii 1997; Churazov et al. 2001; Ingram 2015). The propagating fluctuations model naturally explains, for example, the lagging of ``hard'' (higher energy) X-ray variability behind ``softer'' (lower energy) bands (Kotov et al. 2001; Ingram \& van der Klis 2013). Excitations sourced in the outer, cooler, disc regions first produce softer flares before subsequently propagating inwards, sourcing harder flares in the hotter innermost disc regions. 

In this model, the power spectral shape of the broad band aperiodic noise depends on both the noise generating process and the response of the accretion flow. The magneto-rotational instability (MRI: Hawley \& Balbus 1991; Balbus \& Hawley 1998) is the underlying noise generator, which produces variability everywhere in the disc due to the interactions between magnetic fields and a differentially rotating gas. The response of an accretion flow to intrinsic fluctuations is governed by a diffusion equation (LyndenBell \& Pringle 1974). Solving this diffusion equation for a $\delta$-function perturbation (i.e. calculating the Green’s function) is an important step in calculating the resulting power spectrum of the mass accretion rate at radii which differ from where the noise originated.  

On the technical level, simplified Green's functions were used for modelling the propagating fluctuations by Lyubarskii (1997) and Kotov et al. (2001), where the fluctuations were assumed to evolve in an additive manner. The more physical model of {\it multiplicative} fluctuations in the accretion flow were first considered by Ingram \& van der Klis (2013) and Ingram \& Done (2011),  but in each of these cases only inward propagation was considered.  However, fluctuations in (for example) an accretion flow's surface density do not simply evolve inwards towards the central object; they must conserve the flow's total angular momentum. This means that some material must propagate outwards in the flow, soaking up the angular momentum of the material propagating inwards, and fluctuations in the inner disc must therefore effect the properties of the outer disc.  

This important conceptual point was first analysed in detail by Mushtukov et al. (2018), who developed a framework for analysing the combined effects of inward and outward propagating fluctuations in an accretion flow. Mushtukov et al. (2018) used this new framework to show that this additional outward propagation has potentially important observational effects,  including that propagating fluctuations can give rise not only to hard time lags as previously shown, but may also produce {\it negative} lags (softer bands lagging harder bands) at high frequencies, a routinely observed effect which had previously been attributed to reprocessing.

The Mushtukov et al. (2018)  framework involves computing the Fourier transform of the Green's function solutions of the classical thin disc equations (which we shall henceforth call the Fourier-Green's functions). This Fourier-Green's integral has a long history in the literature, having been first written down in the original Lyubarskii (1997) paper, it  has since reappeared in various different forms in a number of subsequent works, but has only ever been solved numerically.  

In this work we present two important advances in the theory of propagating fluctuations. The first is deriving the exact solution of the Fourier-Green's integral, which turns out to be surprisingly simple in its final form.  With the exact analytical solutions of the Fourier integral now at hand, various  properties of these solutions may be derived. In section 2, for example,  we demonstrate that high frequency variability in the mass accretion rate is suppressed as $\exp(-\Delta f^{1/2})$, where $\Delta(x, x')$ is a function of the magnitude of the difference between the two disc locations $x$ and $x'$. The high and low frequency asymptotic behaviour of the power spectrum resulting from mass accretion rate variability is also determined, and related to the intrinsic variability in the disc surface density/alpha parameter.  On a practical level, knowledge of these exact solutions will rapidly speed up, and improve the accuracy of, the process of fitting analytical models of accretion variability to observational data;  the numerical cost of Fourier transforming thin disc Green's functions had previously been substantial. 

The second, and potentially more important,  development is in presenting the first analysis of the Fourier-Green's  solutions of the general relativistic thin disc equation. In this paper we solve the Fourier-Green's integral for  a general Kerr metric, under the assumption that the dynamical disc stress vanishes at the ISCO.  These solutions depend implicitly on the central black hole's spin through their dependence on the spacetime's ISCO radius, and for the first time the effects of the black hole's spin on the observed variability structure of an accretion flow can be examined.  In the later sections of this paper we demonstrate that a number of observable variability properties of black hole discs are relatively strong functions of black hole spin, and that observations of the aperiodic variability of black hole accretion discs may now, at least in principle,  offer a new avenue to directly determine black hole spins.  

The layout of this paper is the following. In section 2 we derive and analyse the formal solutions of the Fourier-Green's integral. In section 3 we specialise the analysis specifically to the solutions of the Newtonian disc equation, while in section 4 we discuss the corresponding relativistic solutions. In section 5 we introduce the Mushtukov et al. (2018) framework for relating these results to directly observable quantities, before showing in section 6 that the black hole spin imprints strong signals onto the observed aperiodic variability structure of observed light curves. We conclude in section 7, with some technical results presented in Appendices. 

The reader interested in the application of this analysis for observational constraints on black hole spins may wish to skip directly to section 5. 
\section{Thin disc Green's functions in the Fourier domain: general solution and properties}\label{secF}

\subsection{The  disc evolution  equations }
The evolution of the disc surface density is described by a diffusion-type equations which fundamentally arise from the turbulent transportation of angular momentum within the disc.  The evolution equation for a Newtonian theory of gravity was first analysed in detail by Lynden-Bell and Pringle (1974), while the relativistic equation was derived in Balbus (2017).  We will introduce the general relativistic form of the evolution equation at this point, as it is simpler to take the Newtonian limit ($r \gg GM/c^2$) than the alternative.

The coordinates used to describe the relativistic thin disc equation are the cylindrical Boyer-Lindquist representation of the Kerr metric: $r$ (radial), $\phi$ (azimuthal), and $z$ (vertical).   The governing equation describes the evolution of the azimuthally-averaged, height-integrated disc surface density $\Sigma (r, t)$.   The contravariant four velocity of the disc fluid is $U^\mu$; the covariant counterpart is $U_\mu$.  The specific angular momentum corresponds to $U_\phi$, a covariant quantity.     There is an anomalous stress tensor present, $\W$, due to low level disk turbulence, which is a measure of the correlation between the fluctuations in $U^r$ and $U_\phi$ (Eardley \& Lightman 1975, Balbus 2017).  This is, as the notation suggests, a mixed tensor.    $\W$ serves both to transport angular momentum as well as to extract the free-energy of the disc shear, which is then thermalised and radiated from the disc surface, both assumed to be local processes.    

Under these assumptions the governing disc equation can be expressed in the following compact form 
\beq\label{27q}
{\partial \zeta\over \partial t} =  { \W\over (U^0)^2}{\partial\ \over \partial r} \left({U^0\over U'_\phi} \left[   {\partial \zeta \over \partial r}\right] \right).
\eeq
here the primed notation $'$ denotes an ordinary derivative with respect to $r$, and 
\beq
\zeta \equiv {r \Sigma \W \over U^0} .
\eeq 
The Newtonian limit corresponds to $U^0 = 1, U_\phi' = \sqrt{GM/4r}$. Naturally one must specify a functional form of the disc's turbulent stress tensor $\W$ to derive solutions of this equation, for the remainder of this paper we shall consider stress parameterisations of the form 
\beq
\W = w \left({r\over r_0}\right)^\mu, 
\eeq
which is a popular and analytically tractable choice. 
\subsection{Solution of the Fourier integral}
The Green's function solutions of both the Newtonian (Lynden-Bell and Pringle 1974) and general relativistic (Mummery 2023) thin disc equations are of the general form
\beq
G(x, x_0, t) = {q(x) \over t} \exp\left({- g(x)^2 - g(x_0)^2 \over 4 t}\right) I_{\nu} \left({g(x)g(x_0)\over 2 t} \right) ,
\eeq
where $G(x, x_0, t)$ describes the evolution of the variable $\zeta$ for an initial delta-function spike located at $t = 0, x=x_0$. {In this expression $x$ is a radial coordinate, which is typically normalised by some characteristic scale, either the ISCO (in the relativistic case), or the initial radius $r_0$ of the spike. }
In this expression $I_\nu$ is the modified Bessel function of the first kind, of order $\nu$.  The index $\nu$ is related to the stress parameterisation through 
\beq
\nu = 1/(3-2\mu).
\eeq
For the particular case of the Newtonian Green's functions these functions are particularly simple 
\beq
q(x) \propto x^{1/4}, \quad g(x) \propto x^{1/4\nu} . 
\eeq
{The functions $g(x), g(x_0)$ as defined here have dimensions of $\sqrt{\rm time}$. Physically, the amplitude of these functions correspond to the (square root of the) timescale with which a perturbation at $x$ propagates to the inner disc edge. This introduces a natural scale into these expressions, which will be discussed further in the following sections.   }
It will be useful to note however that this solution is in fact a Laplace-mode superposition, of the form (Gradshteyn and Ryzhik et al. 2007, Lynden-Bell and Pringle 1974, Balbus 2017, Mummery 2023)
\beq
G(x, x_0, t) = \int_0^\infty q(x) J_\nu(\sqrt{s} g(x)) J_\nu(\sqrt{s} g(x_0)) \exp({-st}) \, {\rm d}s .
\eeq
{The function denoted $J_\nu$ is an ordinary Bessel function of the first kind. }The Fourier transform of $G(x, x_0, t)$, denoted $\widetilde G(x, x_0, f)$ is defined by  the complex integral
\beq
\widetilde G(x, x_0, f) \equiv \int_0^\infty G(x, x_0, t) \exp(-2\pi i f t) \, {\rm d} t ,
\eeq
where we have used the fact that $G(x, x_0, t<0) = 0$. When written in terms of the Laplace mode superposition, this integral becomes 
\begin{multline}
\widetilde G(x, x_0, f) = q(x)\int_0^\infty \Bigg[ \int_0^\infty  J_\nu(\sqrt{s} g(x)) J_\nu(\sqrt{s} g(x_0)) \\ \exp({-st}) \, {\rm d}s  \Bigg]  \exp(-2\pi i f t) \, {\rm d} t .
\end{multline}
As both integrals converge, we can  swap the order of integration 
\begin{multline}
\widetilde G(x, x_0, f) = q(x)\int_0^\infty \Bigg[ \int_0^\infty  \exp(-st - 2\pi i ft) \, {\rm d}t \Bigg] \\  J_\nu(\sqrt{s} g(x)) J_\nu(\sqrt{s} g(x_0))  \, {\rm d}s  
\end{multline}
which is more easily solved. Performing the $t$ integral leaves
\beq
\widetilde G(x, x_0, f) = q(x) \int_0^\infty { J_\nu(\sqrt{s} g(x)) J_\nu(\sqrt{s} g(x_0)) \over s + 2\pi i f} \, {\rm d}s  .
\eeq
By making the substitution $u = \sqrt{s}$, this integral becomes 
\beq
\widetilde G(x, x_0, f) = 2 q(x) \int_0^\infty { u J_\nu(u g(x)) J_\nu( u g(x_0)) \over u^2 + \beta^2} \, {\rm d}u  ,
\eeq
where 
\beq
\beta \equiv (1 + i) \sqrt{\pi f}.
\eeq
When written in this form the solution of the integral is a standard result, which can be found in the text of Gradshteyn and Ryzhik et al. (2007) 
\beq
\widetilde G(x, x_0, f) = 2 q(x) 
\begin{cases}
& I_\nu(\beta g(x)) K_\nu(\beta g(x_0)) , \quad x< x_0, \\
\\
& I_\nu(\beta g(x_0)) K_\nu(\beta g(x)) , \quad x> x_0 .
\end{cases}
\eeq  
In this expression $K_\nu$ is the modified Bessel function of the second kind.   The Green's function solutions for the mass accretion rate, denoted $G_{\dot M}$, are of interest in understanding the variability properties of black hole discs, as it is often assumed that variability in the mass accretion rate is directly communicated into variability in the locally emitted flux (e.g., Lyubarskii 1997, Ingram \& van der Klis 2013, Ingram \& Done 2012, Mushtukov et al, 2018).  The mass accretion rate Green's function has the following form (we again write this generally so as to consider both the Newtonian and relativistic solutions simultaneously) 
\beq
G_{\dot M} (x, x_0, t) = p(x) {\partial \over \partial x} G(x, x_0, t) .
\eeq
In the Newtonian limit the function $p(x)$ is simple 
\beq
p(x) \propto x^{1/2}. 
\eeq
In the Fourier domain
\begin{align}
\widetilde G_{\dot M} (x, x_0, f) &\equiv \int_0^\infty G_{\dot M}(x, x_0, t) \exp(-2\pi i f t) \, {\rm d} t,  \nonumber \\ 
&=   \int_0^\infty p(x) {\partial \over \partial x} \left[G(x, x_0, t)\right] \exp(-2\pi i f t) \, {\rm d} t, \nonumber \\ 
&=  p(x) {\partial \over \partial x}\widetilde G(x, x_0, f),
\end{align}
where in going to the final line we have used the fact that the $x$ derivative and $t$ integral commute.  We therefore have the general solution 
\beq\label{general_def}
{1\over 2} \widetilde  G_{\dot M}=  p(x) 
\begin{cases}
&  K_\nu(\beta g(x_0)) \,  {\partial_x} \left[ q(x) I_\nu(\beta g(x))\right] , \quad x< x_0, \\
\\
& I_\nu(\beta g(x_0)) \,  {\partial_x} \left[ q(x)  K_\nu(\beta g(x)) \right] , \quad x> x_0 ,
\end{cases}
\eeq  
where we use the notation $\partial_x \equiv \partial/\partial x$. As we shall demonstrate in section \ref{secN}, this equation further simplifies for the particular case of the Newtonian disc equations. Before we specialise to either the Newtonian or relativistic regimes, we analyse the asymptotic properties of these Fourier-Green's function.  

\subsection{Asymptotic properties}
The asymptotic ($f\rightarrow\infty$ and $f \rightarrow 0$) properties of $\widetilde G_{\dot M}$ can be determined from this general formula. {These asymptotic limits should be understood as the limiting behaviour at Fourier frequencies which are significantly larger (or smaller) than the characteristic  accretion frequency associated with radius $x$. The characteristic accretion frequency for a Newtonian solution with perturbation at $r_0$ has the following form (Lynden-Bell and Pringle 1974)  }
\beq\label{Newtf0}
f_{0} = {1 \over t_{\rm acc}(r_0)} = {(3 - 2\mu)^2 \over 2} \sqrt{w^2 \over GM_{\rm BH} r_0^3} .
\eeq
{(We derive the relativistic analogue of this expression in Appendix \ref{full_GR}). The amplitude of the functions $g(x)$ are $1/\sqrt{f_0}$.  }
\subsubsection{High frequency Fourier modes}
The large frequency limit ({$f \gg f_0$)} of the two Bessel functions are the following 
\beq
\lim_{\beta \rightarrow \infty} K_\nu(\beta g) = \sqrt{\pi \over 2\beta g} \exp(-\beta g) ,
\eeq
and 
\beq
\lim_{\beta \rightarrow \infty} I_\nu(\beta g) = \sqrt{1 \over 2\pi \beta g} \exp(+\beta g) .
\eeq
We therefore have the following simple result, valid in both of the $x<x_0$ and $x>x_0$ regimes: 
\begin{multline}
 \widetilde G_{\dot M}(x, x_0, f\rightarrow \infty)  \sim {p(x) q(x) g'(x) \over \sqrt{g(x)g(x_0)}} \\ \exp(-\sqrt{\pi f} (1 + i) \left| g(x) - g(x_0) \right|) ,
\end{multline}
where $|z|$ denotes the absolute value of $z$, and $'$ denotes a derivative with respect to $x$. We see here that high frequency Fourier modes are exponential suppressed, with a suppression scale which depends on the inverse of the absolute magnitude of the difference between any two disc radii. 

{Physically, this corresponds to an exponential suppression of the propagation of modes with frequencies higher than the local accretion frequency.  It is interesting to note that this expression is symmetric in the sign of  $x - x_0$. This  is an important result: at high Fourier frequencies (relative to the local accretion frequency), outward propagation of material is equally as important as the inward propagation of material. This is not true at all Fourier frequencies, as we demonstrate in future sections. }

The fact that the suppression of high-frequency modes is proportional to $\exp(- \Delta f^{1/2})$ is unsurprising, and is a result of the disc angular-momentum diffusion processes. To see this explicitly, consider the simple one-dimensional diffusion equation 
\beq
{\partial \psi \over \partial t} = D {\partial^2 \psi \over \partial x^2} .
\eeq
In terms of the Fourier modes ($\psi = \int_{-\infty}^{+\infty} \widetilde \psi \exp(2\pi i f t) {\rm d} f$), this equation reads 
\beq
2\pi i f \widetilde \psi = D {\partial^2 \widetilde \psi \over \partial x^2} ,
\eeq
with a solution that is well behaved at large $x$ of 
\beq
\widetilde \psi(x, f) = A \exp\left(- \sqrt{\pi f \over D} (1 + i) \, x\right) . 
\eeq
It is clear to see that a high frequency exponential $\exp(-f^{1/2})$ suppression is a generic property of diffusive systems. 

\subsubsection{Low frequency Fourier modes}
The small frequency limit ($f \ll f_0$) of $\widetilde G_{\dot M}$ can also be understood from the asymptotic properties of the Bessel functions $K_\nu$ and $I_\nu$. In addition we can understand a priori the asymptotic properties in the small frequency limit, as  $f \rightarrow 0$  corresponds physically to the integral 
\beq
\widetilde G_{\dot M}(x, x_0, f\rightarrow 0) \rightarrow \int_0^\infty G_{\dot M}(x, x_0, t) \, {\rm d}t .
\eeq
The integral over all times of the mass accretion rate at radius $x$, initiated by a perturbation at radius $x_0$, must equal (as all of the disc material is eventually accreted) to 
\beq
\widetilde G_{\dot M}(x, x_0, f\rightarrow 0) \rightarrow 
\begin{cases}
-M_d, \,\,\, \quad x < x_0, \\
\\
0, \,\,\,\,\,\,\,\quad \quad x > x_0. 
\end{cases}
\eeq
To prove this more rigorously, consider the small frequency limits of the two Bessel functions:
\beq
\lim_{\beta \rightarrow 0} K_\nu(\beta g) = {\Gamma(\nu) \over 2} \left({\beta g \over 2}\right)^{-\nu} +  {\Gamma(-\nu) \over 2} \left({\beta g \over 2}\right)^{+\nu} + {\cal O}((\beta g )^{2-\nu}) 
\eeq
and 
\beq
\lim_{\beta \rightarrow 0} I_\nu(\beta g) = {1 \over \Gamma(\nu+1)} \left({\beta g \over 2}\right)^{+\nu} + {\cal O}((\beta g)^{2+\nu})
\eeq
where $\Gamma(n)$ is the usual gamma function.  It will transpire that we are required to keep the first two terms in the expansion of $K_\nu$ to compute the correct $f \rightarrow 0$ limit of these solutions. For $x < x_0$ (inward propagating modes) we have
\beq
\left| \widetilde G_{\dot M}(x < x_0, f\rightarrow 0) \right| \sim {p(x) \over \nu} 
 \left({1 \over g(x_0)}\right)^\nu {\partial \over \partial x} \left[ q(x) (g(x))^\nu \right] ,
\eeq
i.e., a constant which is independent of $f$. With the proper normalisation of $q(x)$ and $g(x)$ this constant will be equal to the disc mass. For outward propagating modes ($x > x_0$) we have 
\begin{multline}
\left| \widetilde G_{\dot M}(x> x_0, f\rightarrow 0) \right| \sim {p(x) \over \nu} 
\left({ g(x_0)}\right)^\nu {\partial \over \partial x} \left[ { q(x) \over (g(x))^{\nu}} \right]  \\  + p(x) {\Gamma(-\nu) \over \Gamma(\nu+1)} \left({ g(x_0) \over 4 }\right)^\nu |\beta|^{2\nu} {\partial \over \partial x} \left[ { q(x)  (g(x))^{\nu}} \right] + {\cal O}(|\beta|^2),
\end{multline}
which means
\beq\label{lowfreq}
\left| \widetilde G_{\dot M}(x > x_0, f\rightarrow 0) \right| \sim 
\begin{cases} 
f^\nu , \quad { q(x) \propto (g(x))^{\nu}}, \quad \nu < 1,  \\
\\
f^1 , \quad { q(x) \propto (g(x))^{\nu}}, \quad \nu > 1,  \\
\\
f^{0} , \quad {\rm otherwise} .
\end{cases}
\eeq
Unsurprisingly, for the particular case of a Newtonian disc model $q(x) \propto x^{1/4}$ and $g(x) \propto x^{1/4\nu}$, and the amplitude of the outward propagating low frequency modes goes to zero as a power-law of frequency in the limit of $f \rightarrow 0$.  

{At low Fourier frequencies (relative to the local accretion frequency), only the inwardly  propagating  material significantly contributes to the local variability of the accretion rate. As conventional approaches typically neglect outward propagation, they are likely to be most accurate at low frequencies.    }

\subsection{Discontinuity of the mass accretion rate Fourier-Green's function at $x = x_0$}
As can be seen from the above analysis, the properties of the mass accretion rate Fourier-Green's functions must be discontinuous at the location of the perturbation $x = x_0$. This can be proved rather generally by using the Wronskian of the modified Bessel functions  (Abramowitz \& Stegun 1965):
\beq
{\cal W}\left[I_\nu, K_\nu \right] \equiv I_\nu(z) {{\rm d} \over {\rm d } z} K_\nu(z) -  K_\nu(z) {{\rm d} \over {\rm d } z} I_\nu(z) = -{1 \over z} ,
\eeq
which implies a magnitude of the discontinuity of 
\begin{multline}
\widetilde G_{\dot M}(x \rightarrow x_0, x_0 > x, f) - \widetilde G_{\dot M}(x \rightarrow x_0, x_0 < x, f) \\ = p(x) q(x) {{\rm d} \over {\rm d} x} \ln g(x) .
\end{multline}
Physically this discontinuity results from the eventual accretion of all of the disc material through the inner disc edge.  The preferred mass flow direction (inwards) fundamentally breaks the $x < x_0$, $x > x_0$ symmetry of the mass accretion rate Fourier-Green's  functions.

\subsection{The mass accretion rate power spectrum}\label{ps_lims}
As derived by Mushtukov {\it et al}. (2018) the power spectrum of the mass accretion rate at a particular radius, denoted $S_{\dot m}(x, f)$, is given by 
\beq\label{local_power_spec}
S_{\dot m}(x, f) = \int_{x_{\rm in}}^{x_{\rm out}} \left({1 \over x'}\right)^2 l(x') \left| \widetilde G_{\dot M}(x, x', f) \right|^2 S_\Sigma(x', f) \, {\rm d}x',
\eeq
where $l(x')$ is the radial scale over which  the initial perturbations can be considered as coherent with one another, and  $S_\Sigma(x, f)$ is the power spectrum of the initial surface density perturbations at radius $x$. {In general it is possible that fluctuations in certain regions of the disc might be coherent over a larger radial scale than in other regions, and so in general $l(x')$ is a function of radius. In this work however we shall assume that all disc radii have the same coherence length. }  Mushtukov showed numerically that at high and low frequencies the power  $P(f) = f S_{\dot m}(f)$ behaved like a power-law in frequency, assuming that the input surface density power spectrum $S_\Sigma$ was Lorentzian:
\beq\label{Lorentzian}
S_\Sigma(x, f) = {2 p \over \pi} {f_{\rm br} \over f^2 + f_{\rm br}^2} ,
\eeq 
where $p$ and $f_{\rm br}$ are the total power and break frequency (simply parameters of the model) at radius $x$ respectively. 

{As the focus of this work is on single epoch variability the parameter $p$, the amplitude of the locally driven power in the surface density perturbations, is treated   as a simple constant. However, in a real physical system $p$ has an important property: it scales quadratically with the local average (background) mass accretion rate $\dot M^2$. This is a property of the propagating fluctuations model, which was first shown by Lyubarskii (1997), and was expanded  upon by Mushtukov {\it et al}. (2018). This behaviour of $p$ is important for understanding the variability properties of accreting sources at different stages of their evolution, as it ultimately results in the well known linear flux-rms relation (Uttley \& McHardy 2001, Uttley et al. 2005; see Mushtukov et al. 2018 for a proof of this statement).    }

{Finally, in addition to the leading order term presented here, there is formally an additional non-linear term present in equation \ref{local_power_spec}, which arise from local fluctuations of the surface density superimposing  on top of existing variability in the accretion flow. Mushtukov et al. (2018) found that the inclusion of these non-linear terms alters the quantitative properties of the local power spectrum at the $1\%$ level, and do not alter the qualitative properties of the theory. The amplitude of deviation does however grow with the injected variability amplitude, and so this simplification should be kept in mind. These non-linear effects are beyond the scope of the present work, and will not be considered further.  }

In the following sub-sections we rigorously prove various numerical results presented in Mushtukov {\it et al}. (2018).   

\subsubsection{High frequency power spectrum }
For $f \rightarrow \infty$ the inward and outward propagating modes both have transfer functions which behave like
\begin{multline}
\left| \widetilde G_{\dot M}(x, x_0, f\rightarrow \infty) \right| \sim {p(x) q(x) g'(x) \over \sqrt{g(x)g(x_0)}} \\ \exp(-\sqrt{\pi f} \left| g(x) - g(x_0) \right|) ,
\end{multline}
thus 
\begin{multline}
S_{\dot m}(x, f\rightarrow \infty) \sim \left({p(x) q(x) g'(x) \over \sqrt{g(x)}}\right)^2 \int_{x_{\rm in}}^{x_{\rm out}} \left({1 \over x'}\right)^2 {l(x') \over g(x')}  \\ \exp(-2\sqrt{\pi f} \left| g(x) - g(x') \right|) \, S_\Sigma(x', f) \, {\rm d}x',
\end{multline}
which, provided that the high frequency fall-off of $S_\Sigma(f)$ is weaker than $\exp(-Af^{1/2})$, is an integral which can be performed by Laplace's method. The leading order behaviour is simply 
\beq
S_{\dot m}(x, f\rightarrow \infty) \sim f^{-1/2} S_\Sigma(x, f\rightarrow \infty) . 
\eeq
{This result highlights that at the highest frequencies, the observed variability is dominated by locally driven variability, with exponentially small contributions from distant disc regions.} For the particular case of a Lorentzian   power spectrum for $\Sigma$, we have
\beq
P(f\rightarrow \infty) \equiv f S_{\dot m}(x, f\rightarrow \infty) \sim f^{-3/2}. 
\eeq
This exact behaviour was discovered numerically by Mushtukov et al. (2018). 
\subsubsection{Low frequency power spectrum }
For $f \rightarrow 0$ the inward propagating modes  have transfer functions which become independent of frequency 
\beq
\left| \widetilde G_{\dot M}(x<x_0, x_0, f\rightarrow 0) \right| \sim f^0 ,
\eeq
thus 
\begin{multline}
S_{\dot m}(x, f\rightarrow 0) \sim  \int_{x}^{x_{\rm out}}F_1(x', x) \, S_\Sigma(x', f) \, {\rm d}x' \\ + f^\alpha \int_{x_{\rm in}}^{x} F_2(x', x) \, S_\Sigma(x', f) \, {\rm d}x'   ,
\end{multline}
where $F_1(x, x')$ and  $F_2(x, x')$ are independent  of frequency, and $\alpha \geq 0$. The leading order behaviour of the mass accretion rate power spectrum is therefore given, in the low frequency limit,  simply by that of the initial surface density perturbations at {radii larger than} $x$ (the first integral)
\beq
S_{\dot m}(x, f\rightarrow 0) \sim  \int_{x}^{x_{\rm out}}F_1(x', x) \, S_\Sigma(x', f\to0) \, {\rm d}x' .
\eeq
{As expected from the earlier analysis (section 2.3), only those perturbations initialised at larger radii  $x_0>x$ which then propagate inwards contribute to the power spectrum in the low frequency limit. }

For the particular case of a Lorentzian power spectrum for $a$, we have
\beq
P(f\rightarrow 0) \equiv f S_{\dot m}(f\rightarrow 0) \sim f^{1} ,
\eeq
as found numerically by Mushtukov et al. (2018). 
\section{Newtonian Fourier-Green's functions}\label{secN}
The particular Green's function solutions for a Newtonian theory of gravity have
\beq
q(x) \propto x^{1/4},\quad p(x) \propto x^{1/2}, \quad g(x) \propto x^{1/4\nu}.
\eeq
It transpires that the derivative in equation \ref{general_def} simplifies greatly, a result of the identities 
\beq
{{\rm d} \over {\rm d}z} I_l(z) =  I_{l-1}(z) - {l \over z} I_l(z),
\eeq
and 
\beq
{{\rm d} \over {\rm d}z} K_l(z) =  - K_{l-1}(z) - {l \over z} K_l(z).
\eeq
We therefore have the remarkably simple result
\beq
\widetilde G_{\dot M} = 
\begin{cases}
+ A \beta x^{(1-\nu)/4\nu} K_\nu(\beta \epsilon x_0^{1/4\nu}) I_{\nu - 1}(\beta \epsilon x^{1/4\nu}), \quad  x < x_0, \\
\\
 - A \beta x^{(1-\nu)/4\nu} I_\nu(\beta \epsilon x_0^{1/4\nu}) K_{\nu - 1}(\beta \epsilon x^{1/4\nu}), \quad x > x_0 , 
\end{cases}
\eeq
where (units where $GM = c = w = 1$; see also eq. \ref{Newtf0})
\beq
A =  {x_0^{1/4}   \epsilon M_d }, \quad \epsilon = 2\nu \sqrt{2x_0^\mu} , \quad \mu = {3-1/\nu \over 2} ,
\eeq
and we remind the reader
\beq
\beta \equiv (1 + i) \sqrt{\pi f}.
\eeq
Note that Newtonian gravity is entirely scale free, and $x$ here should be thought of as the disc radius suitably normalised by some arbitrary radial scale in the problem of interest. For relativistic systems the solution is explicitly a function of $r/r_g$, where $r_g = GM/c^2$ is the gravitational radii of the black hole. 

\begin{figure}
\includegraphics[width=\linewidth]{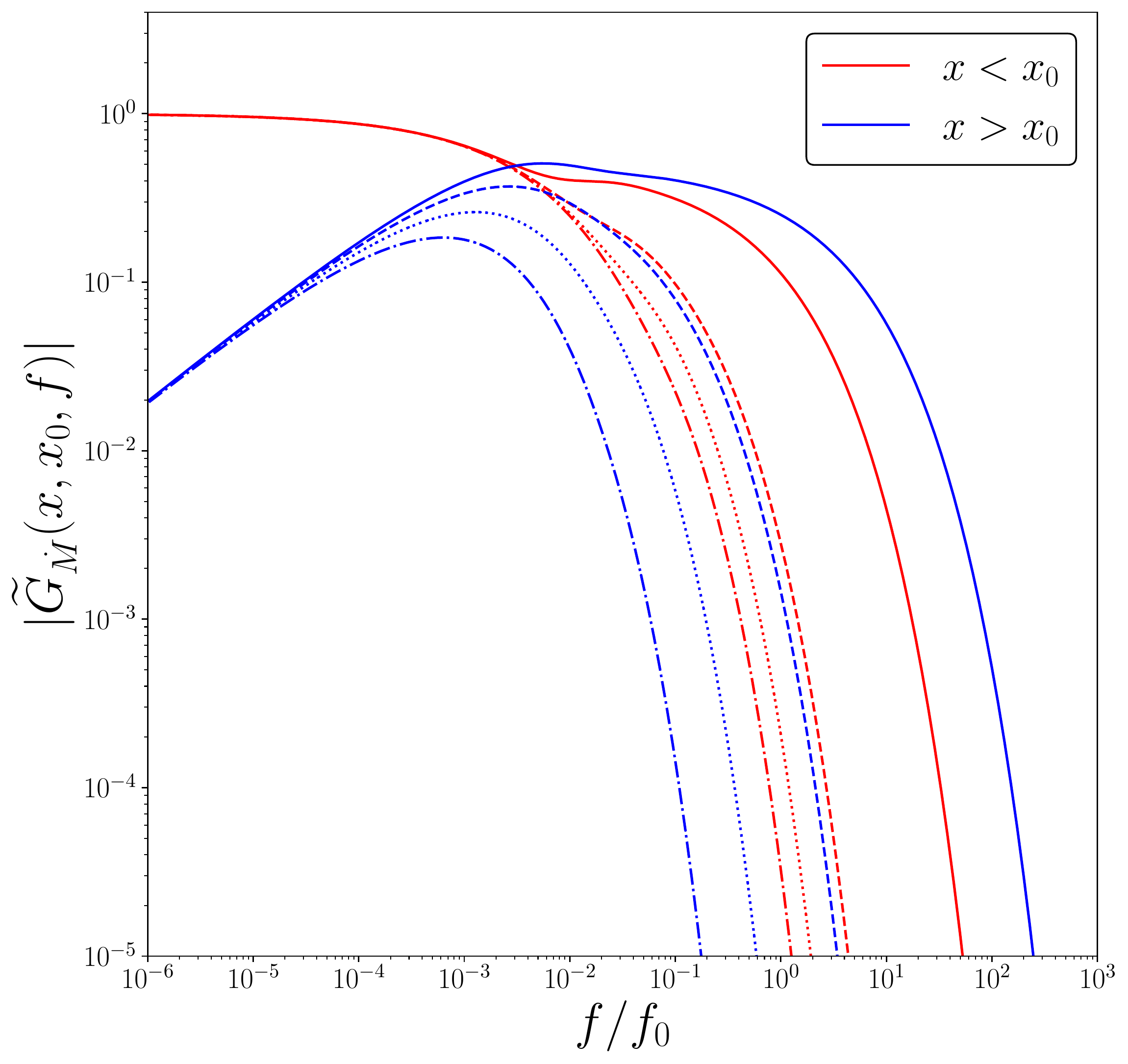}
\caption{The amplitude of the Newtonian Fourier-Green's function of the mass accretion rate. Red curves are for inward propagating modes $x < x_0$, and blue curves show outward propagating modes $x > x_0$. For this figure we take $x_0 = 10$, and the inward propagating modes have $x = 1$ (dot-dashed), $2$ (dotted), $4$ (dashed) and $8$ (solid), while the outward propagating modes have $x = 11$ {(solid)}, $18$ (dashed), $25$ (dotted) and $30$ (dot-dashed). The Fourier-Green's functions are normalised so that the total accreted mass is equal to 1. }
\label{NewtonAmp}
\end{figure}

In Figure \ref{NewtonAmp} we display the amplitude of the Newtonian Fourier-Green's function, as a function of Fourier frequency, for a number of different disc radii $x$ (listed in caption), and $x_0 = 10$. It is clear to see that the properties of the Fourier-Green's functions are as predicted by the asymptotic analysis of the previous section.  At high Fourier frequencies there is an exponential suppression of the Fourier mode amplitude, with modes at disc radii $x$ closer to $x_0$ having substantially higher amplitude at high frequencies when compared to disc radii further from $x_0$ (contrast the solid curves with the dot-dashed curves).   At low Fourier frequencies the  inward propagating modes ($x < x_0$) the Fourier-Green's functions approach $1$ (when normalised by the disc mass), while outward propagating mode ($x > x_0$) approach zero as a power-law in frequency.

\begin{figure}
\includegraphics[width=\linewidth]{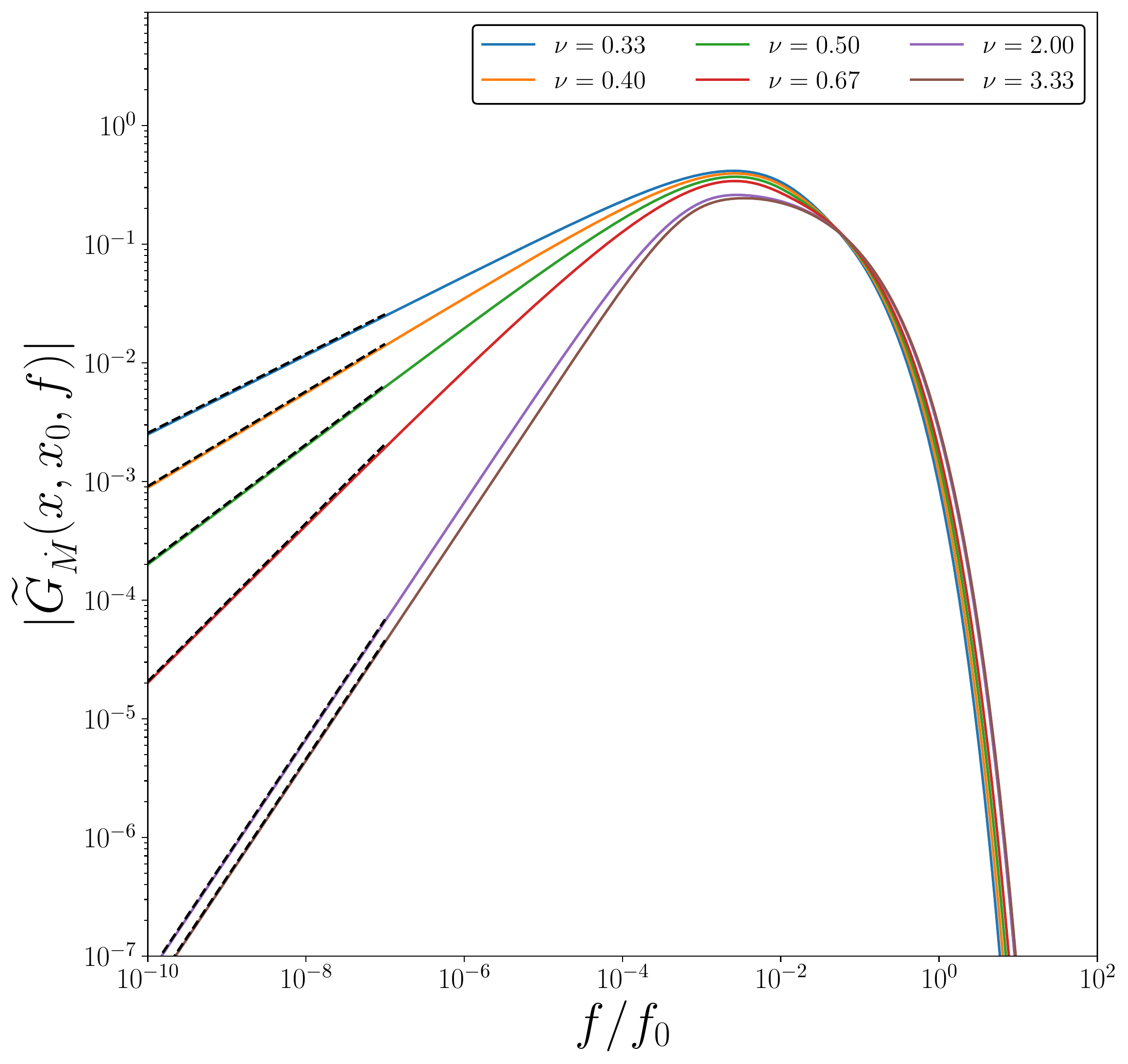}
\caption{The amplitude of the Newtonian Fourier-Green's function of the mass accretion rate, produced with $x_0 = 10$ and $x = 20$, for a variety of different indices $\nu$.   }
\label{NewtonLowFreq}
\end{figure}

\begin{figure}
\includegraphics[width=\linewidth]{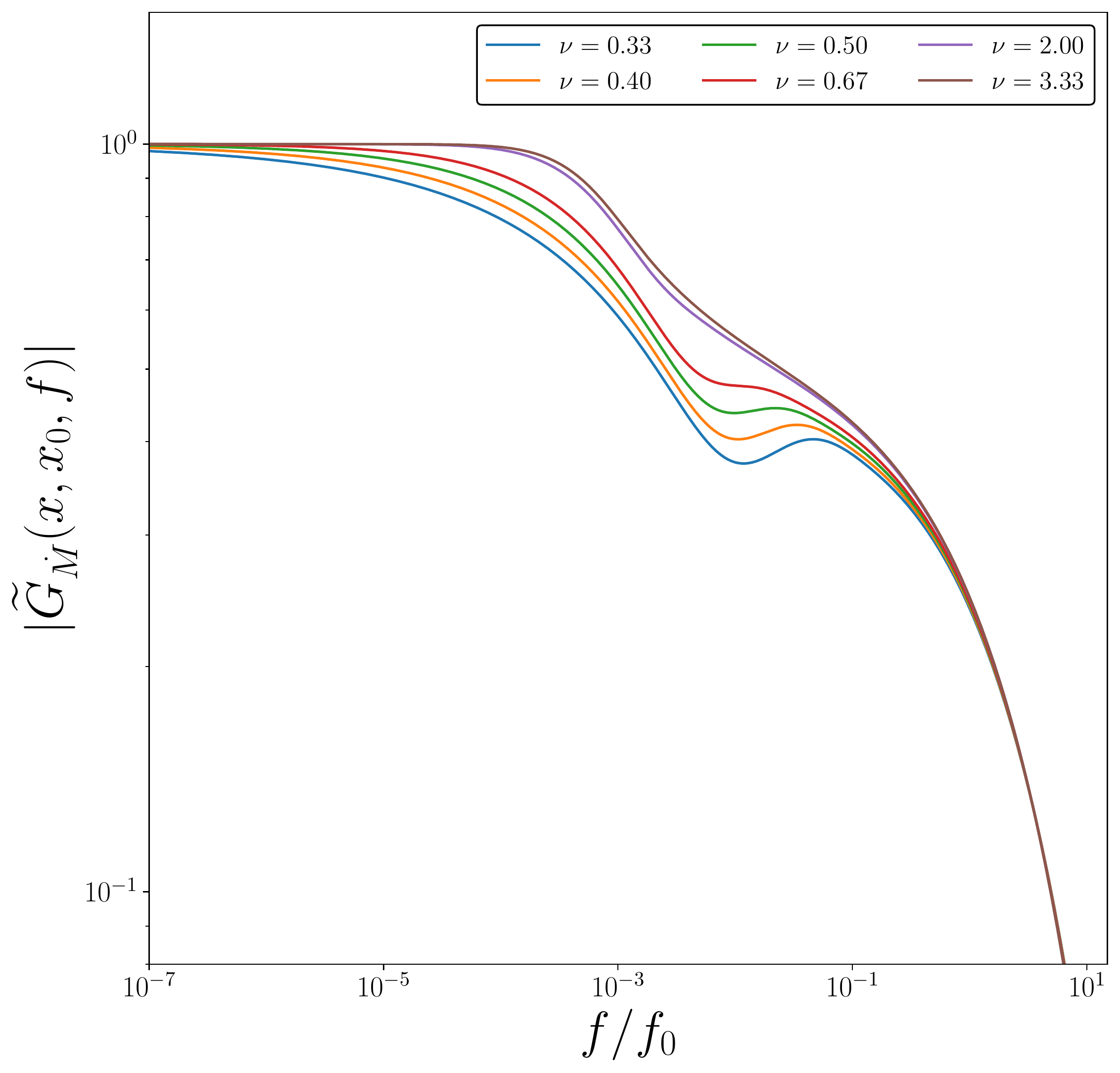}
\caption{The amplitude of the Newtonian Fourier-Green's function of the mass accretion rate, produced with $x_0 = 10$ and $x = 9$, for a variety of different indices $\nu$.  The stress index modifies the properties of the mass accretion rate Fourier-Green's functions at intermediate frequencies.  }
\label{NewtonINdiffStress}
\end{figure}

This low-frequency asymptotic behaviour is further demonstrated  in Figure \ref{NewtonLowFreq}, where we plot the amplitude of the Fourier-Green's functions for different indices $\nu$, for $x=20, x_0= 10$. At low frequencies each amplitude approaches a power-law in frequency, with power-law index given by equation \ref{lowfreq} (black dashed curves). {In Fig. \ref{NewtonINdiffStress} we examine the effects of varying the stress index on the inward propagating Fourier-Green's functions. While less severe than for the outward propagating modes, the stress parameterisation does quantitatively effect the properties of the Fourier-Green solutions of inward propagating modes. }

The angle of the Newtonian Fourier-Green's function is defined as 
\beq
\Phi_{\widetilde G_{\dot M}} = {\rm arg}\left(\widetilde G_{\dot M}\right) \equiv \tan^{-1} \left( {{\cal I} \left[ \widetilde G_{\dot M} \right] \over {\cal R} \left[ \widetilde G_{\dot M} \right]} \right) ,
\eeq
where ${\cal I}[z]$ and ${\cal R}[z]$ denote the imaginary and real parts of the complex variable $z$ respectively, and the appropriate branches of the $\tan^{-1}$ function are chosen so that $-\pi < {\rm arg}(z) < \pi$.  The angle of these Fourier-Green's functions is related to the time lag,  a more useful observable quantity, via 
\beq
t_{\rm lag} = { \Phi_{\widetilde G_{\dot M}}  \over 2\pi f} .
\eeq
{The  angle of the Newtonian Fourier-Green's functions, as a function of disc radius $x$, behave qualitatively as follows.  } At low Fourier frequencies the angle of the Fourier-Green's functions are approximately constant, whereas at high Fourier frequencies the angle of the Fourier-Green's functions cycle through $\pi \rightarrow -\pi$ (or $-\pi \rightarrow \pi$ for $x<x_0$) as a function of increasing  radius.   The angle of the Fourier-Green's functions are discontinuous at $x = x_0$. 

The cycling of the Fourier-Green's function angle at high frequencies is simple to understand analytically, using the results of the proceeding section:
\beq
\widetilde G_{\dot M} (x, x_0, f\rightarrow \infty) \sim C \exp(- i \sqrt{\pi f}  \left| g(x) - g(x_0) \right|) ,
\eeq
where $C$ is purely real, meaning 
\beq
 \tan\left( \Phi_{\widetilde G_{\dot M}}  \right) \rightarrow   \tan\left[- \sqrt{\pi f}  \left( g(x) - g(x_0) \right) \right], 
\eeq
resulting in a cyclic behaviour. 

\begin{figure}
\includegraphics[width=\linewidth]{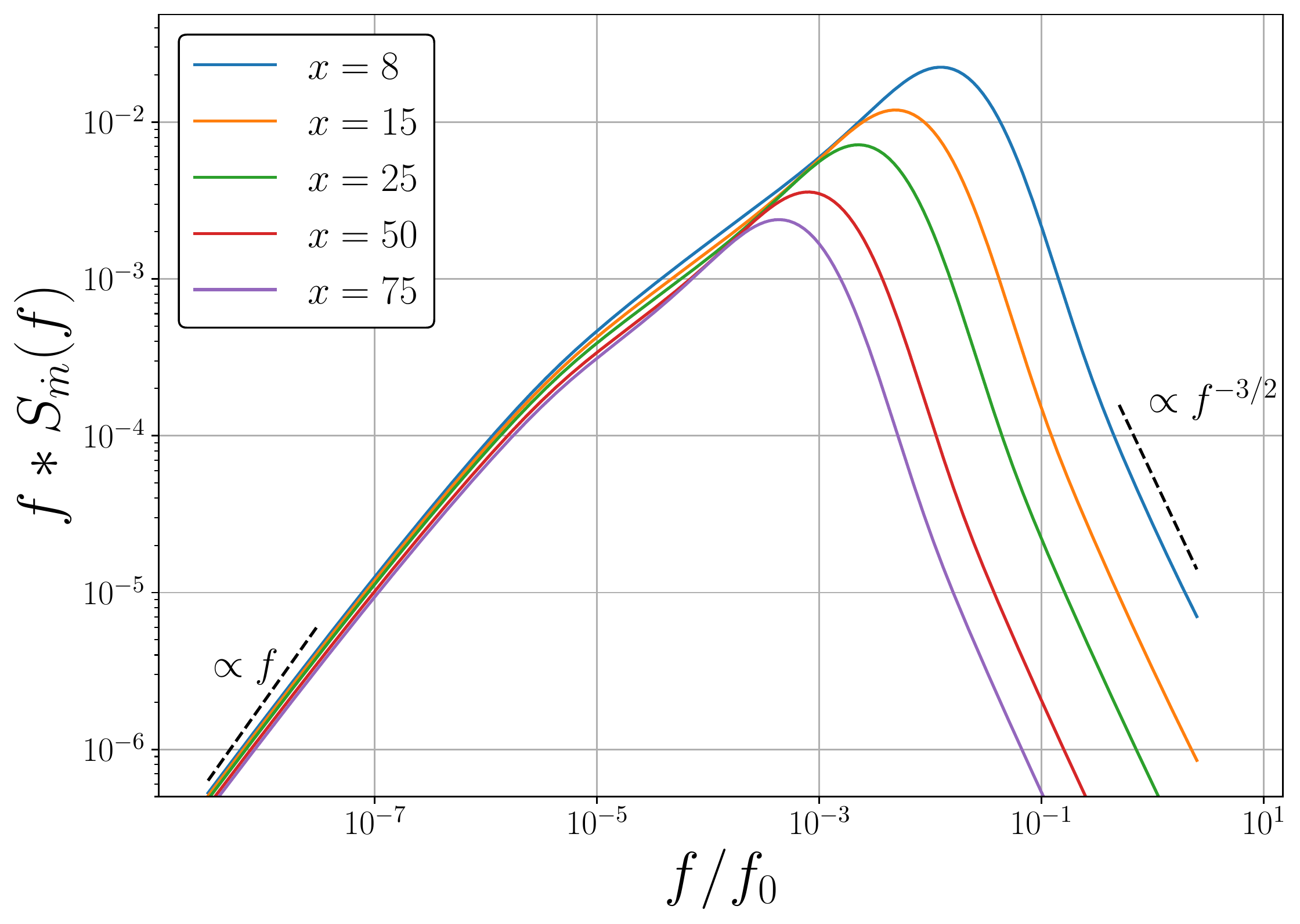}
\caption{The power density spectrum of the mass accretion rate multiplied by the Fourier frequency, at a number of disc radii $x$ (denoted in legend), as a function of Fourier frequency. Displayed as black dashed  curves are the high and low frequency asymptotic results derived in section \ref{ps_lims}. The amplitude of the accretion rate power density spectrum increases with decreasing disc radius, a result which is particularly true at high Fourier frequencies.     }
\label{NewtonPDS}
\end{figure}

The mass accretion rate power density spectrum $S_{\dot m}(x, f)$, defined in section \ref{ps_lims}, is shown in Figure \ref{NewtonPDS}. In Figure  \ref{NewtonPDS} we take an accretion rate fluctuation coherence length $l(x') = 1$, and in the Laplacian input for $S_\Sigma$ we take $p = 1$, and $f_{\rm br} = \sqrt{GM/r^3}$, the Keplerian frequency at disc radius $r$.  Also displayed as black dashed  curves are the high and low frequency asymptotic results derived in section \ref{ps_lims}. The amplitude of the accretion rate power density spectrum increases with decreasing disc radius, a result which is particularly true at high Fourier frequencies.  

The complex cross-spectrum, which measures the correlation between variability at disc radii $x_1$ and $x_2$,   is given by (Mushtukov {\it et al}. 2018) 
\begin{multline}
C_{\dot m}(x_1, x_2, f) = \int_{x_{\rm in}}^{x_{\rm out}} \widetilde G_{\dot M}(x_1, x', f)   \widetilde G^\dagger_{\dot M}(x_2,  x', f) \\ \left({1 \over x'}\right)^2 l(x')   S_\Sigma(x', f) \, {\rm d}x',
\end{multline}
where $z^\dagger$ denotes the complex conjugate of $z$. We define the coherence function 
\beq
{\rm Coh}_{\dot m}(x_1, x_2, f) \equiv {\left| C_{\dot m}(x_1, x_2, f) \right|^2 \over S_{\dot m}(x_1, f) S_{\dot m}(x_2, f)}
\eeq
which is limited to the range $0 < {\rm Coh}_{\dot m}(x_1, x_2, f) < 1$. As its name suggests, the coherence function measures the level of coherence between two variable quantities, in this case the variability at radii $x_1$ and $x_2$ in the disc. Note that ${\rm Coh} = 1$ corresponds to a fully coherent fluctuations at both radii, while ${\rm Coh} = 0$ represents incoherent fluctuations.

\begin{figure}
\includegraphics[width=\linewidth]{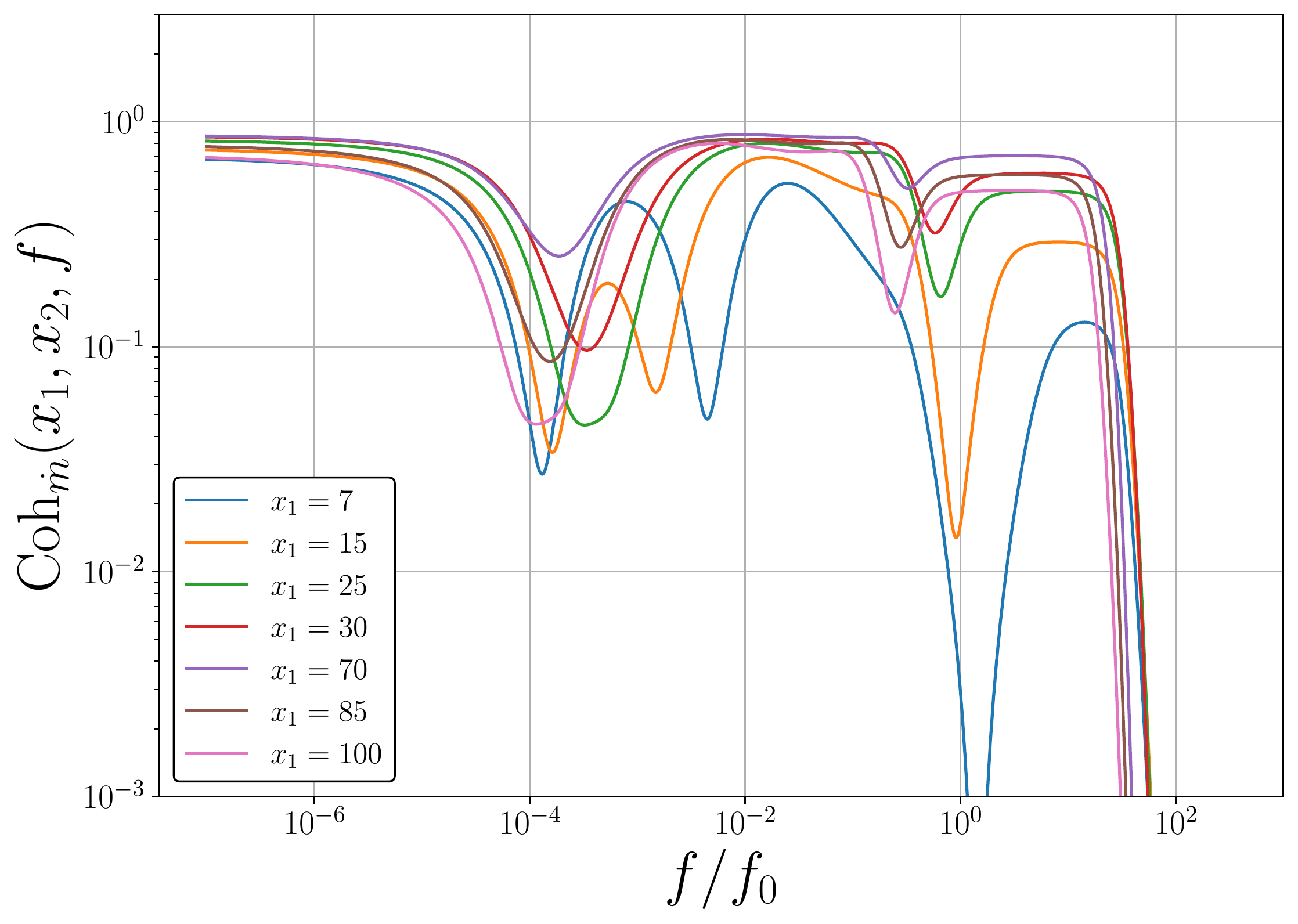}
\caption{The coherence function of the mass accretion rate, between disc radii $x_1$ (displayed on legend), and $x_2 = 50$.  At high enough Fourier frequencies the coherence between any two disc radii decays exponentially. At low Fourier frequencies the coherence between disc radii tends to 1.     }
\label{NewtonCoh}
\end{figure}

\begin{figure}
\includegraphics[width=\linewidth]{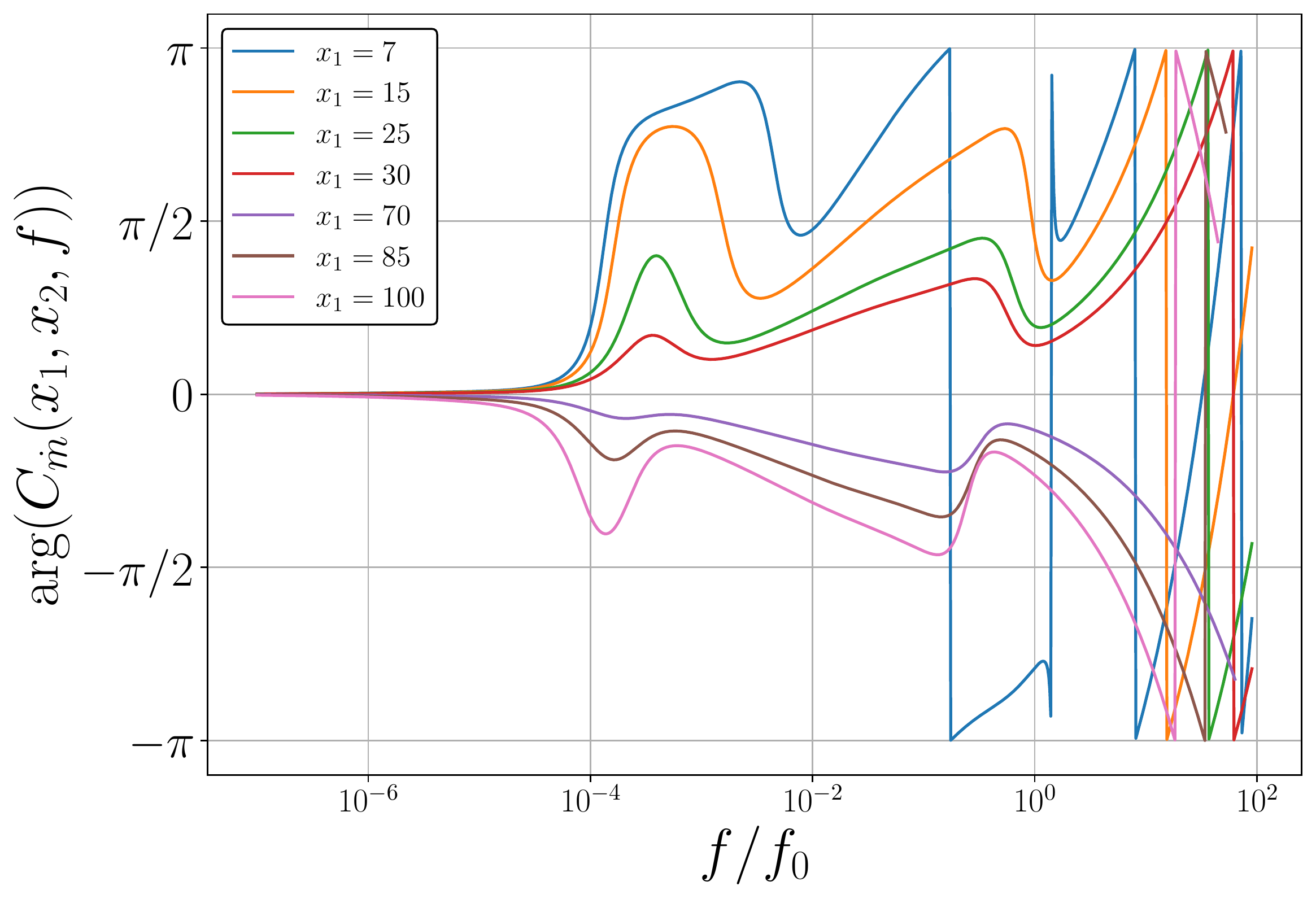}
\caption{The angle of the mass accretion rate cross spectrum between disc radii $x_1$ (displayed on legend), and $x_2 = 50$. At low Fourier frequencies disc variability at radii $x_1$ leads variability at disc radii $x_2 > x_1$, whereas  disc variability at $x_1 > x_2$ lags variability at radii $x_2$. At high Fourier frequencies variability at small radii (e.g., the blue curve) can lag, rather than lead, variability at larger radii.  At the highest Fourier frequencies the cross spectrum cycles from $-\pi \rightarrow \pi$, as a function of increasing frequency. }
\label{NewtonArgC}
\end{figure}

The coherence function is plotted  as a function of Fourier frequency in Figure \ref{NewtonCoh}, for a number of disc radii $x_1$, and $x_2 = 50$.  At high enough Fourier frequencies the coherence between any two disc radii decays exponentially. At low Fourier frequencies the coherence between disc radii tends to unity.   Clearly the coherence function is a complicated function of Fourier frequency at intermediate frequency scales.    

In Figure \ref{NewtonArgC} we plot the angle of the mass accretion rate cross spectrum between disc radii $x_1$ (displayed on legend), and $x_2 = 50$. At low Fourier frequencies disc variability at radii $x_1$ leads variability at disc radii $x_2 > x_1$, whereas  disc variability at $x_1 > x_2$ lags variability at radii $x_2$. At high Fourier frequencies variability at small radii (e.g., the blue curve) can lag, rather than lead, variability at larger radii.  At the highest Fourier frequencies the cross spectrum cycles from $-\pi \rightarrow \pi$, as a function of frequency.

In this section we have calculated a number of the properties of the Newtonian Fourier-Green's functions, showing that their properties are exactly as predicted by the analytical analysis of the previous section. For the remainder of this paper we shall focus on the properties of relativistic discs. 

\section{Relativistic Fourier-Green's functions}\label{secGR}
The Fourier-Green's functions of the previous section are solutions of the Newtonian disc equation. A large number of observed variable disc systems however are those discs evolving around black holes, and must therefore be described by a relativistic theory.   Analytical solutions of the relativistic disc equations have recently been derived by Mummery (2023), and are discussed below. 
\subsection{Relativistic Green's functions in the time domain}

\begin{figure}
\includegraphics[width=\linewidth]{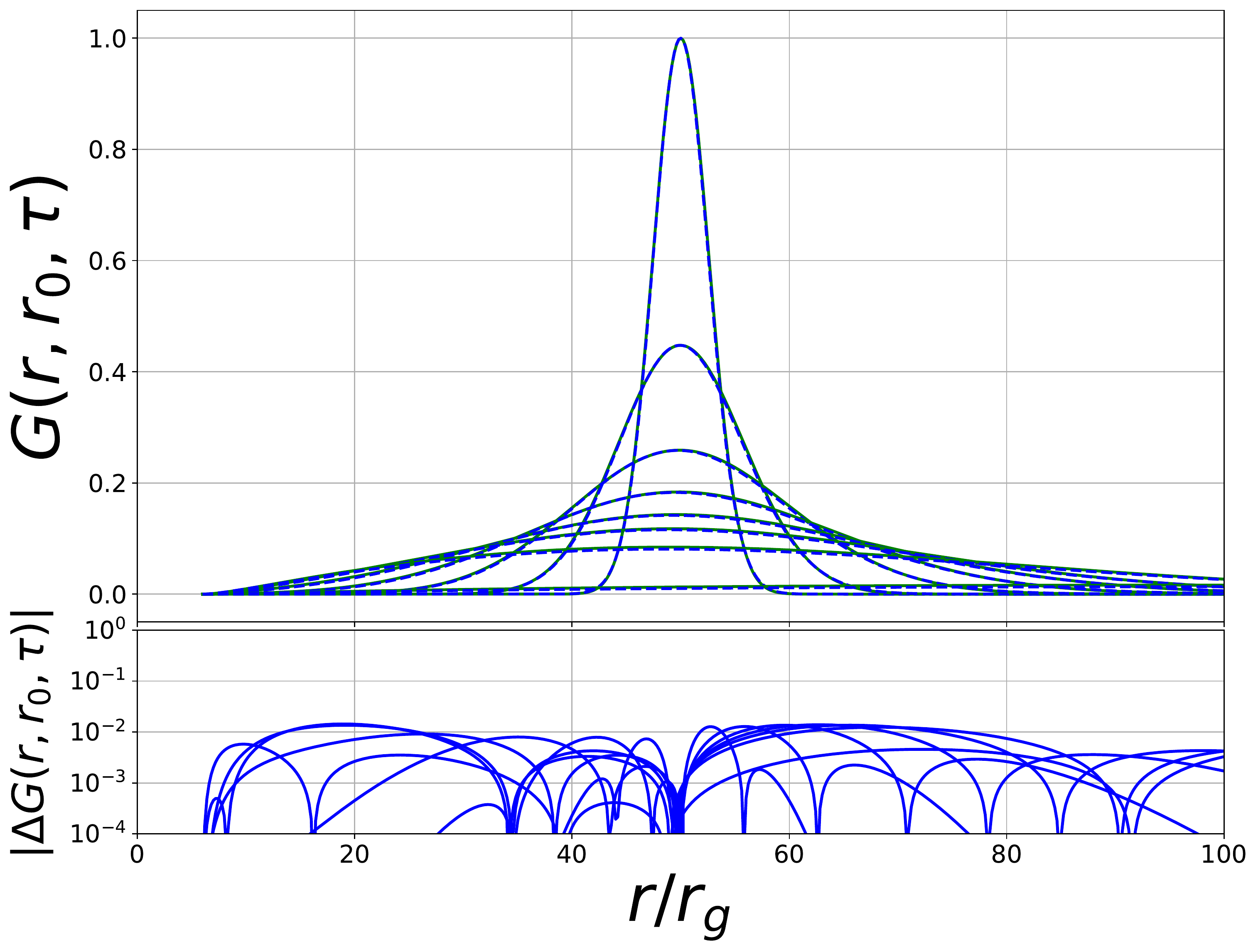}
\caption{Upper: The Green's function solution of the variable $y \equiv r \Sigma \W$, for a Kerr black hole with spin $a = 0$. The blue dashed curves are the numerical solutions of the full general relativistic disc equations, while the green solid curves are the analytical solution of Mummery (2023). The initial radius was $r_0 = 50r_g$ and the curves are plotted at dimensionless times $t/t_{\rm visc} = 0.003, 0.015, 0.045, 0.09, 0.15, 0.225, 0.45$ and $4.5$.  Lower: The absolute value of the difference between the numerical and analytical Green's function solutions.  To allow a proper comparison at each time both the numerical and analytical Green's functions are renormalised to have a peak amplitude of $1$. This Figure is reproduced from Mummery (2023).   }
\label{GFR}
\end{figure}

\begin{figure}
\includegraphics[width=\linewidth]{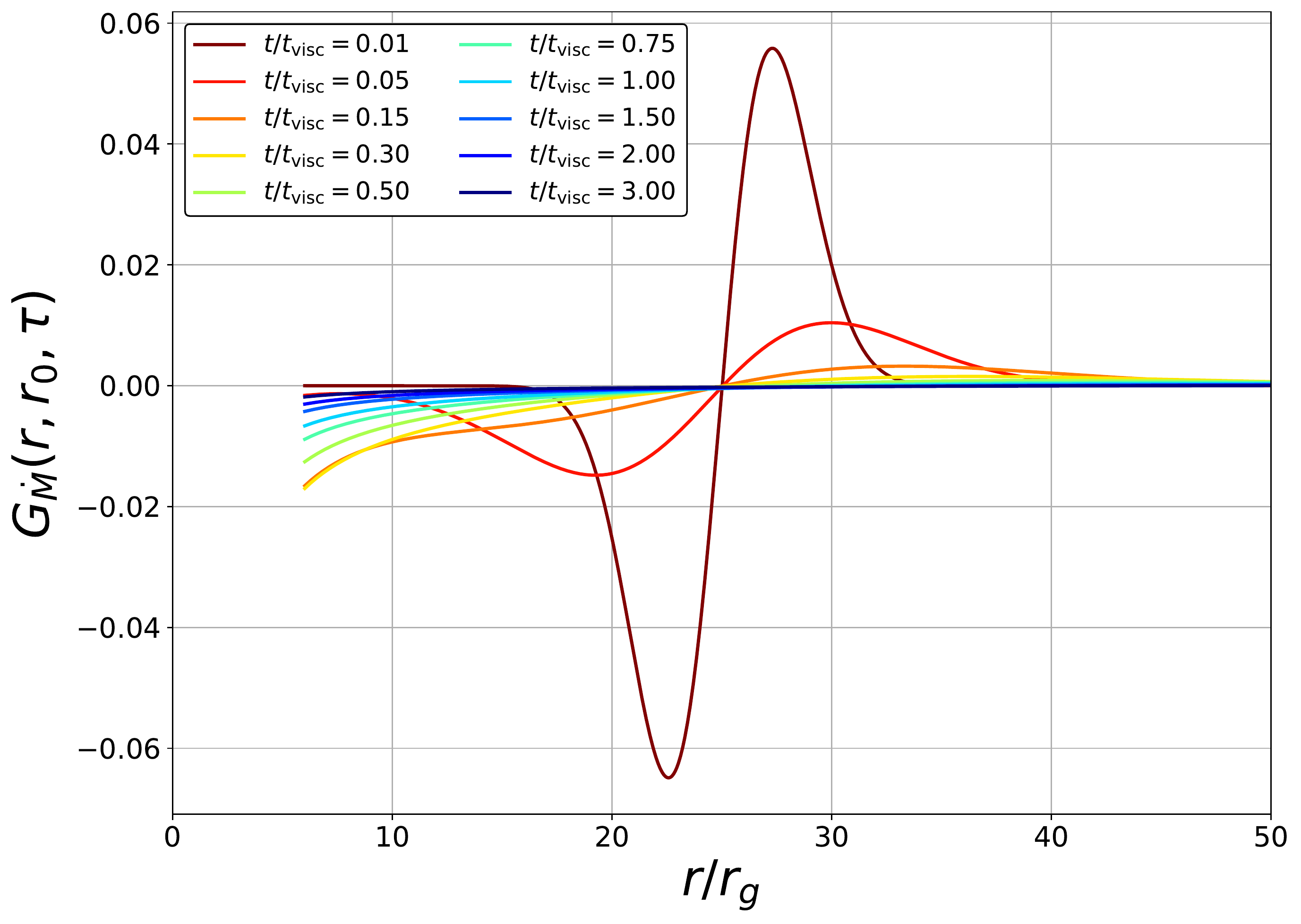}
\caption{The Green's function solution of the mass accretion rate for a Schwarzschild black hole  ($a = 0$). The initial radius was $r_0 = 25r_g$ and the curves are plotted at the dimensionless times denoted in the legend.  The normalisation of the accretion rate was chosen so that the time-integrated ISCO accretion rate was equal to 1. This Figure is reproduced from Mummery (2023).   }
\label{GFR2}
\end{figure}

Balbus (2017) demonstrated that in full general relativity  the governing disc equation can be expressed in the following compact form 
\beq\label{27}
{\partial \zeta\over \partial t} =  { \W\over (U^0)^2}{\partial\ \over \partial r} \left({U^0\over U'_\phi} \left[   {\partial \zeta \over \partial r}\right] \right).
\eeq
here the primed notation $'$ denotes an ordinary derivative with respect to $r$, and 
\beq
\zeta \equiv {r \Sigma \W \over U^0} .
\eeq 
Only two of the orbital components of the disc's flow appear in this equation. These are the time dilation factor 
 \begin{equation}
U^0 = \frac{1+a\sqrt{{r_g}/{r^3}}}{\left({1 - {3r_g}/{r} + 2a\sqrt{{r_g}/{r^3} } }\right)^{1/2}} ,
\end{equation}
and the circular orbit angular momentum gradient 
 \beq\label{rel_ang_mom}
 U_\phi ' = \frac{\sqrt{GM} \left( a\sqrt{r_g} + r^{{3}/{2}} \right) \left( r^2 - 6r_g r - 3a^2 + 8a\sqrt{r_g r}\right)}{2r^4 \left(  1 - {3r_g}/{r} + 2a\sqrt{{r_g}/{r^3}}   \right)^{{3}/{2}}}.
 \eeq
 In these expressions, $a$ is the black hole's angular momentum parameter (having dimensions of length), $M $ is the black hole's mass, $r_g = GM/c^2$ the gravitational radius, and $G$ and $c$ are Newton's constant and the speed of light respectively. 
 
 Clearly, the relativistic disc equation (\ref{27}) is extremely algebraically complex, and in fact  the general relativistic thin disc equation does not have exact solutions that can be written in terms of elementary functions. However, in a recent work Mummery (2023)  derived the leading order general relativistic  Green's function solution, valid for the case of a vanishing stress at the innermost stable circular orbit (ISCO). The Mummery (2023) Green's function solution has the same functional form as the Newtonian solutions discussed above, but with (note that these solutions are in units where $G = M = w = 1$, see Appendix \ref{full_GR} for the relevant solutions presented in physical units)
\begin{multline}
g(x) = {x^\alpha \over 2 \alpha} \sqrt{1 - {2\over x}}\left[1  - {x^{ - 1} \over { (\alpha - 1)}} {}_2F_1\left(1, {3\over 2}-\alpha; 2-\alpha; {2\over x}\right) \right] \\ + {2^{\alpha - 2} \over \alpha (\alpha - 1)}\sqrt{\pi} {\Gamma(2-\alpha)  \over  \Gamma({3/ 2} - \alpha)} , 
\end{multline}
\beq
q(x) = x^{1/4} \sqrt{x^{-\alpha} g(x)} \exp\left({1 \over 2 x}\right) \left[1 - {2\over x}\right]^{5/4 - 3/8\alpha} ,
\eeq
and 
\beq
p(x) = {x^{1/2} \exp\left(-{1/ x}\right) \over 1 - {2/ x}} ,
\eeq
where 
\beq
x \equiv {2 r \over r_I}, \quad \alpha = {1\over 4\nu} , \label{spindepfun}
\eeq 
and ${}_2F_1(a, b; c; z)$ is the hypergeometric function.  For a full description of the approximations employed in deriving this solution see Mummery (2023).  Note that in the $x\to \infty$ limit the above solutions revert to their Newtonian form. 

While the above solutions are not formally exact, in Figure \ref{GFR} we plot the numerically (blue dashed curve) and analytically (green solid curve) computed $r\Sigma \W$ profiles, assuming an initial radius of $r_0 = 50r_g$ and Kerr angular momentum parameter $a = 0$.  The curves are plotted at dimensionless times $t/t_{\rm visc} = 0.003, 0.015, 0.045, 0.09, 0.15, 0.225, 0.45$ and $4.5$, the curves at later times are identifiable through their decreasing peak amplitude.  It is remarkable how accurately the analytical Green's function solutions described above reproduces the properties of the full numerical solutions.

In Figure \ref{GFR2}  we show the Green's function of the mass accretion rate, plotted as a function of radius, for the spin parameter $a = 0$, at a number of different dimensionless times denoted on each plot. The initial radius in both cases was taken to be $r_0 = 25r_g$.  A value of $\dot M(r, t) < 0$ denotes mass inflow (towards the ISCO), while $\dot M(r, t) > 0$ denotes mass outflow (some mass must move outwards within the disc so as to conserve the total angular momentum of the flow).  The normalisation of the accretion rate was chosen so that the time-integrated ISCO accretion rate was equal to 1. 

\subsection{Relativistic Green's functions in the frequency domain}
With this solution in hand,  we may write down the Fourier-Green's function solutions of the relativistic disc equation  by substituting the above definitions into equation \ref{general_def}. The full expressions for the relativistic Fourier-Greens functions are rather complex, and are presented in Appendix \ref{full_GR}.  These solutions have the same gross properties in the Fourier domain as their Newtonian analogues, but they differ in the details, as we now discuss. 

\begin{figure}
\includegraphics[width=\linewidth]{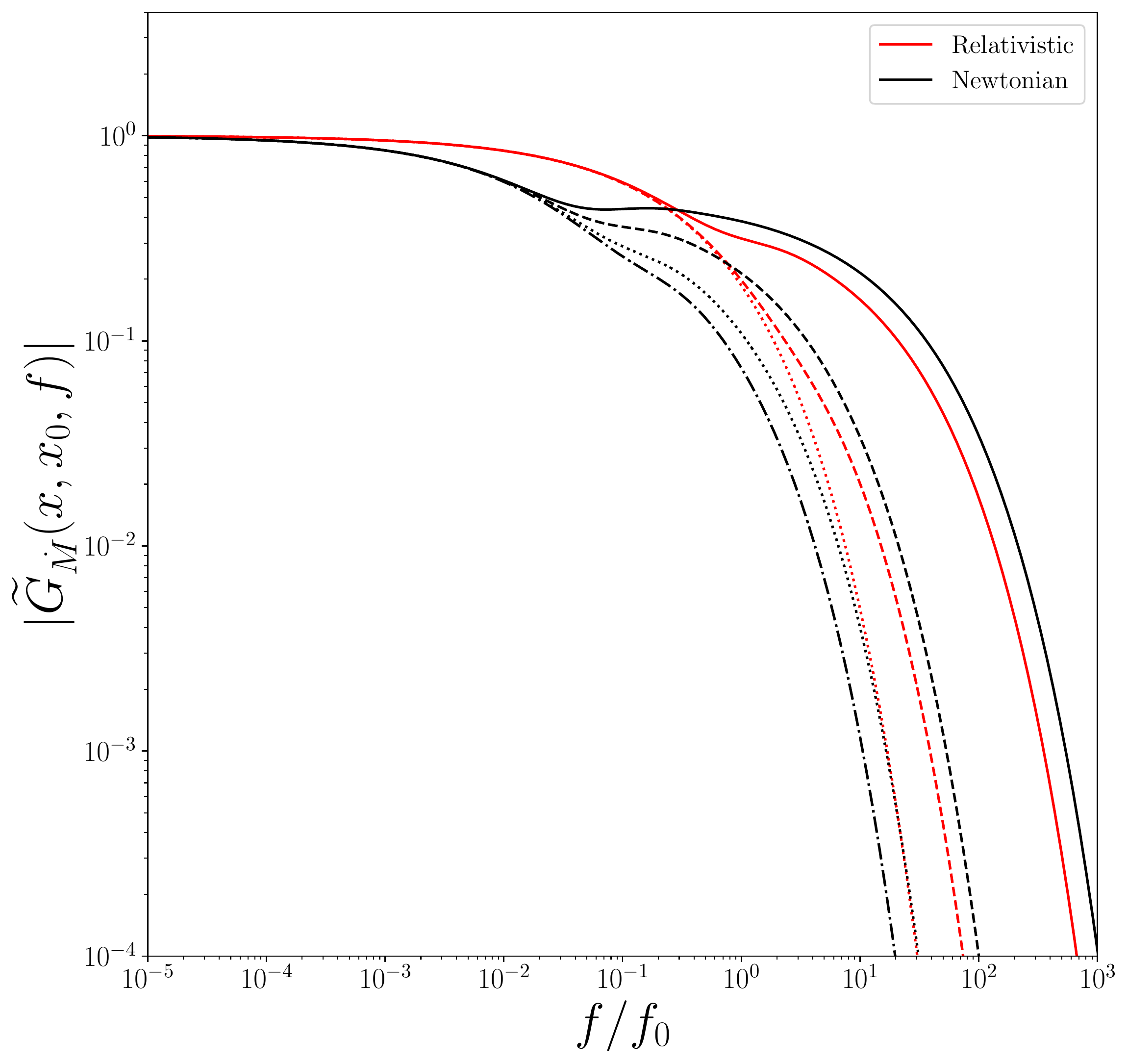}
\caption{The amplitude of the general relativistic Fourier-Green's functions of the mass accretion rate (red), for inward propagating modes $x < x_0$. This calculation was for a Kerr black hole with $a=0.5$.  Newtonian modes are displayed in black.  For this figure we take $r_0 = 10 r_g$, and the inward propagating modes have $r/r_g = 4$ (dot-dashed), $5$ (dotted), $7$ (dashed) and $9$ (solid). The Fourier-Green's functions are normalised so that the total accreted mass is equal to 1. We note that the relativistic Fourier-Green's modes are more strongly suppressed at high frequencies, but are larger at intermediate frequencies.  }
\label{DD1}
\end{figure}

\begin{figure}
\includegraphics[width=\linewidth]{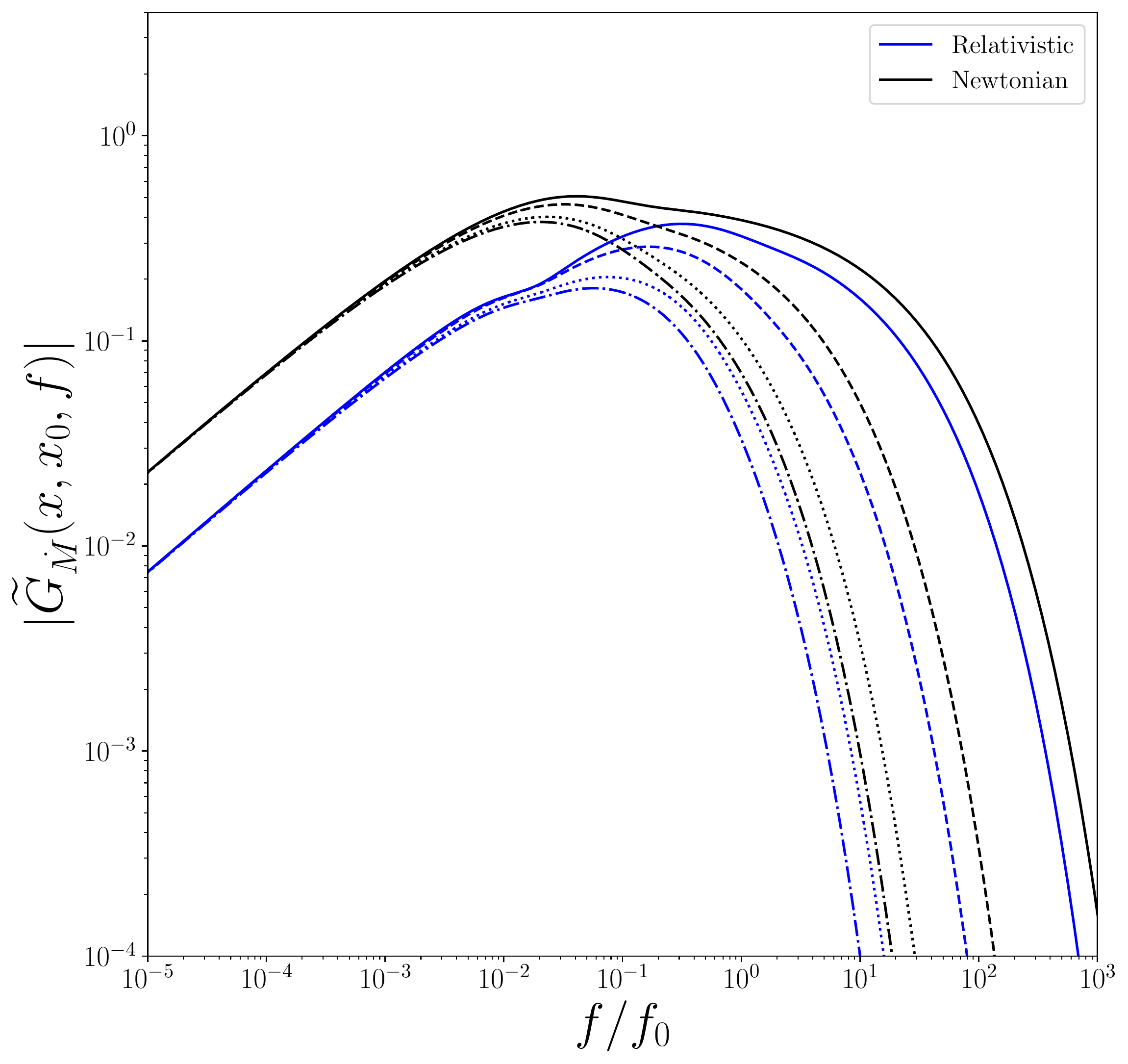}
\caption{The amplitude of the general relativistic Fourier-Green's functions of the mass accretion rate (blue), for outward propagating modes $x > x_0$. This calculation was for a Kerr black hole $a=0.5$.  Newtonian modes are displayed in black.  For this figure we take $r_0 = 10 r_g$, and the outward propagating modes have $r/r_g = 11$ (dot-dashed), $13$ (dotted), $15$ (dashed) and $17$ (solid). The Fourier-Green's functions are normalised so that the total accreted mass is equal to 1. We note that the outward propagating relativistic Fourier-Green's modes are more strongly suppressed at all frequencies, when compared to Newtonian modes.}
\label{DD2}
\end{figure}

\begin{figure}
\includegraphics[width=\linewidth]{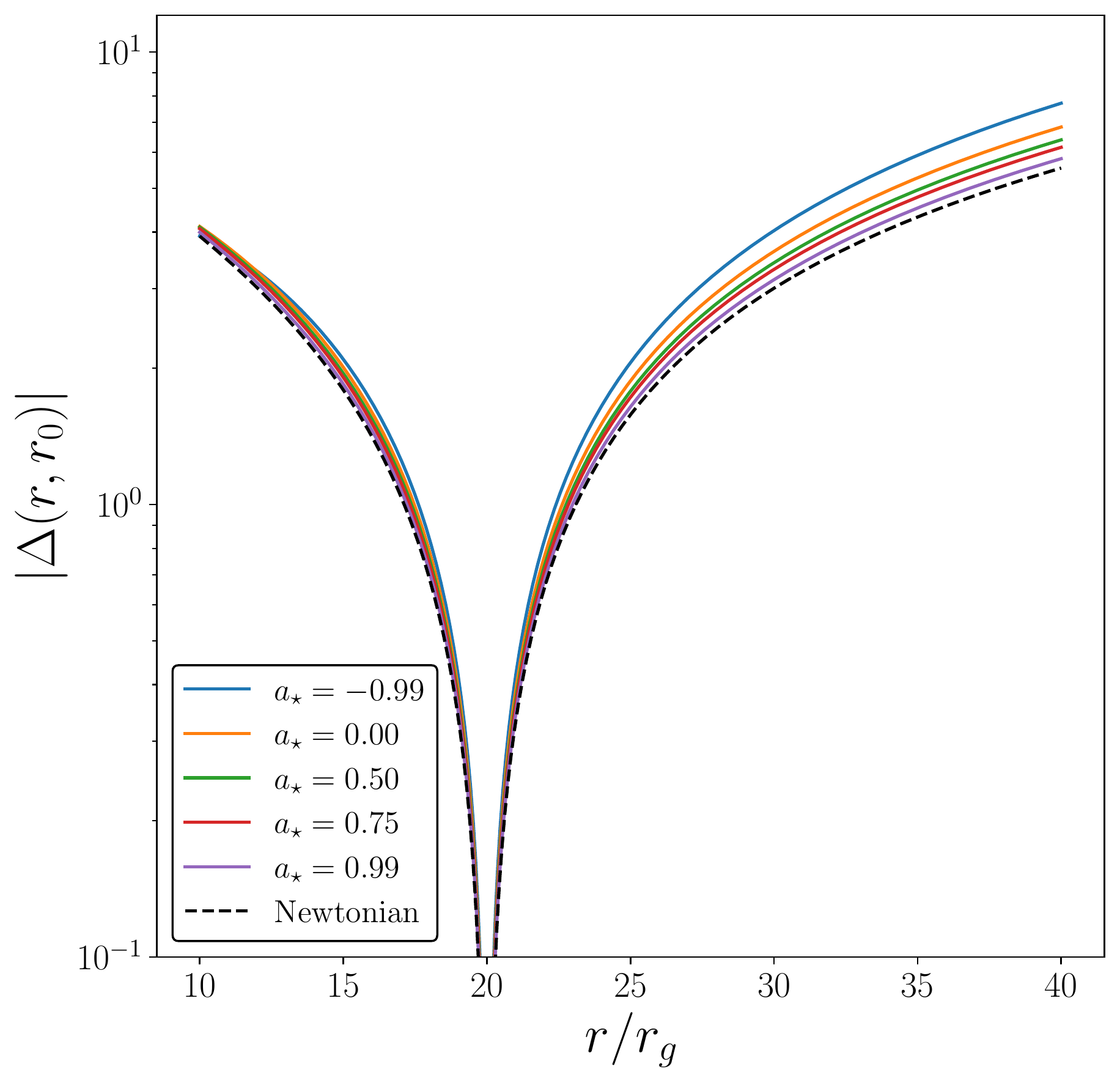}
\caption{The high-frequency  mode suppression function $\Delta(r, r_0) \equiv |g(r) - g(r_0)|$, which suppresses modes as $\exp(-\Delta f^{1/2})$ at high Fourier frequencies, as a function of black hole spin for $r_0 = 20r_g$.  We see that relativistic modes are more strongly suppressed at high Fourier frequencies than Newtonian modes, but that higher black hole spins reduces this suppression.  }
\label{DD3}
\end{figure}

In Figures \ref{DD1} and \ref{DD2} we plot the amplitude of the general relativistic Fourier-Green's functions of the mass accretion rate, for inward (Fig. \ref{DD1}) and outward (Fig. \ref{DD2}) propagating modes.  In black we plot the corresponding Newtonian Fourier-Green's solutions. We note that for inward propagating modes the relativistic Fourier-Green's modes are more strongly suppressed at high frequencies, but are larger at intermediate frequencies, than their Newtonian analogues. In contrast, for outward propagating modes the relativistic Fourier-Green's modes are more strongly suppressed at all frequencies.

It is important to note that the outward propagating relativistic Fourier-Green's modes must be treated with some care. Mathematically this results from the fact that the solutions described in the proceeding sub-section are not exact, but are asymptotic ``leading order'' solutions (see Mummery (2023) for a detailed discussion). Physically care is required because at very late times the exact numerical and analytical solution begin to deviate, and an extremely small fraction of the initial disc mass is not accreted in these solutions.  As a result 
\beq
 q(x) \neq A (g(x))^{\nu}
\eeq
and therefore (following the reasoning outlined in section 2)
\beq
\lim_{f \to 0} \widetilde G_{\dot M}(x > x_0, f)  \to \delta(x, x_0) M_d \neq 0 .
\eeq
The discrepancy is  small, as a result of the high accuracy of these analytical solutions (Fig. \ref{GFR}), and typically only effects extremely small frequencies $f/f_0 \ll 10^{-6}$ to a small degree
\beq
\delta \ll 10^{-4}. 
\eeq
As such, the conclusions derived in the remainder of this paper are not noticeably quantitatively or qualitatively effected by this slight discrepancy. The function $\delta(x, x_0)$ can be written in a closed form (Appendix \ref{full_GR}), and then simply subtracted off the low frequency Fourier-Green's modes. 

\subsection{Differences between the Newtonian and Relativistic Fourier-Green's functions}
One of the main differences between the Newtonian and relativistic Fourier-Green's functions is highlighted in Figures. \ref{DD1} and \ref{DD2}: relativistic Fourier-Green solutions are more strongly suppressed at high Fourier frequencies  than their Newtonian analogues. This can be understood by plotting the high-frequency  mode suppression function $\Delta(r, r_0) \equiv |g(r) - g(r_0)|$ (Fig. \ref{DD3}). High frequency modes are suppressed as as $\exp(-\Delta f^{1/2})$ (see section 2). In Fig. \ref{DD3} we plot $\Delta(r, r_0)$  as a function of black hole spin for $r_0 = 20r_g$.  We see that relativistic modes are more strongly suppressed at high Fourier frequencies than Newtonian modes, but that higher black hole spins reduce this suppression.

\begin{figure}
\includegraphics[width=\linewidth]{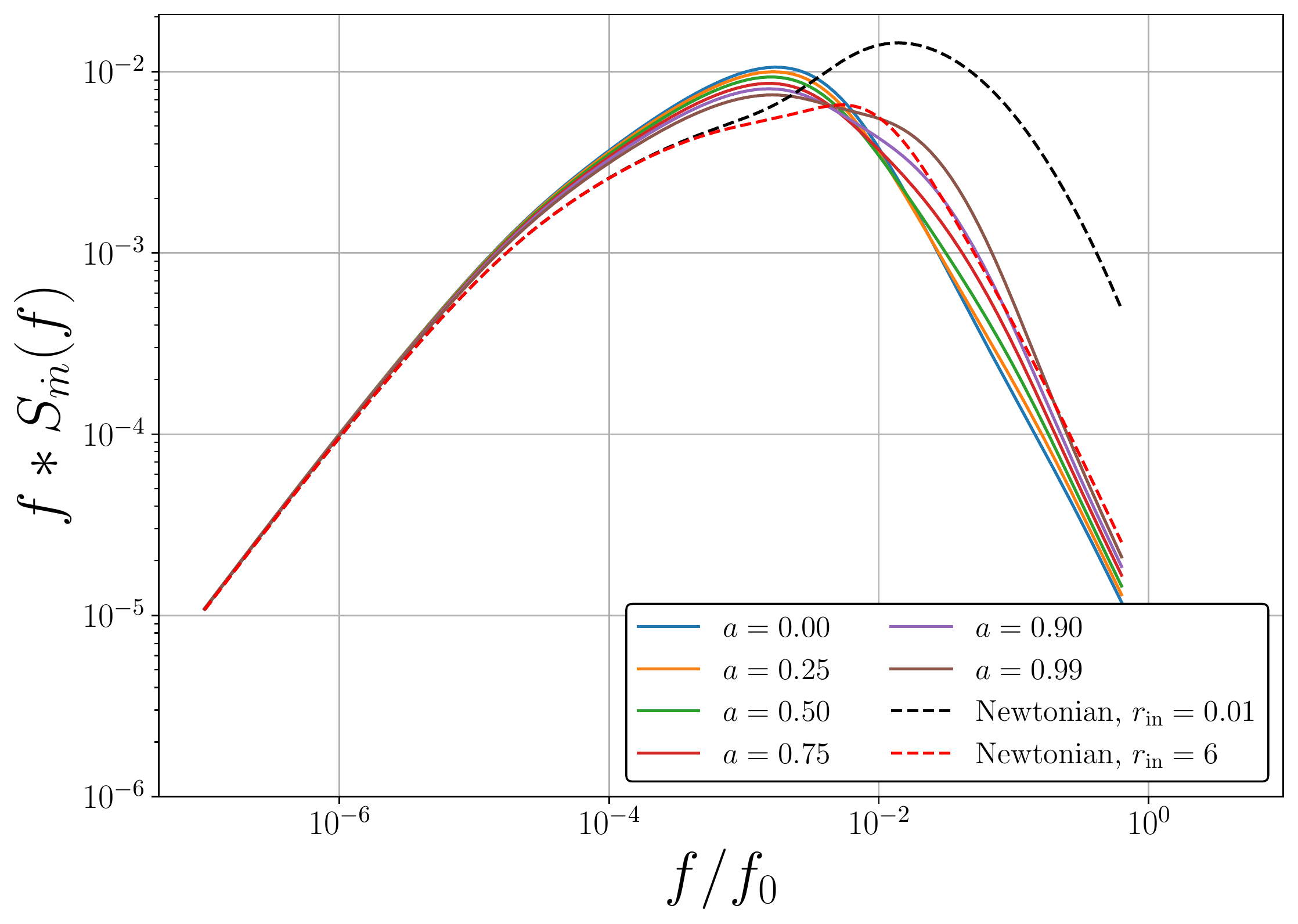}
\caption{The power density spectrum of the mass accretion rate multiplied by the Fourier frequency, at a disc radius $r = 10r_g$ for a number of black hole spins (denoted in legend), as a function of Fourier frequency. Displayed as black and red dashed curves are the equivalent Newtonian solutions with different ``inner radii'' (the radius at which the input power is set to zero, see text). Local relativistic and Newtonian accretion rate power density spectra are noticeably distinct.    }
\label{DD4}
\end{figure}

One interesting result that is highlighted in Figure \ref{DD1} is that the inwards propagating relativistic and Newtonian Fourier-Green's functions differ at intermediate frequencies, with Newtonian modes having a more pronounced ``kink'' at $f \sim 10^{-2} f_0$. This difference propagates through into different properties of the local accretion rate power spectra of relativistic and Newtonian discs (Fig. \ref{DD4}). In Figure \ref{DD4} we compute the local (at radius $r$) mass accretion rate power spectrum using equation \ref{local_power_spec}, for a range of different black hole spins and also for Newtonian discs with different ``inner radii''. To mimic an effective inner disc radii for the Newtonian solutions we set the input power of the surface density variability (see equation \ref{Lorentzian}) to be zero for $r \leq r_{\rm in}$. It is clear to see in Figure \ref{DD4} that the relativistic and Newtonian Fourier-Green's functions result in significantly different local accretion rate power spectra. While this is particularly true for the case of a Newtonian disc with small inner radius $r_{\rm in} = 0.01$ (black dashed curve), this is still true for Newtonian discs with more realistic inner radii (red dashed curve).

\section{Observable quantities }

The previous sections have focused on the properties of the Green's function solutions of the Newtonian and relativistic disc equations in the Fourier domain. The Fourier properties of the local mass accretion rate are of course not by themselves directly observable, and in this section we focus our discussion onto the properties of the emitted disc flux, which is a more readily observable quantity. 

The problem now becomes one of determining how mass accretion rate fluctuations are observed in the light curves of accreting sources. In a purely steady state flow the local energy available to be radiated is directly proportional to the local mass accretion rate, and in the literature it is generally assumed that the {\it local variability} in the accretion rate is directly proportional to the variability in the local emission (e.g., Lyubarskii 1997, Ingram \& van der Klis 2013, Ingram \& Done 2012, Mushtukov et al, 2018). This is unlikely to hold much beyond small fluctuations in $\dot M$, as the locally radiated energy in a time dependent flow is not directly proportional to the accretion rate, but instead to the energy liberated by the local disc shear. In a time dependent accretion flow the local mass accretion rate can in fact be negative (see Fig.  \ref{GFR2}), while the locally emitted flux will of course remain positive.

In this work, in common with the literature,  we will also make the assumption that the variability in the local accretion rate is directly proportional to the variability in the locally emitted photon flux. We stress however that it is important to bear in mind that this assumption can only hold for small variations in both the photon flux and mass accretion rate. 

The emission in a certain band (which we shall call ``hard'' $h$, or ``soft'' $s$) is therefore assumed to be correlated with the local accretion rate, with some ``emissivity profile''  $s(r)$ and $h(r)$ (e.g., Lyubarskii 1997, Ingram \& van der Klis 2013, Ingram \& Done 2012, Mushtukov et al, 2018) which specifies the constant of proportionality between the two quantities. {Much of the variable emission in a typical observation of an X-ray binary system is sourced from a hot ``corona''. The corona is a population of hot electrons, located close to the black hole,  whose existence is inferred from the observation of a power law spectral component resulting physically from photons being Compton up-scattered as they pass through the electrons.  The geometry of this corona is currently poorly understood, and could in principal  be located above the disc,  or as part of a hot ``inner flow''.  The advantage of working with emissivity profiles, as opposed to any particular physical model for the emission, is that under the assumption that  coronal emission also scales locally with the mass accretion rate (perhaps as a result of variability in the seed photon field), variability in both thermal and non-thermal emission components may be modelled with additional  degrees of freedom. We discuss potential extensions to this analysis in section 7.     }

It is common in the literature to assume that these emissivity profiles are given by power laws of disc radius $r$ (although there is no compelling justification for this), and in this work we shall parameterise our emissivity profiles with the following functional form  
\begin{align}
s(r) &= s_0 \left({r \over r_I}\right)^{-\gamma_s} \left({1 - \sqrt{r_I \over r} } \right),\\
h(r) &= h_0 \left({r \over r_I}\right)^{-\gamma_h}  \left({1 - \sqrt{r_I \over r} } \right) .
\end{align}
Here $r_I$ is the black hole's ISCO radius, $\gamma_{h, s}$ are phenomenological emissivity indices, and the final term $1-\sqrt{r_I/r}$ enforces the vanishing ISCO stress condition used in deriving the relativistic Green's functions. The constants $s_0$ and $h_0$ are arbitrary scaling factors included for dimensional reasons which we shall simply set equal to 1 for the remainder of this paper.  We require $\gamma_h > \gamma_s \geq 2$, so that the hard emission is produced at radii interior to the soft emission ($\gamma_h > \gamma_s$) and that the flux observed in either band falls off with radius at least as quickly as the total liberated photon flux ($\gamma_{h,s} \geq 2$). The first requirement, that $\gamma_h > \gamma_s$, is in effect simply the definition  of the ``hard'' and ``soft'' bands: the flux from the inner regions is emitted at higher photon energies, and will thus contribute more to harder (higher energy) bands. As the disc cools at larger radii, the relative contribution of that disc region to harder bands will be suppressed more than its contribution to softer bands.   

In the following three subsections we shall present formal expressions for three readily observable quantities: the  power density spectrum of a band, and the cross spectrum and coherence between different bands. It will then be demonstrated that each of these quantities is a relatively strong function of black hole spin, and could in principle be used to constrain the black hole spins of astrophysical sources. 

\subsection{The power density spectrum of a band}
The first observable quantity we consider is the power density spectrum of the flux variability in a given observing band. 

For simplicity we shall quote the results derived by Mushtukov et al. (2018), before discussing the assumptions inherent to the modelling.  The following expression for the power density spectrum of a band (we here denote the band by $h$, for ``hard band'' emission) assumes that the  total luminosity available to be radiated in a given region of the accretion flow is proportional to the local mass accretion rate $\dot M(r, t)$. Further assuming that the local variability of the mass accretion rate ($\dot m$) is small in comparison with the average mass accretion rate, the fluctuations of the flux in some energy band will also be proportional to $ \dot m (r, t)$. Under the above assumptions Mushtukov et al. (2018) derived the following expression for the power density spectrum of band $h$, denoted $S_h(f)$
\beq
S_h(f) = \int_{\cal D} \int_{\cal D} h(r_1) h(r_2) C_{\dot m} (r_1, r_2, f) \, {\rm d}r_1 \, {\rm d}r_2 , 
\eeq
where we use the shorthand 
\beq
\int_{\cal D} f(r) \, {\rm d}r\equiv \int_{r_{\rm in}}^{r_{\rm out}} f(r) \, {\rm d}r ,
\eeq
to indicate an integral over the entire disc surface ${\cal D}$. We remind the reader that 
\begin{multline}
C_{\dot m}(r_1, r_2, f) = \int_{\cal D} \widetilde G_{\dot M}(r_1, r', f) \,  \widetilde G^\dagger_{\dot M}(r_2,  r', f) \\ \left({1 \over r'}\right)^2 l(r')   S_\Sigma(r', f) \, {\rm d}r',
\end{multline}
and so we have in reality a triple integral to compute 
\begin{multline}
S_h(f) = \int_{\cal D} \int_{\cal D} \int_{\cal D}  h(r_1) h(r_2)  \widetilde G_{\dot M}(r_1, r', f) \,  \widetilde G^\dagger_{\dot M}(r_2,  r', f) \\ \left({1 \over r'}\right)^2 l(r')   S_\Sigma(r', f) \, {\rm d}r' \, {\rm d}r_1 \, {\rm d}r_2  .
\end{multline}
{Note that, as argued in Mushtukov et al. (2018), as the amplitude of the local surface density variability scales with $\dot M^2$, integrating $S_h(f)$ over all frequencies, and then square rooting, one recovers the linear rms-flux relationship. }

The physical reason behind the power density spectrum being related to a {\it triple} integral is the following. The flux in a given energy band is, per our assumptions, given by an integral of the local mass accretion rate, with a weighting function $h(r)$, over the entire disc $F_h(t) \propto \int_{\cal D} h(r') \dot M(r', t) \, {\rm d}r'$.   The power density spectrum corresponds to the square of the Fourier-transformed flux $S_h(f) \equiv \widetilde F_h(f) \widetilde F^\dagger_h(f)$, which introduces the second integration. However, variability in the accretion rate at a given radius $r$ is caused by the integrated contributions of surface density fluctuations {\it at all disc radii}, with differing levels of correlation encapsulated by $C_{\dot m}(r_1, r_2, f)$. Summing each of these individual contributions introduces the third integral over the disc surface.

\begin{figure}
\includegraphics[width=\linewidth]{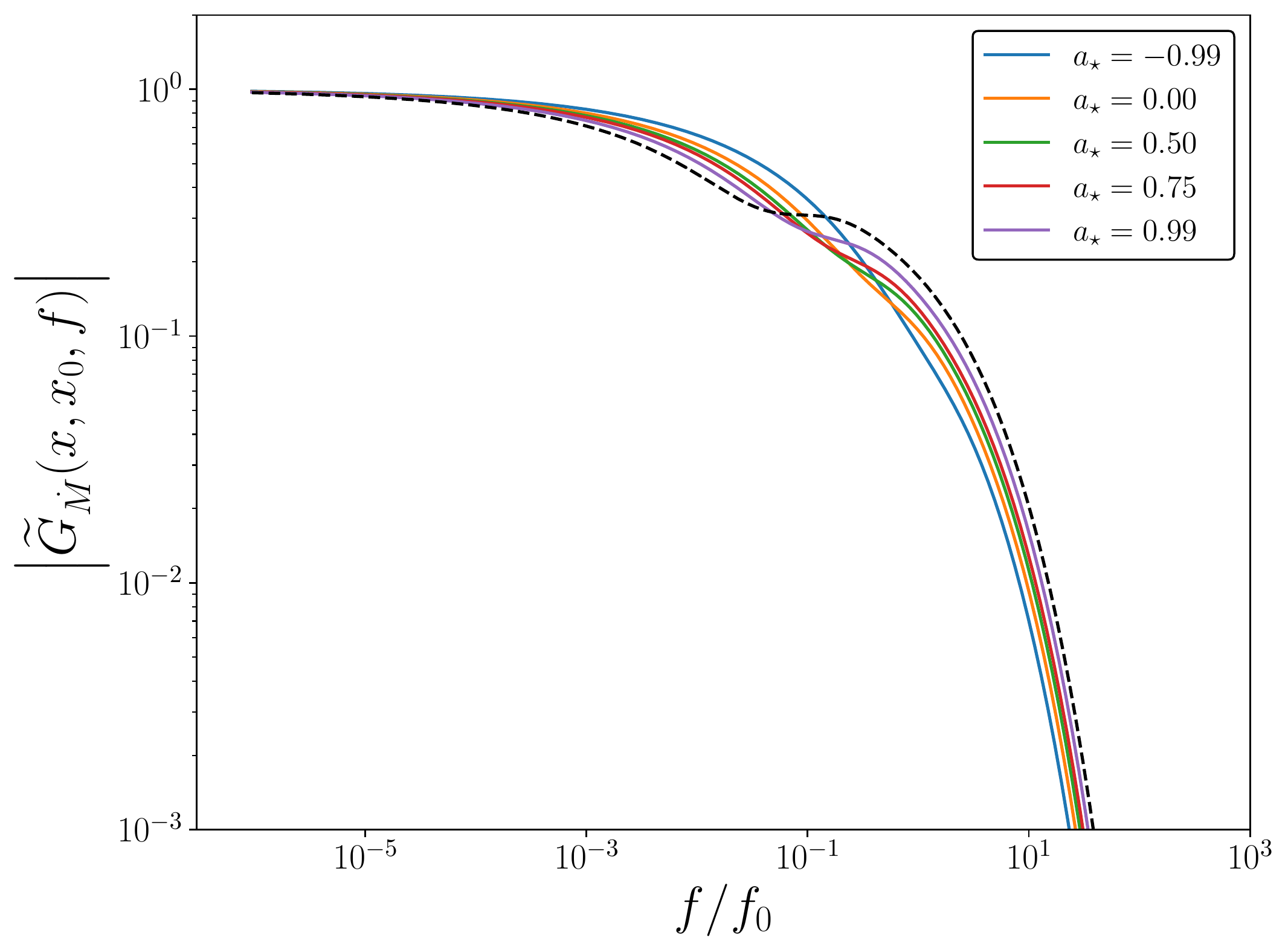}
\caption{The amplitude of the general relativistic Fourier-Green's functions of the mass accretion rate, for inward propagating modes $r < r_0$, for black holes of different Kerr spin parameters.  Newtonian modes are displayed in black.  For this figure we take $r_0 = 20 r_g$, and the inward propagating modes have $r/r_g = 15$. }
\label{SD1}
\end{figure}

\begin{figure}
\includegraphics[width=\linewidth]{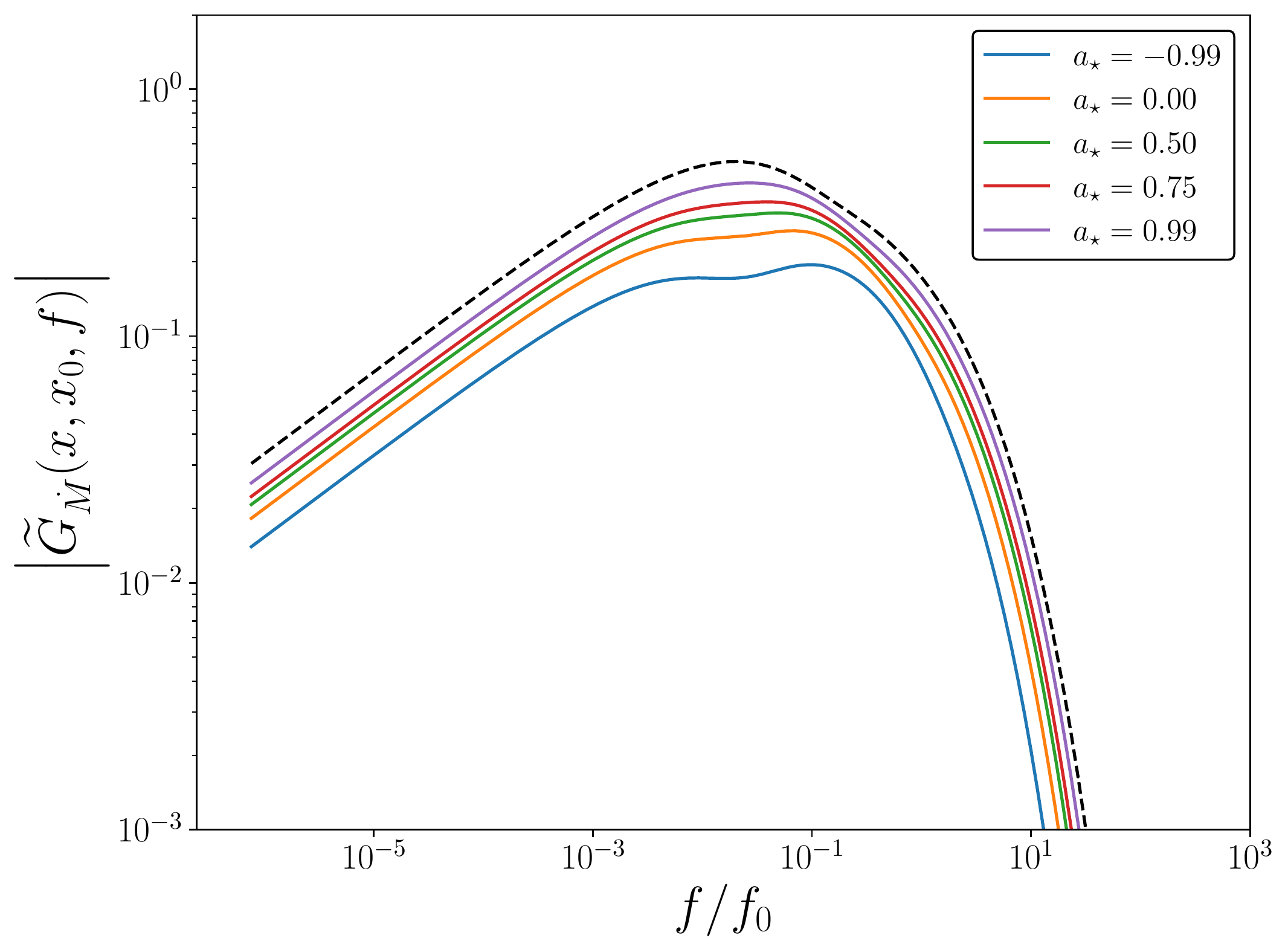}
\caption{The amplitude of the general relativistic Fourier-Green's functions of the mass accretion rate, for outward propagating modes $r > r_0$, for black holes of different Kerr spin parameters.  Newtonian modes are displayed in black.  For this figure we take $r_0 = 15 r_g$, and the inward propagating modes have $r/r_g = 20$. }
\label{SD2}
\end{figure}

\begin{figure}
\includegraphics[width=\linewidth]{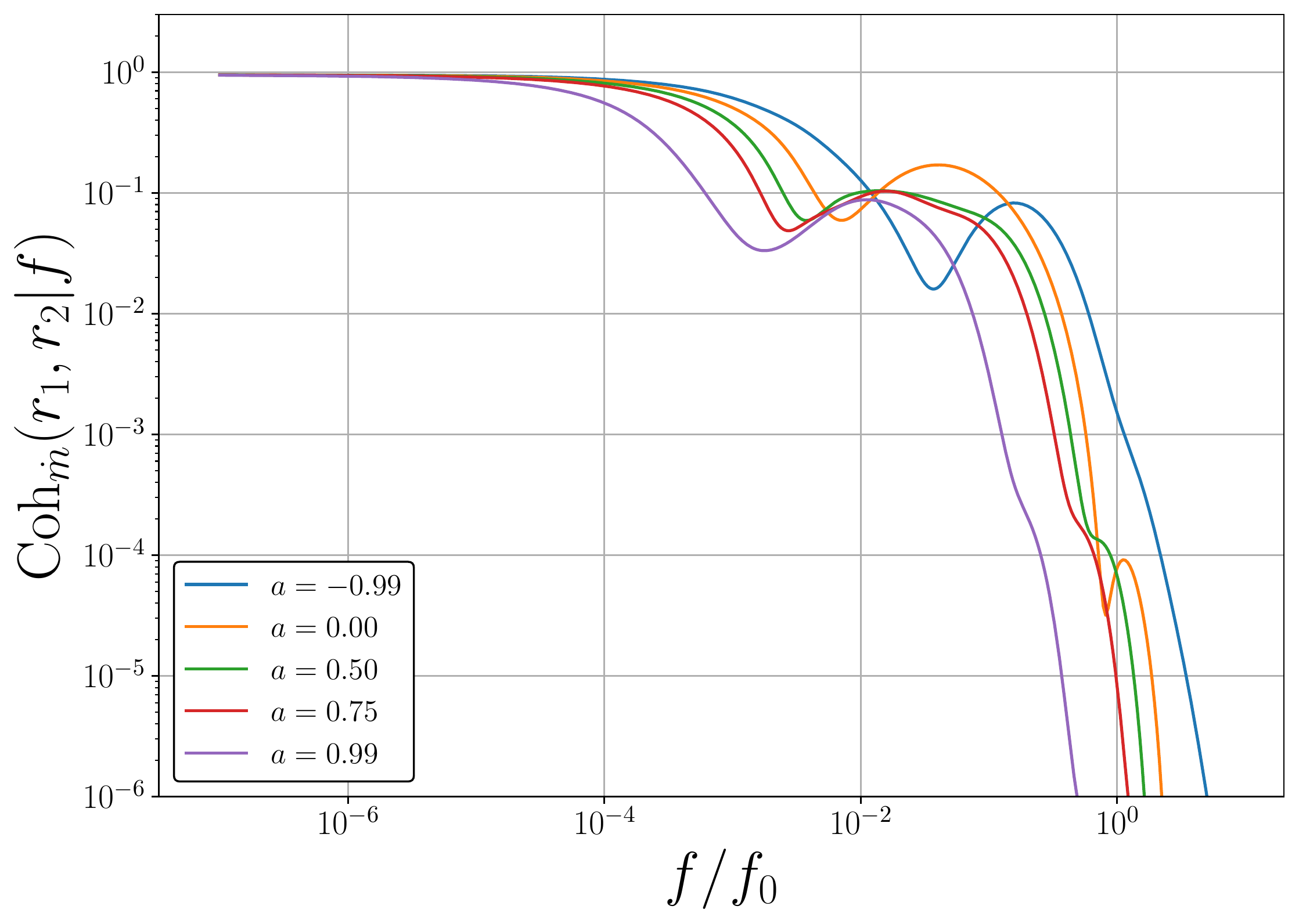}
\includegraphics[width=\linewidth]{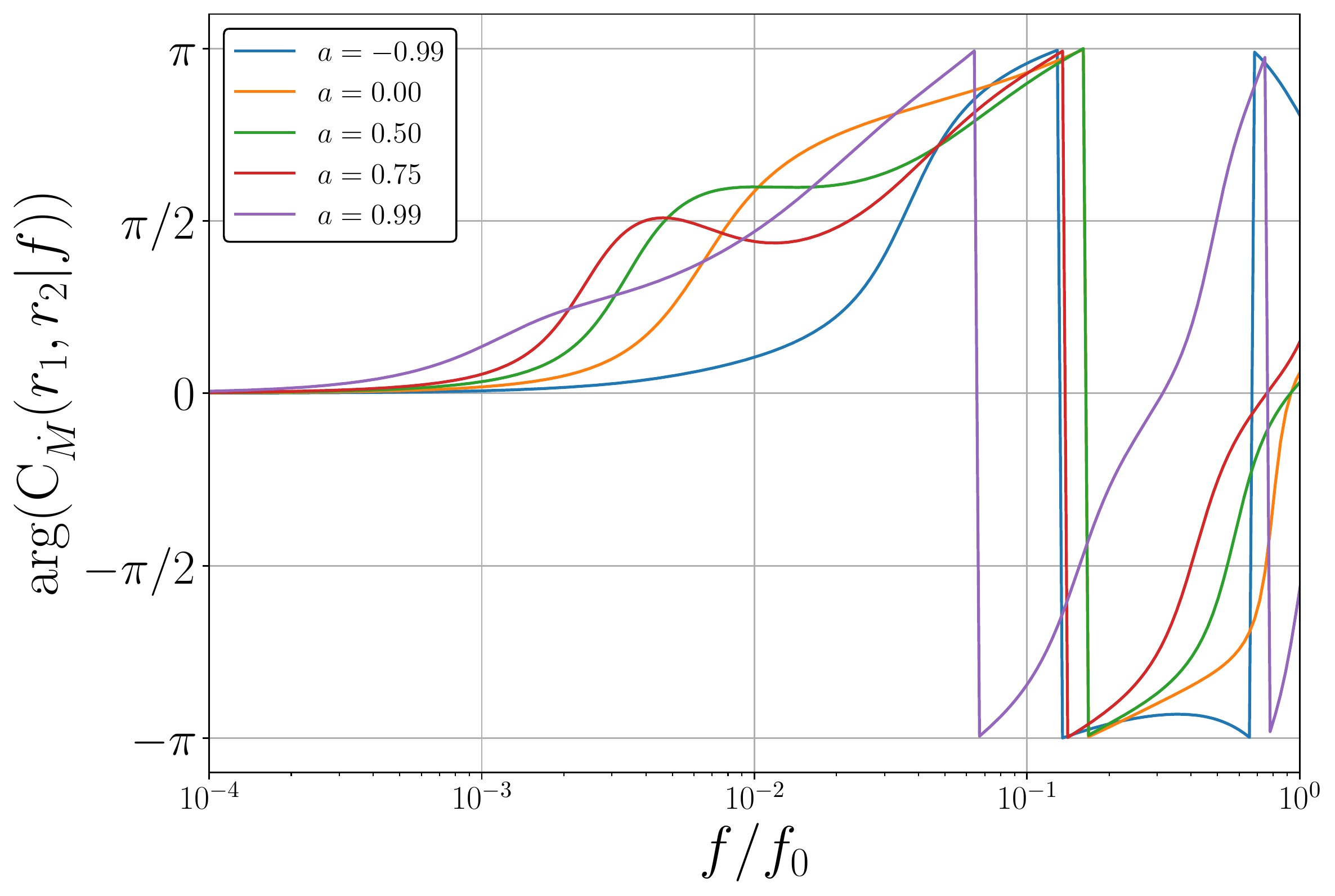}
\caption{{\it Upper:} The coherence function of the mass accretion rate at $r_1 = 10r_g$ and $r_2 = 15r_g$ for Kerr metrics with different spin parameters (denoted on legend). {\it Lower:} The phase of the mass accretion rate cross spectrum between disc radii $r_1=10r_g$ and $r_2 = 15r_g$. At higher frequnecies than displayed in this lower plot the phase wraps between $-\pi$ and $\pi$.  Both the coherence and phase of the mass accretion rate are relatively sensitive to the Kerr metric spin parameter.  }
\label{SD3}
\end{figure}

\begin{figure*}
\includegraphics[width=0.45\linewidth]{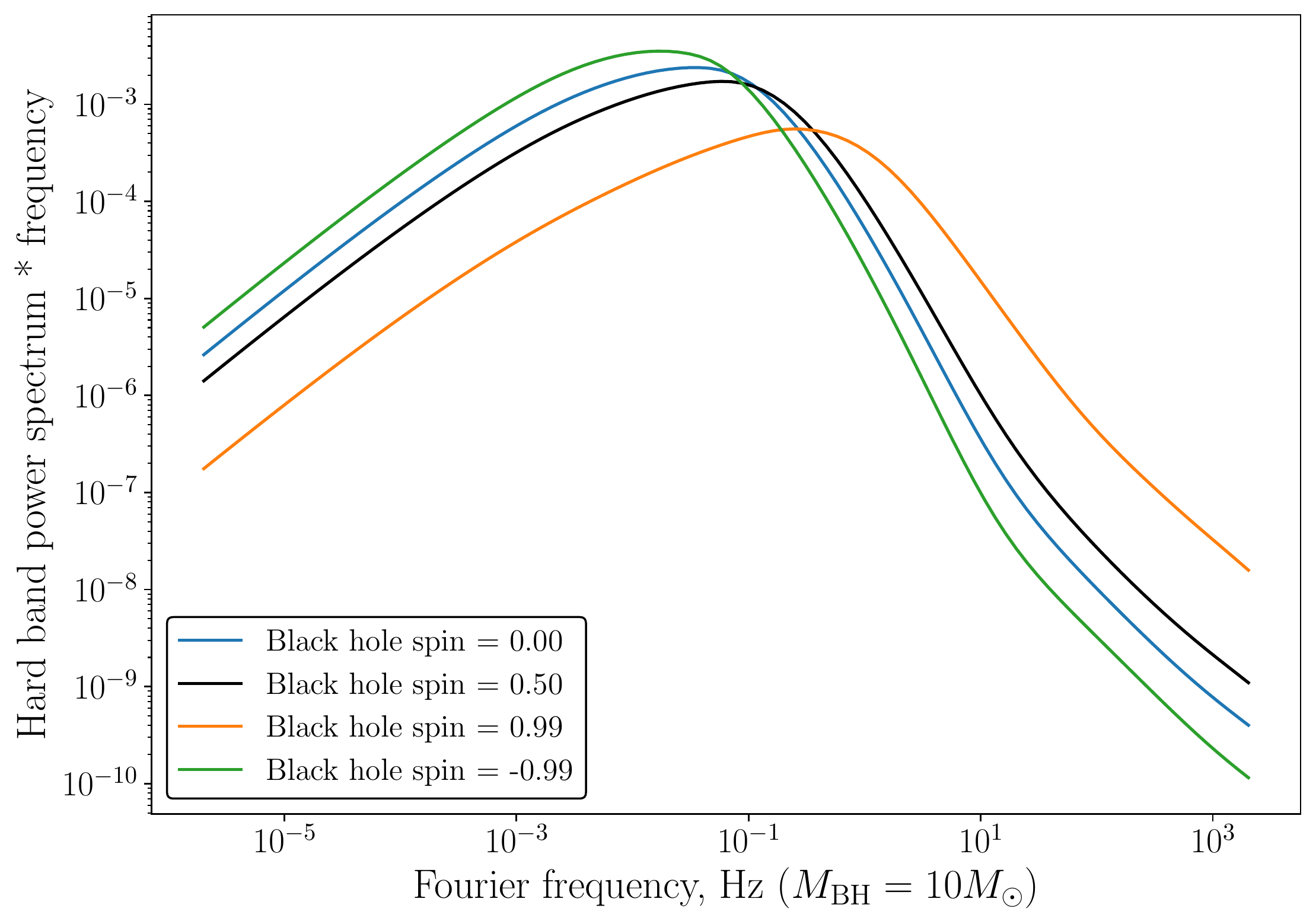}
\includegraphics[width=0.45\linewidth]{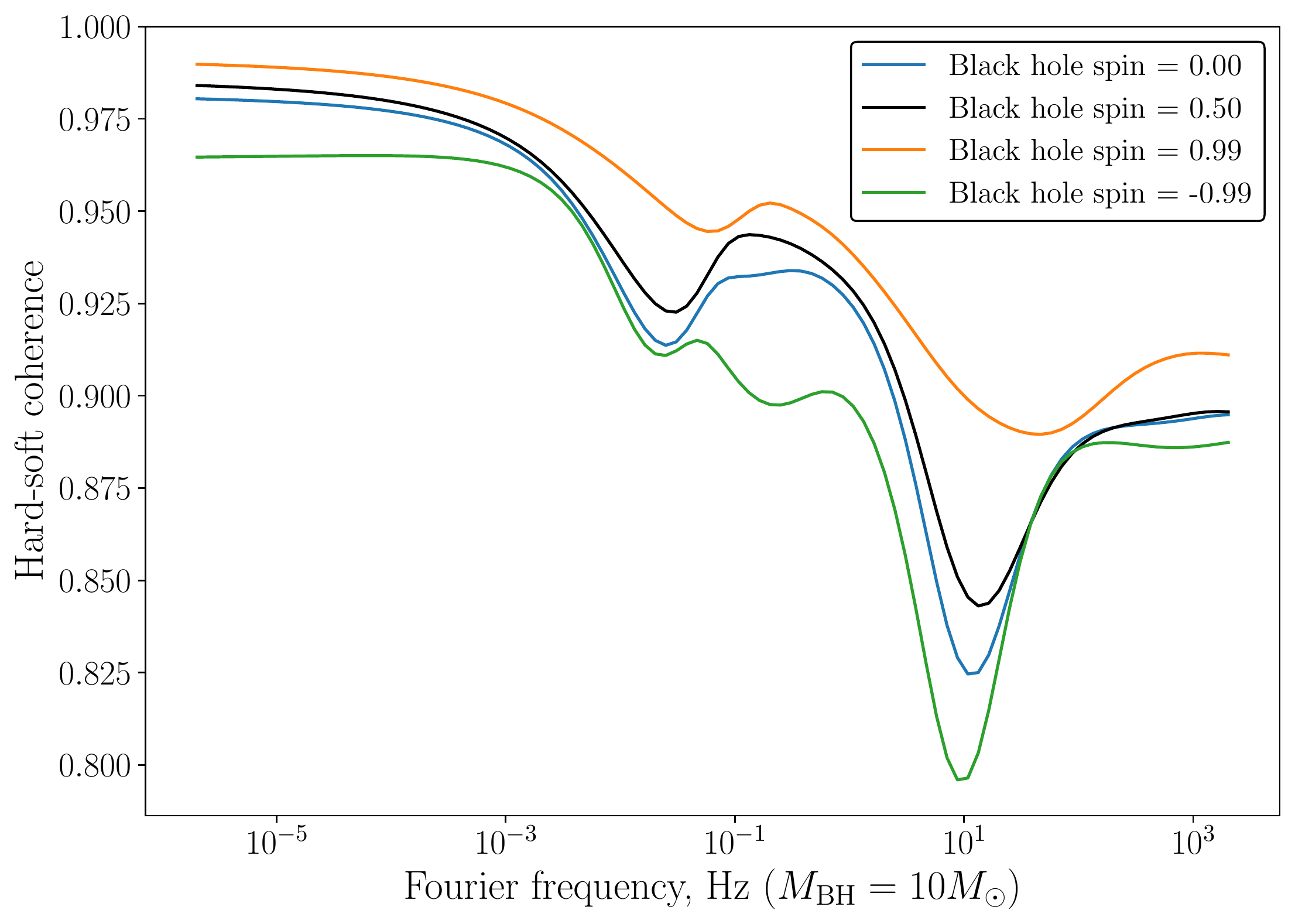}
\includegraphics[width=0.45\linewidth]{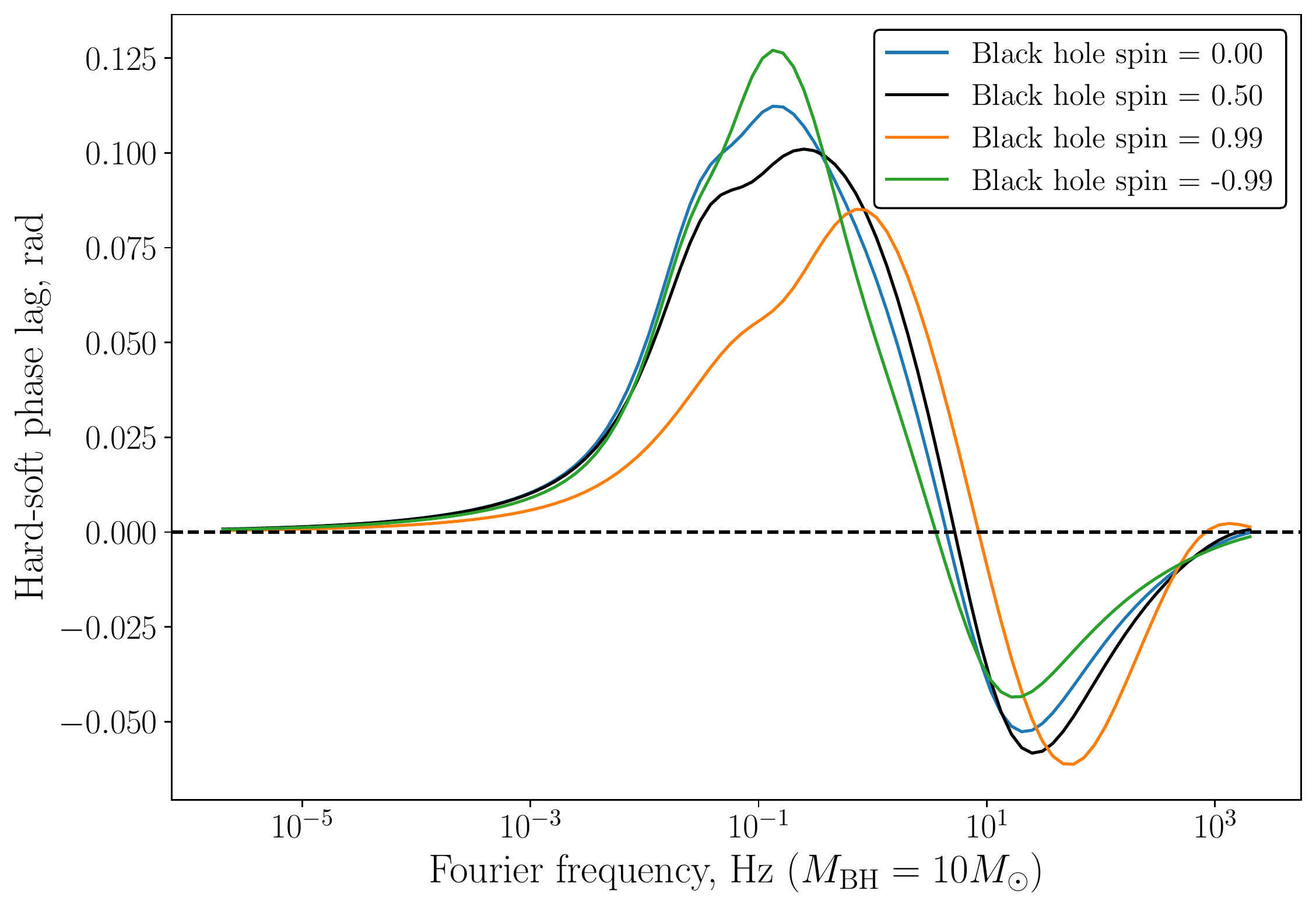}
\includegraphics[width=0.45\linewidth]{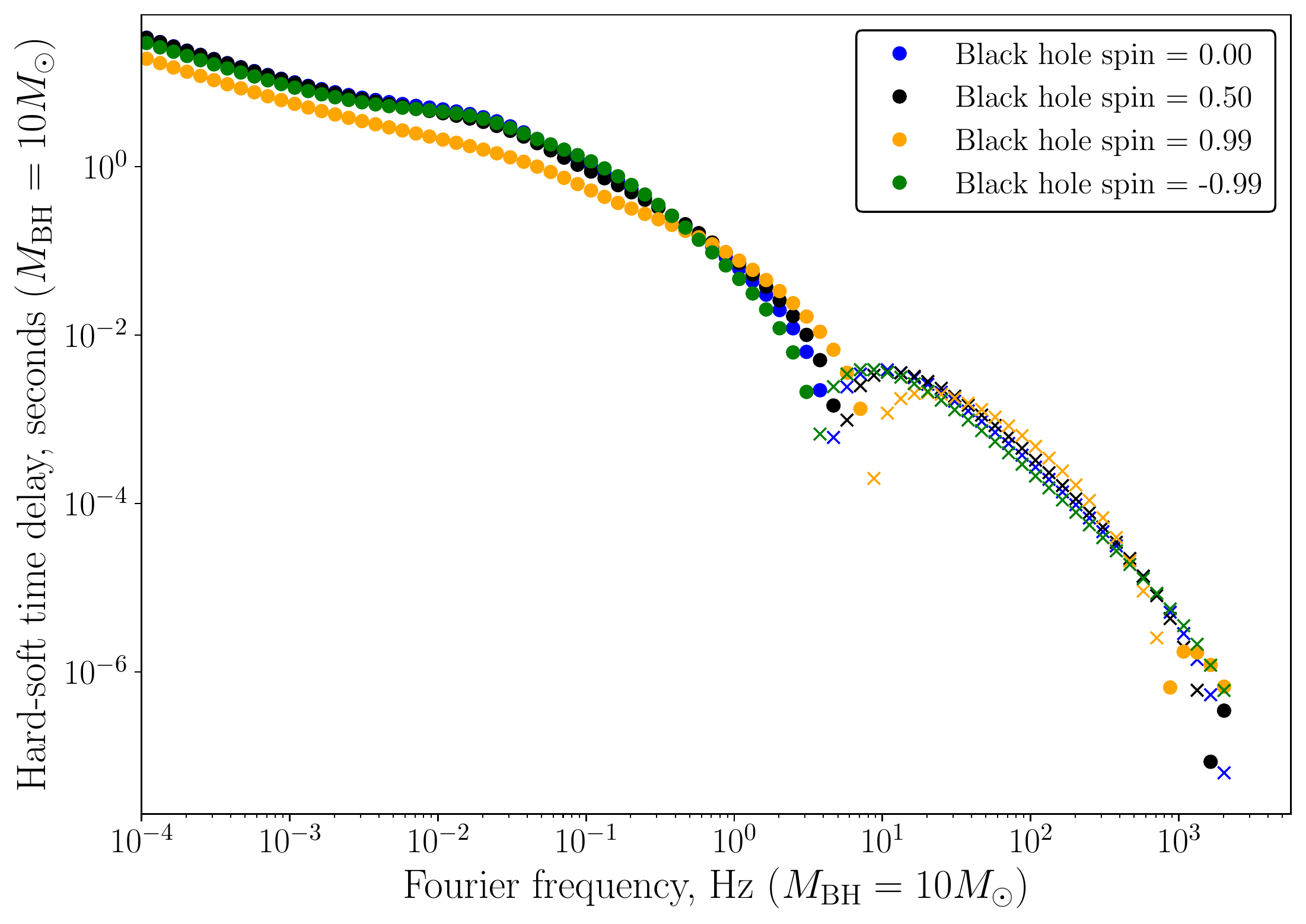}
\caption{ The power density spectrum of the hard band  (upper left), the hard-soft coherence (upper right), the hard-soft phase lag (lower left) and time delay (lower right) for emissivity parameters $\gamma_s = 3$, $\gamma_h = 5$ and four different black hole spins ($a= 0$, blue; $a = 0.5$, black; $a = +0.99$, orange; and $a = -0.99$, green). For the time delay plot we denote by solid dots  positive lags (hard lags soft), and by  crosses negative lags (hard leads soft). The units of the power spectrum are arbitrary and would be set in a physical system by the input power in the surface density variability, the units of frequency and time delay are scaled to a system with black hole mass $M_{\rm BH} = 10M_\odot$, and disc parameters $\alpha = H/R = 0.1$.   }
\label{K1}
\end{figure*}

\begin{figure*}
\includegraphics[width=0.45\linewidth]{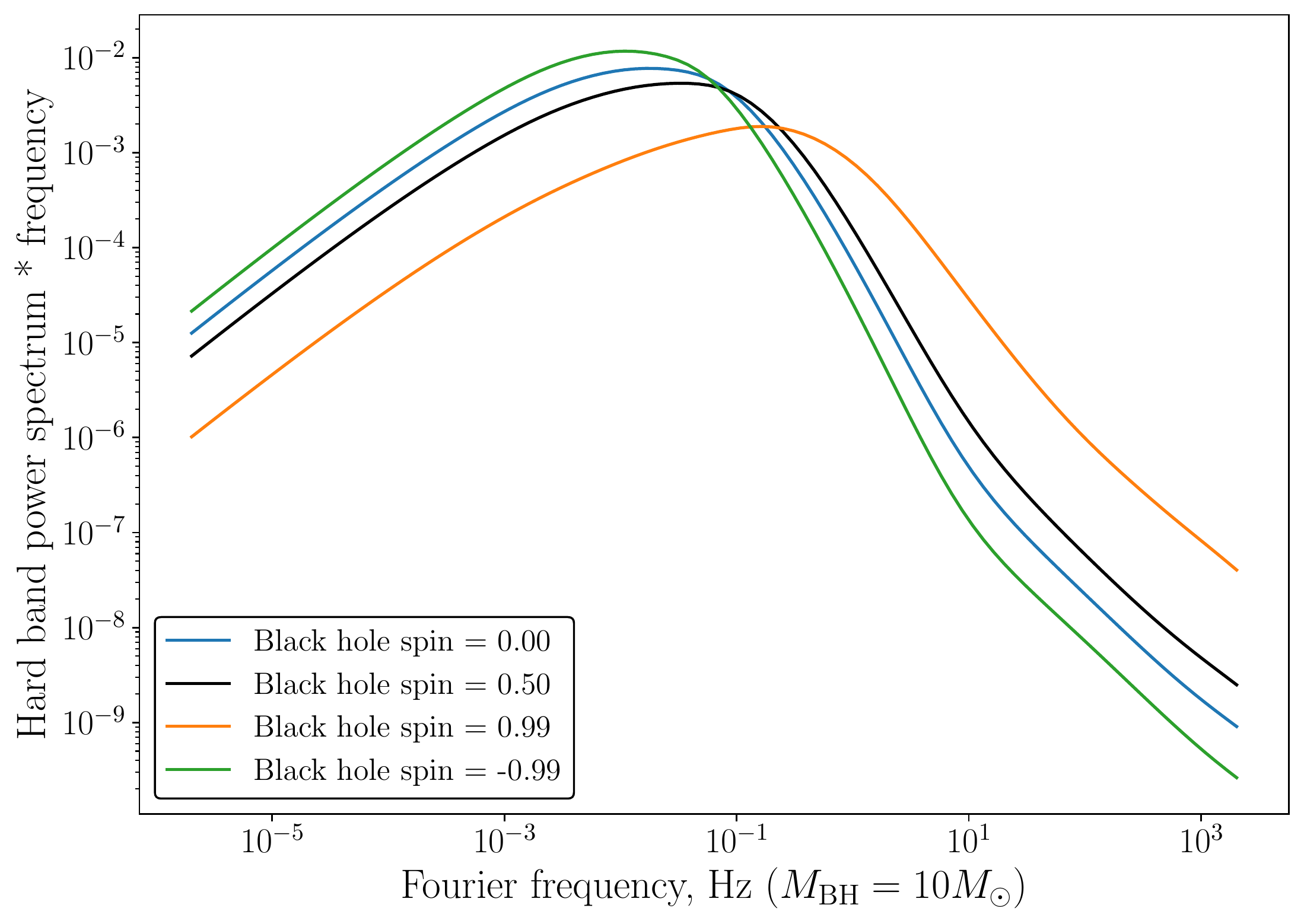}
\includegraphics[width=0.45\linewidth]{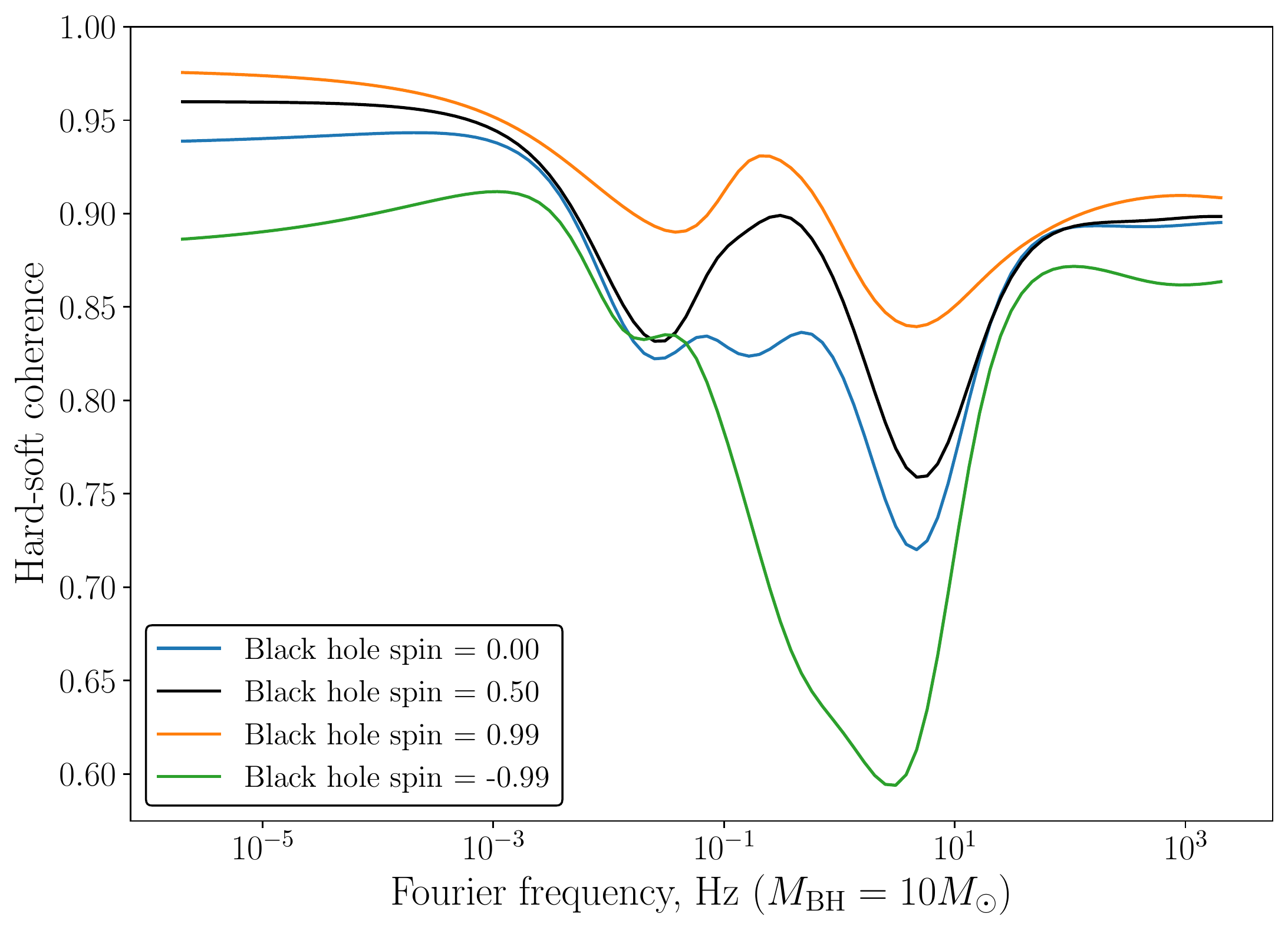}
\includegraphics[width=0.45\linewidth]{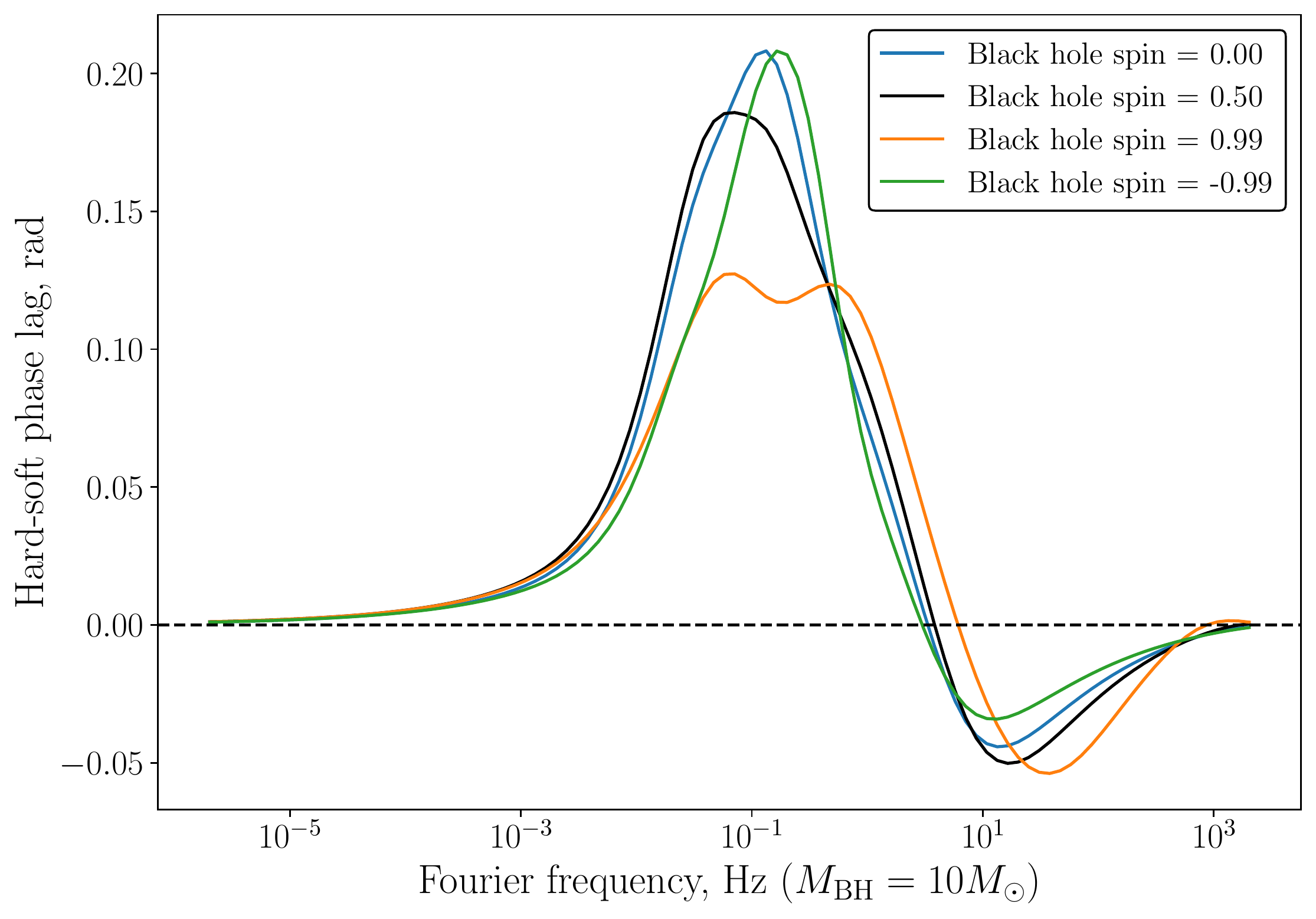}
\includegraphics[width=0.45\linewidth]{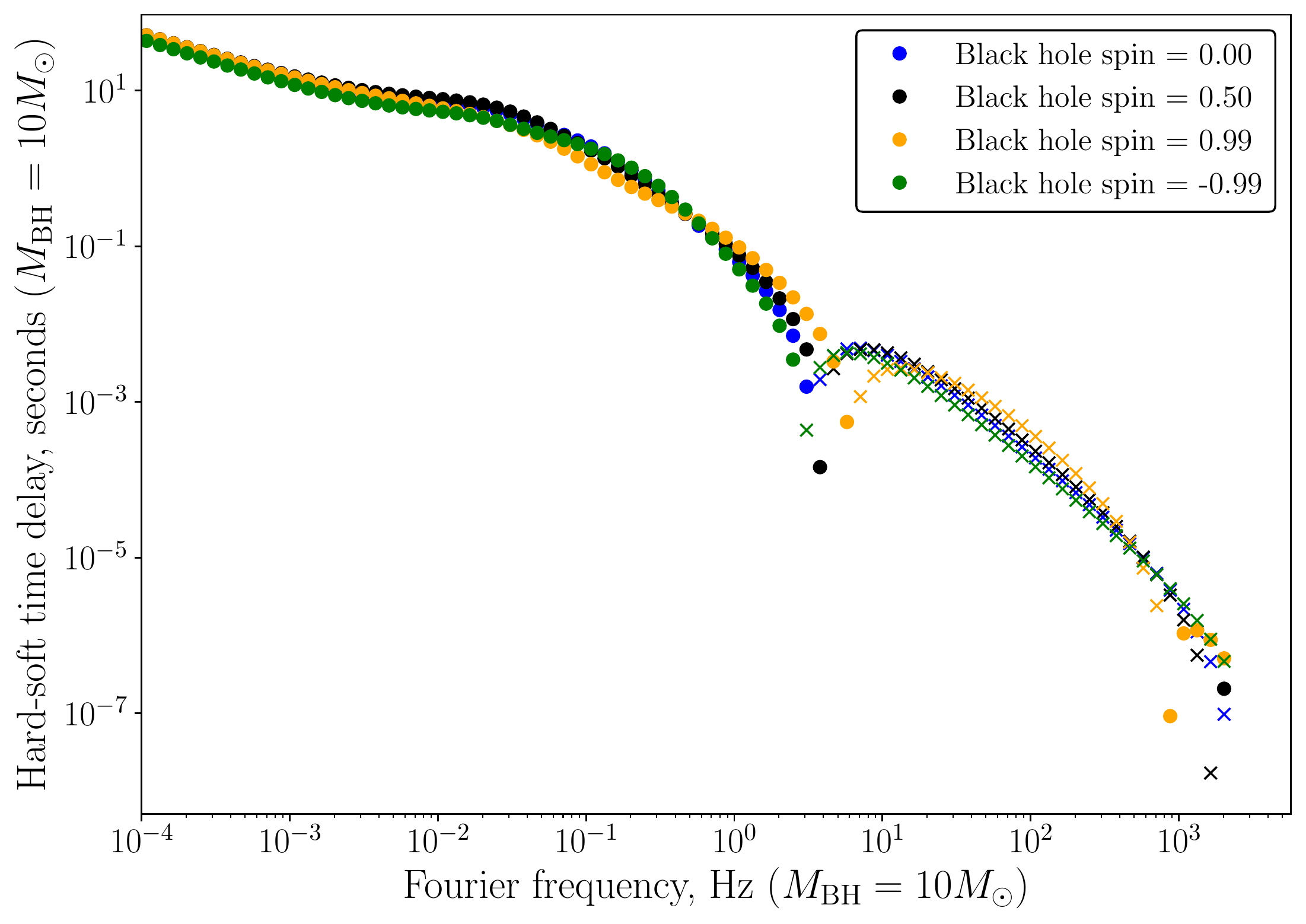}
\caption{The power density spectrum of the hard band  (upper left), the hard-soft coherence (upper right), the hard-soft phase lag (lower left) and time delay (lower right) for emissivity parameters $\gamma_s = 2$, $\gamma_h = 3.5$ and four different black hole spins ($a= 0$, blue; $a = 0.5$, black; $a = +0.99$, orange; and $a = -0.99$, green). For the time delay plot we denote by solid dots  positive lags (hard lags soft), and by  crosses negative lags (hard leads soft). The units of the power spectrum are arbitrary and would be set in a physical system by the input power in the surface density variability, the units of frequency and time delay are scaled to a system with black hole mass $M_{\rm BH} = 10M_\odot$, and disc parameters $\alpha = H/R = 0.1$.    }
\label{K2}
\end{figure*}

\begin{figure*}
\includegraphics[width=0.45\linewidth]{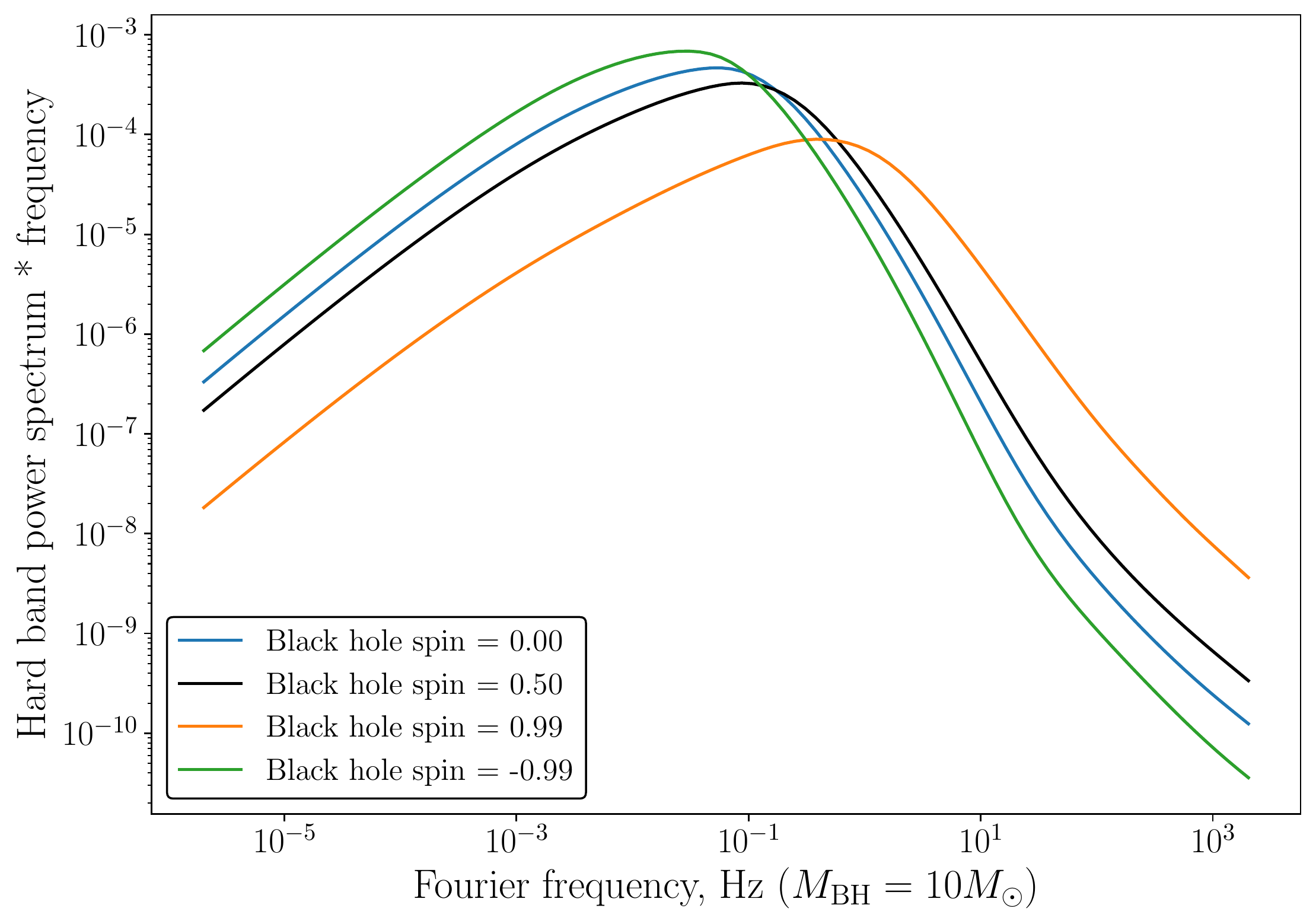}
\includegraphics[width=0.45\linewidth]{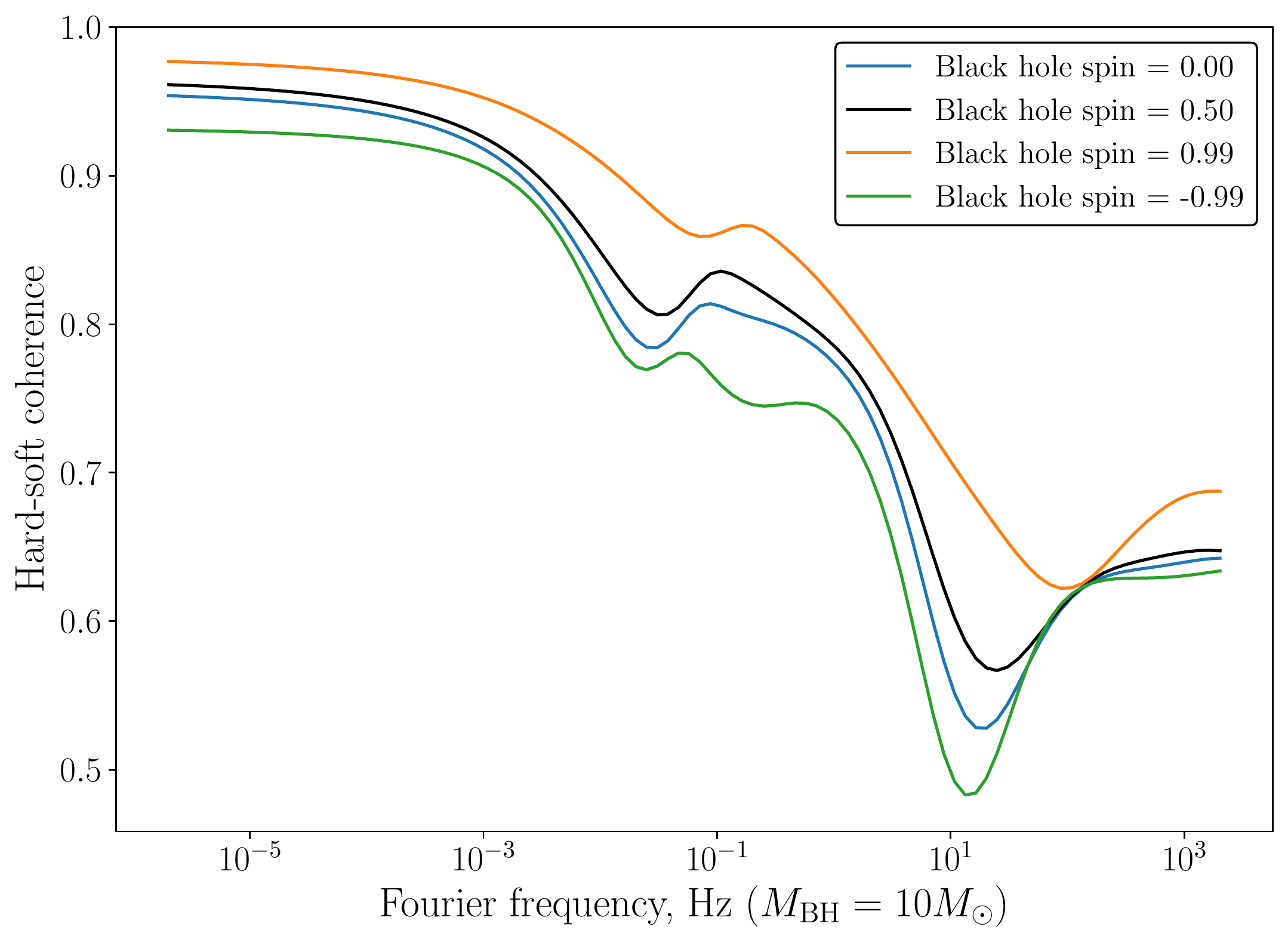}
\includegraphics[width=0.45\linewidth]{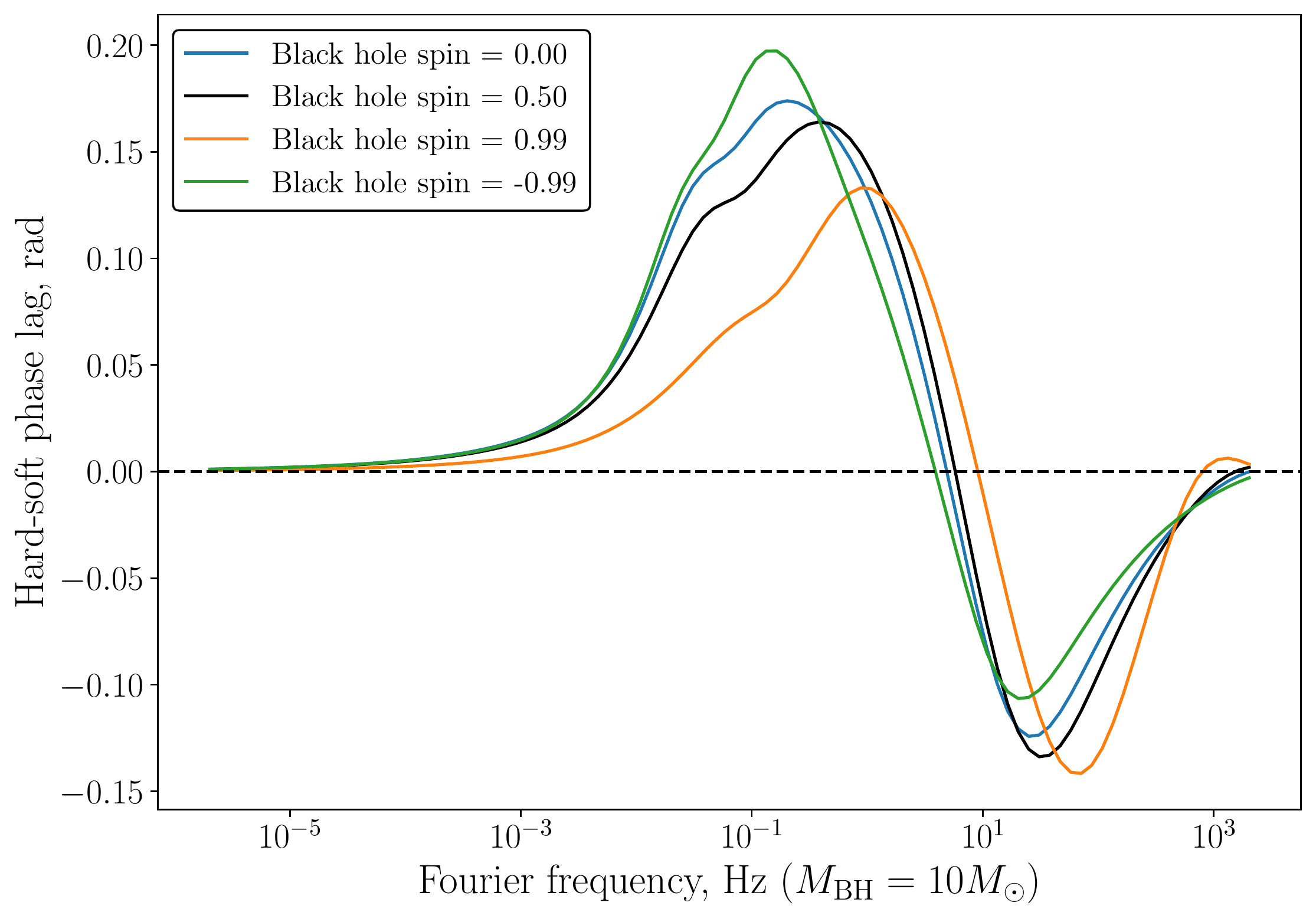}
\includegraphics[width=0.45\linewidth]{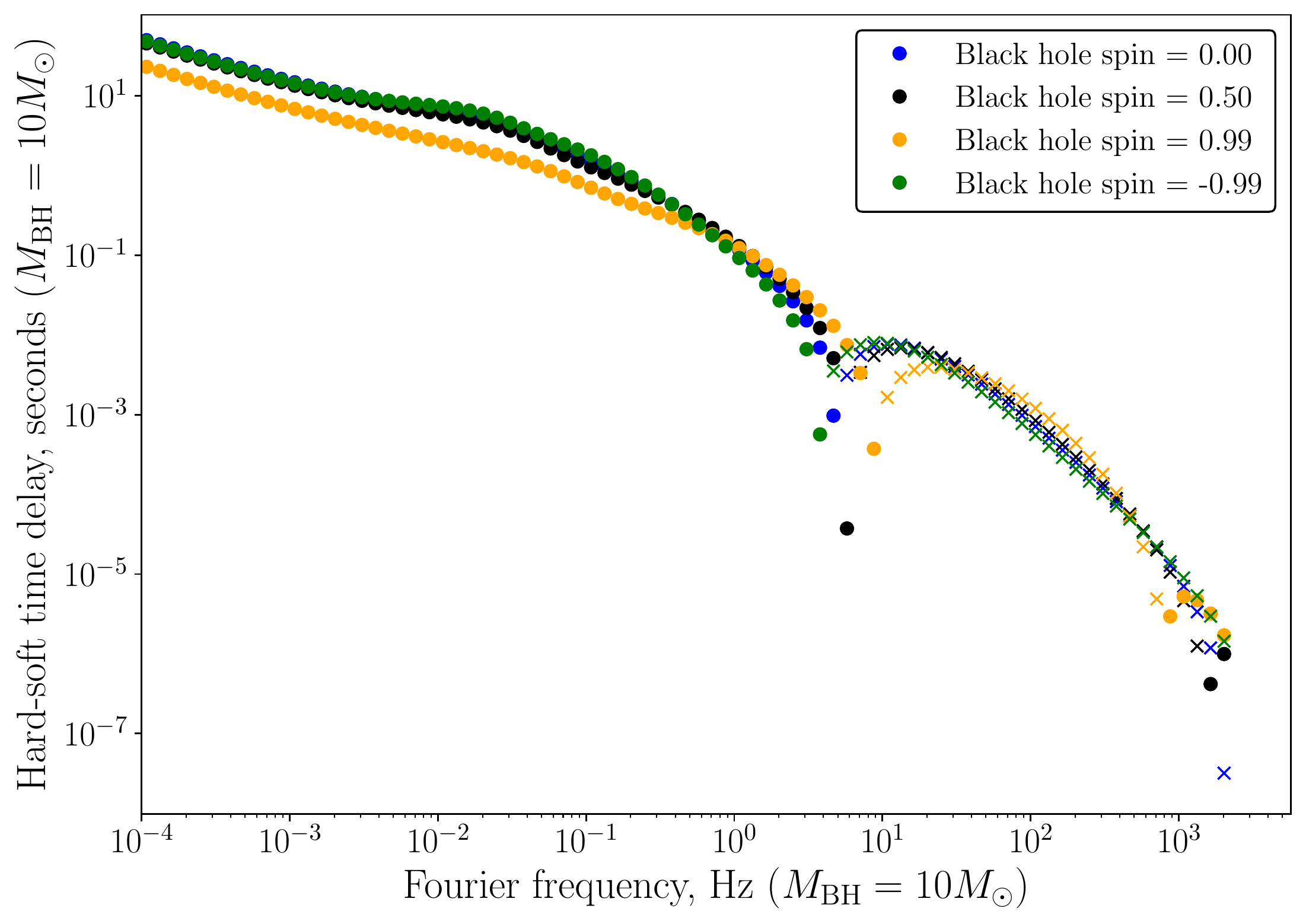}
\caption{The power density spectrum of the hard band (upper left), the hard-soft coherence (upper right), the hard-soft phase lag (lower left) and time delay (lower right) for emissivity parameters $\gamma_s = 3$, $\gamma_h = 8$ and four different black hole spins ($a= 0$, blue; $a = 0.5$, black; $a = +0.99$, orange; and $a = -0.99$, green). For the time delay plot we denote by solid dots  positive lags (hard lags soft), and by  crosses negative lags (hard leads soft). The units of the power spectrum are arbitrary and would be set in a physical system by the input power in the surface density variability, the units of frequency and time delay are scaled to a system with black hole mass $M_{\rm BH} = 10M_\odot$, and disc parameters $\alpha = H/R = 0.1$.   }
\label{K3}
\end{figure*}

\begin{figure*}
\includegraphics[width=0.45\linewidth]{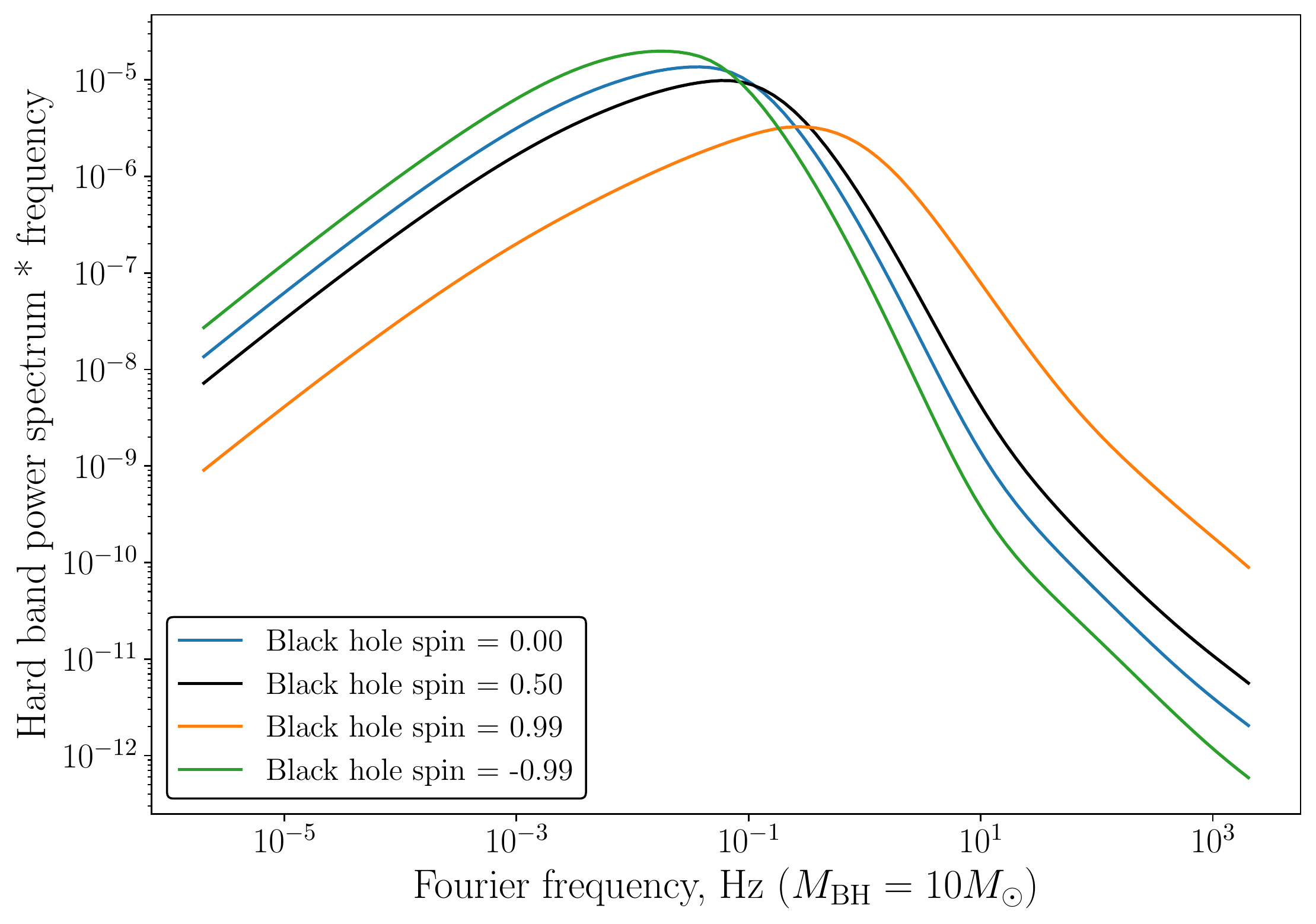}
\includegraphics[width=0.45\linewidth]{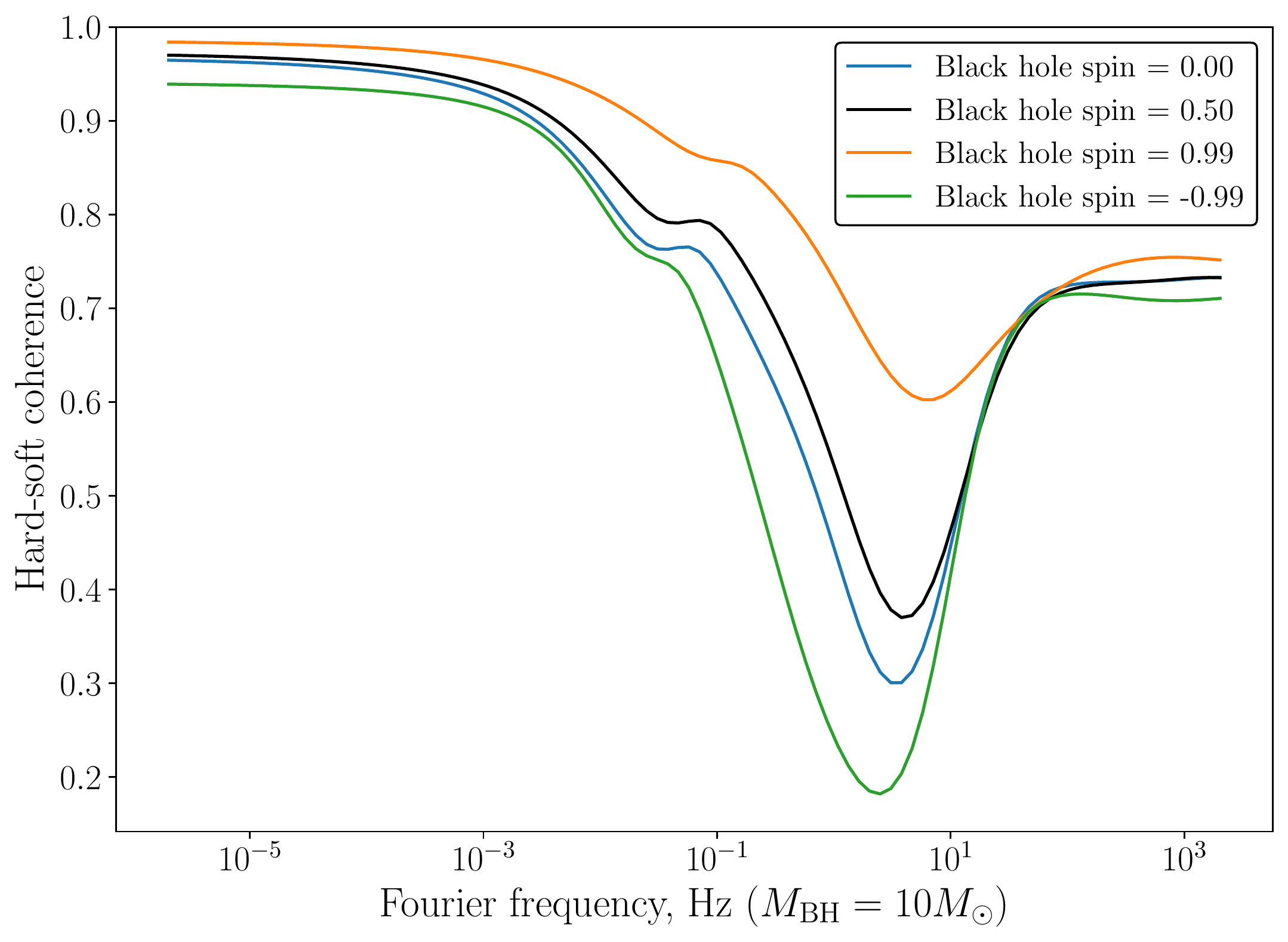}
\includegraphics[width=0.45\linewidth]{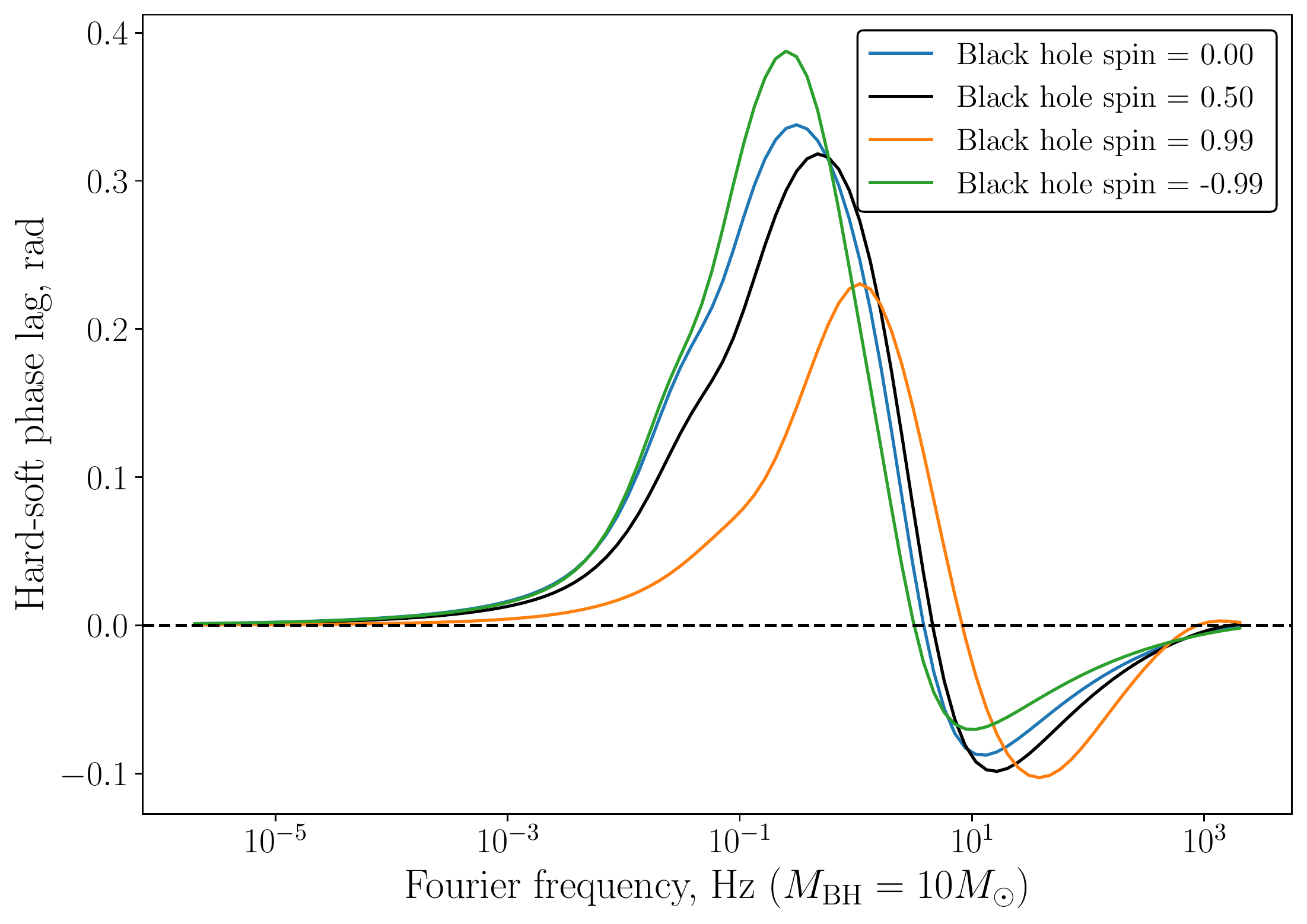}
\includegraphics[width=0.45\linewidth]{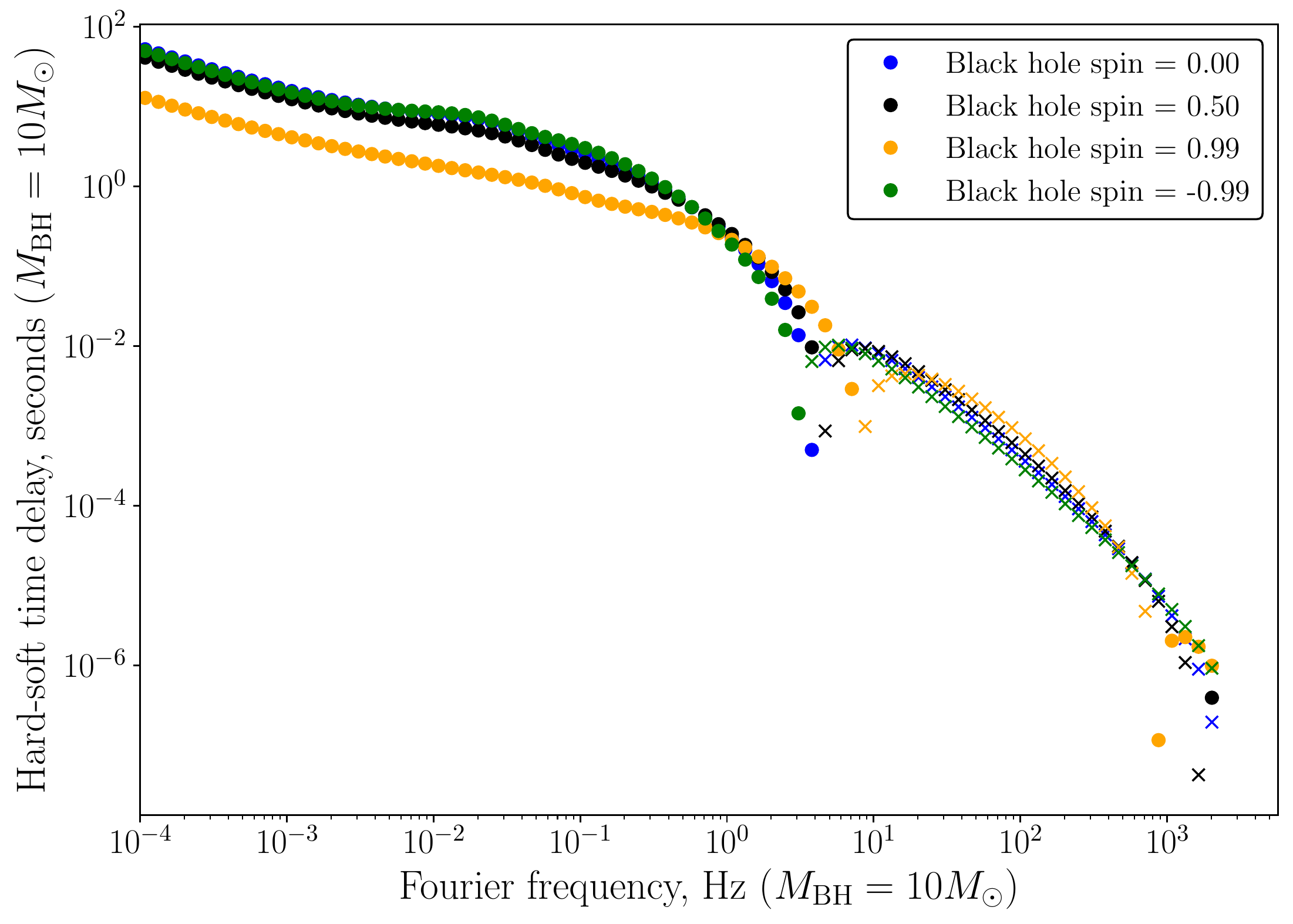}
\caption{The power density spectrum of the hard band  (upper left), the hard-soft coherence (upper right), the hard-soft phase lag (lower left) and time delay (lower right) for exponential emissivity profiles (see text), and four different black hole spins ($a= 0$, blue; $a = 0.5$, black; $a = +0.99$, orange; and $a = -0.99$, green). For the time delay plot we denote by solid dots  positive lags (hard lags soft), and by  crosses negative lags (hard leads soft). The units of the power spectrum are arbitrary and would be set in a physical system by the input power in the surface density variability, the units of frequency and time delay are scaled to a system with black hole mass $M_{\rm BH} = 10M_\odot$, and disc parameters $\alpha = H/R = 0.1$.   }
\label{K4}
\end{figure*}

\subsection{The cross spectrum between bands}
Observations of an accreting system may be contemporaneously taken at numerous different photon energies (or ``bands''). A natural observational question is then how strongly  is the variability in the emission observed across a ``soft'' band correlated with the variability in the emission observed across a ``hard'' band.  This correlation is quantified by the hard-soft cross spectrum, a quantity  given explicitly by (Mushtukov et al. 2018)
\beq
C_{h, s}(f) = \int_{\cal D} \int_{\cal D} h(r_1) s(r_2) C_{\dot m} (r_1, r_2, f)\, {\rm d}r_1 \, {\rm d}r_2 . 
\eeq
The physical cause of this triple disc integral is identical to that of the power density spectrum.  The hard-soft cross spectrum is a complex quantity, and its phase encapsulates an important physical quantity, namely the phase-lag in fluctuations in the hard sate emission with respect to the soft state emission. This phase lag is explicitly equal to 
\beq
\tan \Phi_{h,s}(f) = {{\cal I}[C_{h, s}(f) ] \over {\cal R}[C_{h, s}(f) ] } , 
\eeq
where ${\cal I}[z]$ and ${\cal R}[z]$ represent the real and imaginary parts of $z$ respectively. 
This quantity can also be equivalently  expressed as a time lag between hard and soft fluctuations:
\beq
t_{{\rm lag}\, h,s }(f) = {\Phi_{h,s}(f) \over 2\pi f} .
\eeq
\subsection{The coherence between bands}
The final observable quantity we shall discuss in this paper is the coherence of the fluctuations in hard and soft observing bands, a quantity explicitly given by  
\beq
{\rm Coh}_{h, s}(f) = {|C_{h, s}(f)|^2 \over S_h(f) S_s(f)} .
\eeq
The coherence function satisfies $0 \leq {\rm Coh}_{h, s} \leq 1$, where ${\rm Coh} = 1$ corresponds to a fully coherent fluctuations in both bands, while ${\rm Coh} = 0$ represents incoherent fluctuations.

\section{The back hole spin dependence of observable quantities }

{The relativistic Fourier-Green's functions derived in this paper are functions of the black holes spin through their explicit dependence on the ISCO radius of the black hole (eq. \ref{spindepfun}). This explicit spin dependence can be seen in Figures \ref{SD1}, \ref{SD2} and \ref{SD3}, where we plot the amplitude of the relativistic Fourier-Green's functions,  the phase and coherence of mass accretion rate fluctuations  for a number of different black hole spins. An interesting result seen in Figs. \ref{SD1} and \ref{SD2} is that the higher the black hole spin is made the closer to the Newtonian Fourier-Green's function (black dashed curves) the solution becomes. In Fig. \ref{SD3} we highlight that the black hole spin quantitatively effects the properties of the local mass accretion rate. }

In this section we demonstrate how this spin dependence is also present in the integrated observable properties of the photon flux emitted from the disc surface.  As an explicit example of this result, see Figure \ref{K1}, where we plot the  power density spectrum of the hard band (upper left), the hard-soft coherence (upper right), the hard-soft phase lag (lower left) and time delay (lower right) for four different black hole spins ($a= 0$, blue; $a = 0.5$, black; $a = +0.99$, orange; and $a = -0.99$, green).    For the time delay plot we denote by solid dots  positive lags (hard lags soft), and by  crosses negative lags (hard leads soft). The units of the power spectrum are arbitrary and would be set in a physical system by the input power in the surface density variability, the units of frequency and time delay are scaled to a system with black hole mass $M_{\rm BH} = 10M_\odot$, and disc parameters $\alpha = H/R = 0.1$.  To scale these results to any other system the frequency axis should be scaled by a factor 
\beq
N=\left({\alpha \over 0.1}\right) \left( {H/R \over 0.1}\right)^2 \left({10M_\odot \over M_{\rm BH}}\right) ,
\eeq 
while the time-lags should be scaled by $1/N$. 
This particular plot has emissivity parameters $\gamma_s = 3$, $\gamma_h = 5$. For the input surface density fluctuations we take a correlation length $l(r) = 1r_g$, an integrated power $p = 1$, and a break frequency equal to one percent of the local Keplerian frequency $f_K = \sqrt{GM/r^{3}}$ (see equation \ref{Lorentzian}).

It is clear from Fig. \ref{K1} that the black hole spin has a substantial effect on the observed variability properties of a black hole accretion flow. This is a result of fundamental theoretical importance, and the key observational result of this paper. 

As can be seen in the upper left panel of Fig. \ref{K1}, the power density spectra of the accretion variability predicted by this theory are qualitatively simple. At high and low Fourier frequencies the observed power density spectra are given by simple power-laws of frequency. The precise power-law indices of the high and low frequency slopes are determined by the input surface density variability profile, as discussed and derived in section 2. It is interesting to note that the power density spectrum of a given band peaks at higher Fourier frequencies for larger black hole spins, but generally with a smaller magnitude.  

The upper-right panel of Fig. \ref{K1} displays the coherence of the hard-soft variability. At low Fourier frequencies the coherence function tends to a near-unity constant for each black hole spin, but at larger Fourier frequencies the coherence is in general an extremely non-trivial function frequency. It is interesting to note that at roughly the frequency at which the phase lags turn from positive to negative (lower left panel) the hard-soft variability becomes increasingly incoherent. We note two properties of the coherence as a function of black hole spin: the hard-soft variability is both inherently more coherent (larger ${\rm Coh}_{h, s}$), and is inherently smoother as a function of Fourier frequency, for larger black hole spins. 

A simple interpretation of this spin dependence of the coherence is that the emission from Kerr discs with higher spins primarily originates from a region of the disc which has a smaller radial extent when compared to lower spins. This is because the emission from highly-spinning Kerr systems is dominated by the hottest and very innermost regions, which are physically close together. The coherence between two radii is a strong function of their separation, with larger separations having significantly lower coherence, and smaller separations a correspondingly larger coherence.

In the lower two panels we plot the phase and time lags associated with the hard and soft bands.  As has been discussed by many authors, it is natural that the variability in ``hard'' bands lag the variability in  ``soft''  bands (solid dots, Fig. \ref{K1}), a result of the fluctuations excited in the outer disc regions propagating inwards towards the hotter inner regions where the hard flux is generated. For these Fourier frequencies the time lags are roughly given by the typical time of propagation of fluctuations from outer radii, where the soft flux is predominantly produced, to inner radii, where the ``hard'' flux is predominantly produced. However, negative phase lags (when the soft energy band variability lags hard energy variability) are possible at high frequencies (see lower left and right panels of Fig. \ref{K1}). The negative lags are a result of the fact that the variability of the mass accretion rate at the inner radii can affect the variability at the outer radii through outward propagating modes (as argued first by Mushtukov et al. 2018). 

{These negative phase lags are only present at high Fourier frequencies, a result which can be understood with respect to the analytical analysis of section 2. At high Fourier frequencies the Fourier-Green's functions of inward and outward propagation are symmetric and therefore equally important. At low frequencies however outward propagation is suppressed (as a power-law in frequency), and inward propagation dominates resulting in positive lags.    }

We note that the predicted negative phase lags are comparable to the observed negative lags in stellar mass black hole systems (Uttley et al. 2011; De Marco et al. 2015) and AGN (Zoghbi et al. 2010; Walton et al. 2013; Alston et al. 2014). It is interesting that the fundamental diffusive accretion process in discs can in principle play a key role in the generation of negative time lags. 

The precise quantitative results of Fig. \ref{K1} of course depend on the precise assumptions inherent to the modelling, namely the emissivity indices $\gamma_s$ and $\gamma_h$. While varying these parameters changes the numerical values of each of the observed quantities (power spectrum, coherence and phase lags), the qualitative spin dependence of the different quantities remains unchanged. This is demonstrated in Figs. \ref{K2} and \ref{K3}  where we again plot the  power density spectrum of the hard band (upper left), the hard-soft coherence (upper right), the hard-soft phase lag (lower left) and time delay (lower right) for four different black hole spins ($a= 0$, blue; $a = 0.5$, black; $a = +0.99$, orange; and $a = -0.99$, green).  In Fig. \ref{K2} we take $\gamma_s = 2$, $\gamma_h = 3.5$, while in Fig. \ref{K3} we take $\gamma_s = 3$, $\gamma_h = 8$.  

The following ``rules of thumb'' appear to describe well the spin-dependence of the observed flux variability, independent of the choice of precise emissivity profile    
\begin{itemize}
\item{The magnitude of the maximum hard-soft phase lag decreases with increasing black hole spin  }
\item{The magnitude of the maximum {\it negative} hard-soft phase lag increases with increasing black hole spin  }
\item{ The frequency at which the maximum {\it negative} hard-soft phase lag occurs increases with increasing black hole spin}
\item{ The power density spectrum of a given band peaks at higher Fourier frequencies for larger black hole spins } 
\item{ The power density spectrum of a given band peaks with smaller magnitude for larger black hole spins } 
\item{ The hard-soft variability is inherently more coherent (larger ${\rm Coh}_{h, s}$) for larger black hole spins }
\item{ The hard-soft coherence is inherently smoother as a function of Fourier frequency for larger black hole spins }
\end{itemize}

{The reason that the frequencies at which various key observational properties occur increase with black hole spin is simply as a result of the reducing ISCO radius of the more rapidly rotating Kerr spacetime. These smaller radii have associated with them larger orbital and accretion frequencies, which are then observable  in the variability signatures of these systems.  }

These rules of thumb are robust, and are not even dependent on the chosen functional form of the emissivity profiles. In Fig. \ref{K4} we once again plot  the  power density spectrum of the hard band (upper left), the hard-soft coherence (upper right), the hard-soft phase lag (lower left) and time delay (lower right) for four different black hole spins ($a= 0$, blue; $a = 0.5$, black; $a = +0.99$, orange; and $a = -0.99$, green). In this Figure however we take emissivity profiles given by 
\begin{align}
s(r) &= s_0 \exp \left(-{r \over r_I}\right)  \left({1 - \sqrt{r_I \over r} } \right) ,\\
h(r) &= h_0 \exp \left(- 3 {r \over r_I}\right)  \left({1 - \sqrt{r_I \over r} } \right) .
\end{align}

This could in fact be a more physically reasonable profile, as it may more accurately describe the suppression of hard and soft emission by the Wien-tail of the local blackbody disc emission function $F(E) \propto \exp(-E_{h,s}/kT), E_h > E_s$.  In Figure \ref{K4} we once again see that the precise numerical values of each of the observable quantities depends on the precise emissivity profile chosen, but that the gross spin-dependence of the variability is qualitatively unchanged.

\section{Future extensions to the model }
{The relativistic variability model presented in the previous section employed a number of simplifications. In this section we recap and discuss these simplifications, their physical basis, and how they may be improved upon in future work.  

The principal simplification employed in this analysis is in relating the variability in the mass accretion rate to the variability in the observed photon field. As we have discussed,  in this work we employ phenomenological emissivity profiles.    The advantage of working with emissivity profiles, as opposed to any particular physical model for the emission is that, provided that both thermal and coronal emission scale locally with the mass accretion rate, variability in both thermal and non-thermal emission components may be modelled with the additional degrees of freedom provided by the emissivity profiles. 

However, a more detailed treatment of the local emission is of interest, and we briefly discuss a possible modelling approach below.  Assuming that the disc emits thermally with an effective temperature profile $T_{\rm eff}(r)$,  the flux at an observed energy $E$ is proportional to 
\beq
F \propto \int_{\cal D} {2 \pi r \over \exp\left(E / k T_{\rm eff}(r)\right) -1 } \, {\rm d} r, 
\eeq
where this expression neglects the effects of gravitational lensing, gravitational redshifts and the Doppler boosting of radiation.  The variable flux $\delta F$ is given, assuming a small fluctuation in $\dot M$ and to linear order, by 
\beq\label{deltaF}
\delta F \propto \int_{\cal D} {\delta \dot M \over \dot M} \left({E \over kT_{\rm eff}(r) }\right) {2 \pi r  \exp\left(E / k T_{\rm eff}(r)\right) \over \left(\exp\left(E / k T_{\rm eff}(r)\right) -1 \right)^2} \, {\rm d} r, 
\eeq
where we have assumed that $\delta T/T = \delta \dot M / 4\dot M$. Therefore, to linear order, the thermal flux from an accretion flow can be treated by the same methods developed here, with an ``emissivity profile'' given by eq. \ref{deltaF}.    The treatment of non-thermal radiation would be more complex, but a variable input thermal flux of the form given above could in principle be propagated through a Comptonising electron population. 

At this level of detail however additional relativistic effects will likely become important. This would include energy-dependent delays in the propagation of photons through the spacetime of Kerr black holes, with photons emitted deeper in the gravitational well of the black hole following paths to the observer more severely warped by gravity. In addition, the effects of gravitational and Doppler energy shifting of photons will likely modify the results presented here, at the quantitative level.   These higher order effects will all themselves be sensitive to the black hole's spin parameter, and a more detailed treatment of photon propagation and its effects on observed variability are of real interest.  

A further improvement of the analysis presented here would be in improving the treating of the input surface density fluctuations. In this work we have assumed that the   input surface density perturbations are described by a Lorentzian profile with phenomenological parameters $p$ and $f_{\rm br}$ (eq. \ref{Lorentzian}). While physically motivated, it would be of interest in future works to calibrate this model input with numerical analyses of the fundamental disc equations (e.g. Hogg and Reynolds 2016, Turner and Reynolds 2021).      }

\section{Conclusions} 
In this paper we have presented two important advances in the theoretical framework for describing aperiodic variability from accreting sources (the so-called theory of propagating fluctuations).  First, we present the exact analytical solutions of the Fourier integral of the Green's functions of the classical thin disc equations. With analytical solutions now at hand, various asymptotic properties of these solutions may be derived. In section 2 we demonstrated that high frequency variability in the mass accretion rate is suppressed as $\exp(-\Delta f^{1/2})$, where $\Delta(x, x')$ is a function of the magnitude of the difference between the two disc locations $x$ and $x'${, and corresponds physically to the (square root of) the accretion propagation time between $x$ and $x'$}. The high and low frequency asymptotic behaviour of the power spectrum of variability are also determined, and related to the intrinsic variability in the disc surface density/alpha parameter. 

{We have demonstrated that the power spectrum of the local mass accretion rate spectrum is, at high Fourier frequencies, dominated by  locally  driven variability, with exponentially small contributions from distant disc regions. At high Fourier frequencies the inward and outward propagation of material are equally important, with the Fourier-Green's functions symmetric in $x-x'$ in this limit.  At low Fourier frequencies however the variability is dominated by perturbations sourced at radii which are more distant from the central object, which then propagate inwards. Outward propagation is suppressed at low frequencies, as a power law in frequency.    }

In addition, these exact solutions will rapidly speed up the process of fitting analytical models of accretion variability to observational data;  the numerical cost of Fourier transforming thin disc Green's functions had previously been substantial. 

The second key development is in presenting the first analysis of the Fourier-Green's function solutions of the general relativistic thin disc equation. In this paper we have presented the Fourier-Green's function solutions valid for in a relativistic theory of gravity, under the assumption that the dynamical disc stress vanishes at the ISCO.  These solutions depend implicitly on the central black hole's spin through their dependence on the spacetime's ISCO radius. 

 We use this new theoretical development to highlight the Kerr black hole spin dependence of a number of observable variability properties of black hole discs.  The  power density spectrum of the hard band (upper left), the hard-soft coherence (upper right), the hard-soft phase lag (lower left) and time delay (lower right) are displayed in Figures \ref{K1}, \ref{K2}, \ref{K3} and \ref{K4} for four different black hole spins ($a= 0$, blue; $a = 0.5$, black; $a = +0.99$, orange; and $a = -0.99$, green), and a number of different parameterisations of the disc emissivity.  Clearly the black hole spin imparts a strong signal onto the observable variability properties of black hole disc systems. 

While the precise choice of emissivity profile of the hard and soft bands quantitatively effect the system's observed variability properties, the following ``rules of thumb'' appear to describe well the spin-dependence of the observed flux variability, independent of the choice of precise emissivity profile. These rules of thumb may be of use even in systems where a detailed analysis of the variability is not performed. 
\begin{itemize}
\item{The magnitude of the maximum hard-soft phase lag decreases with increasing black hole spin  }
\item{The magnitude of the maximum {\it negative} hard-soft phase lag increases with increasing black hole spin  }
\item{ The frequency at which the maximum {\it negative} hard-soft phase lag occurs increases with increasing black hole spin}
\item{ The power density spectrum of a given band peaks at higher Fourier frequencies for larger black hole spins } 
\item{ The power density spectrum of a given band peaks with smaller magnitude for larger black hole spins } 
\item{ The hard-soft variability is inherently more coherent (larger ${\rm Coh}_{h, s}$) for larger black hole spins }
\item{ The hard-soft coherence is inherently smoother as a function of Fourier frequency for larger black hole spins }
\end{itemize}

The results presented in this paper therefore open up the  possibility of using the aperiodic variability observed from black hole accretion systems to constrain the central black hole's spin, a parameter of fundamental observational and theoretical interest.

\section*{Acknowledgments} 
I would like to thank  Alexander Mushtukov for interesting discussions which initiated this work.  I am particularly grateful to Adam Ingram for extremely illuminating   discussions regarding the propagating fluctuation model. I am grateful to the reviewer, whose detailed report strengthened the analysis in a number of places.  This work was supported by a Leverhulme Trust International Professorship grant [number LIP- 202-014]. For the purpose of Open Access, I have applied a CC BY public copyright licence to any Author Accepted Manuscript version arising from this submission. 

\section*{Data accessibility statement}
No  observational data was used in producing this manuscript. Python scripts which compute the relativistic Fourier-Green functions and make Figures similar to those  in section 6, are available at {\url{https://github.com/andymummeryastro/GR_prop_fluc}}.

\appendix 
\section{The exact form of the relativistic Fourier-Green's functions }\label{full_GR}
The solution of the mass accretion rate Fourier-Green's integral has the following general form 
\beq\label{general_def_app}
{1\over 2} \widetilde  G_{\dot M}=  p(x) 
\begin{cases}
&  K_\nu(\beta g(x_0)) \,  {\partial_x} \left[ q(x) I_\nu(\beta g(x))\right] , \quad x< x_0, \\
\\
& I_\nu(\beta g(x_0)) \,  {\partial_x} \left[ q(x)  K_\nu(\beta g(x)) \right] , \quad x> x_0 ,
\end{cases}
\eeq  
where 
\beq
\beta \equiv (1 + i) \sqrt{\pi f}.
\eeq
The Mummery (2023) Green's function solution is fully described by (in units where $G = c = w = 1$)
\begin{multline}
g(x) = {x^\alpha \over 2 \alpha} \sqrt{1 - {2\over x}}\left[1  - {x^{ - 1} \over { (\alpha - 1)}} {}_2F_1\left(1, {3\over 2}-\alpha; 2-\alpha; {2\over x}\right) \right] \\ + {2^{\alpha - 2} \over \alpha (\alpha - 1)}\sqrt{\pi} {\Gamma(2-\alpha)  \over  \Gamma({3/ 2} - \alpha)} , 
\end{multline}
\beq
q(x) = x^{1/4} \sqrt{x^{-\alpha} g(x)} \exp\left({1 \over 2 x}\right) \left[1 - {2\over x}\right]^{5/4 - 3/8\alpha} ,
\eeq
and 
\beq
p(x) = {x^{1/2} \exp\left(-{1/ x}\right) \over 1 - {2/ x}} ,
\eeq
where 
\beq
x \equiv {2 r \over r_I}, \quad \alpha = {1\over 4\nu} , 
\eeq 
and ${}_2F_1(a, b; c; z)$ is the hypergeometric function. For the purposes of taking the derivative in the above expression it will be helpful to note that (Mummery 2023)  
\beq
g(x) = {1\over 2} \int_2^{x} x'^{\alpha - 1} \sqrt{1 - {2\over x'}} \, {\rm d}x' ,
\eeq
and so 
\beq
{\partial g \over \partial x} = {1\over 2}  x^{\alpha - 1} \sqrt{1 - {2\over x}} .
\eeq
Expanding in full we have 
\beq
{1\over 2}  \widetilde G_{\dot M} = p(x) {\cal A}_\nu {\cal B}_\nu {\partial q \over \partial x} + p(x) q(x) {\cal A}_\nu {\partial B_\nu \over \partial x},
\eeq
where for notational ease we define 
\beq
{\cal A}_\nu \equiv 
\begin{cases}
&  K_\nu , \quad x< x_0, \\
\\
& I_\nu , \quad x> x_0 ,
\end{cases}
\eeq
and 
\beq
{\cal B}_\nu \equiv 
\begin{cases}
&  I_\nu , \quad x< x_0, \\
\\
& K_\nu , \quad x> x_0 .
\end{cases}
\eeq
Therefore 
\beq
{1\over 2 p(x) q(x) }  \widetilde G_{\dot M} =  {\cal A}_\nu {\cal B}_\nu {\partial \ln q \over \partial x} + {\cal A}_\nu {\partial B_\nu \over \partial x} .
\eeq
The Bessel derivative is simplified by noting 
\beq
{\partial B_\nu \over \partial x}  = \beta {\partial g \over \partial x} {{\rm d} {\cal B}_\nu(z) \over {\rm d} z}, \quad z \equiv \beta g(x) . 
\eeq
The following  identities 
\beq
{{\rm d} \over {\rm d}z} I_l(z) =  I_{l-1}(z) - {l \over z} I_l(z),
\eeq
and 
\beq
{{\rm d} \over {\rm d}z} K_l(z) =  - K_{l-1}(z) - {l \over z} K_l(z) ,
\eeq
suffice to give the quantity ${{\rm d} {\cal B}_\nu/ {\rm d} z}$. Combining 
\begin{align}
{\partial g \over \partial x} &= {1\over 2}  x^{\alpha - 1} \sqrt{1 - {2\over x}}, \\ 
{\partial \ln q \over \partial x} &=  {1 - 2\alpha \over 4x} - {1\over 2 x^2} + {10\alpha - 3 \over 4 \alpha x^2 } {1 \over 1 - 2/x} + {1 \over 2g} {\partial g\over \partial x}, 
\end{align}
with 
\beq
 {{\rm d} {\cal B}_\nu \over {\rm d} z} =  \begin{cases}
& + I_{\nu-1} - {\nu \over \beta g} I_\nu  , \quad x< x_0, \\
\\
& -K_{\nu -1} - {\nu \over \beta g} K_\nu , \quad x> x_0 ,
\end{cases}
\eeq
gives 
\begin{equation}
\widetilde G_{\dot M} = 2 p(x) q(x) \left[ {\cal C}_\nu {\partial \ln q \over \partial x} + \beta {\partial g\over \partial x}  {\cal C}_\nu ' \right],
\end{equation}
where 
\beq
{\cal C}_\nu \equiv {\cal A}_\nu {\cal B}_\nu , \quad {\cal C}_\nu ' \equiv {\cal A}_\nu  {{\rm d} {\cal B}_\nu \over {\rm d} z} . 
\eeq
As a final step we must now re-insert the dimensionful quantities $M_d, M_{\rm BH}, G, c$, etc., so that these results may be fit to observational  data. The physical constraint that 
\beq
\lim_{f \to 0} \widetilde G_{\dot M} (x < x_0, f) \to -M_d ,
\eeq
where $M_d$ is the mass content of the perturbation, simplifies this procedure greatly. Taking this limit, and using the fact that 
\beq
\lim_{\beta \rightarrow 0} I_\nu(\beta g) = {1 \over \Gamma(\nu+1)} \left({\beta g \over 2}\right)^{+\nu} + {\cal O}((\beta g)^{2+\nu}) ,
\eeq
and 
\beq
\lim_{\beta \rightarrow 0} K_\nu(\beta g) = {\Gamma(\nu) \over 2} \left({\beta g \over 2}\right)^{-\nu} +  {\Gamma(-\nu) \over 2} \left({\beta g \over 2}\right)^{+\nu} + {\cal O}((\beta g )^{2-\nu}) ,
\eeq
gives the normalisation constant
\beq
N_1 = -{M_d \over 4 p(x)} \left[\nu \left({g(x_0) \over g(x)}\right)^\nu  {1 \over \partial_x q(x) + \nu q(x)/g(x) }   x^{\alpha - 1} \sqrt{1 - {2\over x}}  \right] ,
\eeq
which must be multiplied to the above Green's function. Finally, the combination $\beta g(x)$ within the Bessel functions must be dimensionless, and thus the  substitution 
\beq
\beta \to N_2 \beta ,
\eeq
must be made. The constant $N_2$ is given by 
\beq
N_2 = {1 \over \sqrt{f_0}} = {2\sqrt{2} \nu} \left({GM_{\rm BH} r_I^3 \over w^2}\right)^{1/4}    \sqrt{1-{2 \over x_0}}  x_0^{-\mu/2} \exp\left(-{1\over x_0}\right),
\eeq
which defines a fiducial viscous frequency associated with the Green's function, $f_0$.

\subsection{The low-frequency correction factor $\delta(x, x_0)$} 
As discussed in section 4 of this paper, at very low frequencies $f/f_0 \ll 10^{-6}$ the analytical solutions of this paper differ from the numerical Fourier-Green's solutions of the relativistic disc equations for outward propagating modes.   Mathematically this results from the fact that the solutions derived in Mummery (2023) are not exact, but are asymptotic ``leading order'' solutions (see Mummery (2023) for a detailed discussion). Physically care is required because at very late times the exact numerical and analytical solution begin to deviate, and an extremely small fraction of the initial disc mass is not accreted in these solutions.  As a result 
\beq
\lim_{f \to 0} \widetilde G_{\dot M}(x > x_0, f)  \to \delta(x, x_0) M_d \neq 0 .
\eeq
The discrepancy is  small, as a result of the high accuracy of these analytical solutions (Fig. \ref{GFR}), and typically only effects extremely small frequencies $f/f_0 \ll 10^{-6}$ to a small degree
\beq
\delta \ll 10^{-4}. 
\eeq
The function $\delta(x, x_0)$ can be written in closed form, and is equal to 
\beq
\delta(x, x_0) = \left({g(x_0) \over g(x)}\right)^{2\nu} \left[{2g(x) \partial_x \ln q(x) - \nu x^{\alpha-1}\sqrt{1-2/x} \over 2g(x) \partial_x \ln q(x)  + \nu x^{\alpha-1}\sqrt{1-2/x} } .\right]
\eeq
The discrepancy can then be removed by taking, for example
\beq
\widetilde G_{\dot M}(x > x_0, f) \to \widetilde G_{\dot M}(x > x_0, f) - \delta(x, x_0) \exp\left(-100 \left({f \over f_0}\right) \right),
\eeq
which removes any discrepancies for $ f \ll f_0$ but does not effect the moderate-to-high frequency behaviour of the solutions.  

\section{Stressed Newtonian Fourier-Green's functions}\label{secNS}
In a Newtonian disc system with a central stress (torque at $r = 0$), the Green's function solution of the disc equations is 
\beq
G(x, x_0, t) = {q(x) \over t} \exp\left({- g(x)^2 - g(x_0)^2 \over 4 t}\right) I_{-\nu} \left({g(x)g(x_0)\over 2 t} \right) ,
\eeq
where it is important to note that the index on the Bessel function is now negative. This solution is in once again more usefully written as a Laplace-mode superposition, of the form (Gradshteyn and Ryzhik et al. 2007)
\beq
G(x, x_0, t) = \int_0^\infty q(x) J_{-\nu}(\sqrt{s} g(x)) J_{-\nu}(\sqrt{s} g(x_0)) \exp({-st}) \, {\rm d}s ,
\eeq
and therefore every step of the derivation performed in section 2 still holds. The Fourier-Green's function for a stressed disc is then 
\beq
\widetilde G(x, x_0, f) = 2 q(x) 
\begin{cases}
& I_{-\nu}(\beta g(x)) K_{-\nu}(\beta g(x_0)) , \quad x< x_0, \\
\\
& I_{-\nu}(\beta g(x_0)) K_{-\nu}(\beta g(x)) , \quad x> x_0 .
\end{cases}
\eeq  
Finally, the derivative with respect to $x$ of this function must be taken, and then the result multiplied by $p(x) \propto x^{1/2}$. We  again find a  remarkably simple result
\beq
\widetilde G_{\dot M} = 
\begin{cases}
+ A \beta x^{(1-\nu)/4\nu} K_{-\nu}(\beta \epsilon x_0^{1/4\nu}) I_{1-\nu}(\beta \epsilon x^{1/4\nu}), \quad  x < x_0, \\
\\
 - A \beta x^{(1-\nu)/4\nu} I_{-\nu}(\beta \epsilon x_0^{1/4\nu}) K_{1-\nu}(\beta \epsilon x^{1/4\nu}), \quad x > x_0 , 
\end{cases}
\eeq
where (units where $G = c = w = 1$)
\beq
A =  {x_0^{1/4}   \epsilon M_d }, \quad \epsilon = 2\nu \sqrt{2x_0^\mu} , \quad \mu = {3-1/\nu \over 2} ,
\eeq
and we remind the reader
\beq
\beta \equiv (1 + i) \sqrt{\pi f}.
\eeq
Note that for the case of an inner disc stress the low-frequency limits of the mass accretion rate Fourier-Green's functions are effectively reversed from the non-stressed case, and we find 
\beq
\lim_{f \to 0} \widetilde G_{\dot M}(x, x_0, f) = 
\begin{cases}
f^A, \quad \, \,\,\,\, x < x_0, \quad A > 0, \\
\\
{\rm const}, \quad x > x_0. 
\end{cases}
\eeq
In other words, all of the disc material is eventually expelled to infinity, with none of the matter crossing the boundary at $x = 0$. 

\label{lastpage}
\end{document}